\let\OriginalKernelLabel\label
\documentclass[%
twocolumn,
superscriptaddress,
amsmath,amssymb,
aps,
prd,
longbibliography,
floatfix,
]{revtex4-2}

\usepackage{graphicx}
\graphicspath{{./figures/}}
\usepackage{xcolor}
\usepackage[normalem]{ulem}
\usepackage{bm}
\usepackage[caption=false]{subfig}
\usepackage{microtype}
\allowdisplaybreaks

\usepackage{physics} 
\makeatletter
\IfFormatAtLeastTF{2023-06-01}{%
  \let\label\OriginalKernelLabel
  \let\ltx@label\label
}{}
\makeatother
\usepackage{hyperref}
\hypersetup{colorlinks, linkcolor = [rgb]{0, 0, 0.5}, citecolor = [rgb]{0,0.0,0.5}, urlcolor = [rgb]{0,0.0,0.5}}
\begin{document}

\title{Extraction of transverse-momentum-dependent helicity distributions of the nucleon} 

\newcommand*{\SDU}{Key Laboratory of Particle Physics and Particle Irradiation (MOE), Institute of Frontier and Interdisciplinary Science, Shandong University, Qingdao, Shandong 266237, China}
\newcommand*{\SCNT}{Southern Center for Nuclear-Science Theory (SCNT), Institute of Modern Physics, Chinese Academy of Sciences, Huizhou 516000, China}
\newcommand*{\PKU}{School of Physics, Peking University, Beijing 100871, China}
\newcommand*{\ZZU}{School of Physics, Zhengzhou University, Zhengzhou 450001, China}
\newcommand*{\CHEP}{Center for High Energy Physics, Peking University, Beijing 100871, China}
\newcommand*{\CICQM}{Collaborative Innovation Center of Quantum Matter, Beijing, China}
\newcommand*{\IMP}{Institute of Modern Physics, Chinese Academy of Sciences, Lanzhou, Gansu Province 730000, China}
\newcommand*{\UCAS}{University of Chinese Academy of Sciences, Beijing 100049, China}
\newcommand*{\KLQLPIPP}{Institute of Particle Physics, Central China Normal University, Wuhan 430079, China}
\newcommand*{\NNU}{Department of Physics and Institute of Theoretical Physics,Nanjing Normal University, Nanjing, Jiangsu 210023, China}

\author{Ke~Yang}\email{yangke2025@sdu.edu.cn}\affiliation{\SDU}
\author{Tianbo~Liu}\email{liutb@sdu.edu.cn}\affiliation{\SDU}\affiliation{\SCNT}
\author{Bo-Qiang~Ma}\email{mabq@pku.edu.cn}\affiliation{\ZZU}\affiliation{\PKU}
\author{Peng~Sun}\email{pengsun@impcas.ac.cn}\affiliation{\IMP}\affiliation{\UCAS}
\author{Yuxiang~Zhao}\email{yxzhao@impcas.ac.cn}\affiliation{\SCNT}\affiliation{\IMP}\affiliation{\UCAS}\affiliation{\KLQLPIPP}
\author{Yiyu~Zhou}\email{zyy@impcas.ac.cn}\affiliation{\IMP}\affiliation{\UCAS}

\collaboration{Transverse Nucleon Tomography Collaboration}
\noaffiliation

\begin{abstract}
We present an updated global analysis of the transverse-momentum-dependent
(TMD) helicity distributions of the nucleon from measurements of the
longitudinal double-spin asymmetry $A_{LL}$ in semi-inclusive deep-inelastic
scattering (SIDIS). The analysis incorporates target- and
detected-hadron-mass corrections, next-to-next-to-leading-order matching
in the small-$b_T$ operator product expansion, and an improved treatment of
correlated systematic uncertainties.
The present data constrain the up- and down-quark helicity TMDs,
whereas the sea-quark and gluon distributions remain weakly constrained.
After integration over transverse momenta up to the hard scale,
the up- and down-quark distributions agree with collinear helicity
distributions within uncertainties.
Using the extracted TMDs and the Wandzura-Wilczek-type approximation,
we predict the $\cos\phi_h$ modulation of $A_{LL}$ and find agreement
with the HERMES and CLAS measurements.
We also provide predictions for SIDIS $A_{LL}$ at CLAS12 and for
longitudinal single-spin asymmetries in $W^\pm$ production with
leptonic decays at STAR kinematics.
An impact study with pseudodata for the proposed
Electron-Ion Collider in China (EicC) indicates substantial potential to improve constraints on helicity TMDs at intermediate $x$, including the presently
weakly constrained sea-quark sector.
\end{abstract}

\maketitle


\section{Introduction}
The three-dimensional internal structure of the nucleon remains a central focus of quantum chromodynamics (QCD) and hadronic physics. Considerable progress has been made in determining the longitudinal momentum distributions of quarks and gluons through collinear parton distribution functions (PDFs)~\cite{deFlorian:2008mr,Ball:2013lla,deFlorian:2014yva,Nocera:2014gqa,Harland-Lang:2014zoa,Buckley:2014ana,H1:2015ubc,Dulat:2015mca,Butterworth:2015oua,NNPDF:2017mvq,Cocuzza:2021cbi,Borsa:2024mss,Bertone:2024taw}. These distributions do not resolve the transverse motion of partons. Transverse-momentum-dependent distributions (TMDs) extend this description by resolving correlations between partonic transverse momentum and spin, providing insight into spin-orbit dynamics and their connections to partonic orbital angular momentum~\cite{Brodsky:2002cx,Bacchetta:2008af,She:2009jq,Avakian:2010br,Aidala:2012mv,Deur:2018roz}.

The TMD helicity distribution $g_{1L}(x,k_T^2)$ describes, at longitudinal momentum fraction $x$ and transverse momentum $k_T$, the difference between the number densities of quarks polarized parallel and antiparallel to the longitudinal polarization of the parent nucleon~\cite{Boussarie:2023izj,Collins:2011zzd}. Its transverse-momentum dependence complements the collinear helicity PDF, while the ratio $g_{1L}/f_1$ characterizes the quark polarization relative to the unpolarized density. A qualitative motivation for studying this dependence comes from relativistic spin kinematics: in light-front quark models, Wigner rotations associated with boosts can suppress the longitudinal polarization of quarks with nonzero transverse momentum~\cite{Wigner:1939cj,Melosh:1974cu,Ma:1991xq,Ma:1992sj}. Experimentally, helicity TMDs can be accessed through the transverse-momentum dependence of the longitudinal double-spin asymmetry $A_{LL}$ in semi-inclusive deep-inelastic scattering (SIDIS), where $g_{1L}$ enters the polarized structure function through a convolution with an unpolarized TMD fragmentation function (FF). Measurements with different targets and identified hadrons therefore provide information on both the transverse-momentum dependence and the flavor structure of quark polarization.

Global TMD analyses have advanced substantially over the past decade. Unpolarized TMD PDFs and FFs have been determined through fits to Drell-Yan and SIDIS data~\cite{Scimemi:2019cmh,Bacchetta:2017gcc,Bertone:2019nxa,Bacchetta:2019sam,Bacchetta:2022awv,Bacchetta:2024qre,Moos:2023yfa,Moos:2025sal,Barry:2025glq,Fernando:2025xzv,Zaccheddu:2026pyy,Fernando:2026elp}. The transverse-polarization sector, including the Sivers function, transversity, and the Collins effect, has also been studied intensively~\cite{Bury:2021sue,Zeng:2022lbo,Zeng:2023nnb,Zeng:2024gun,Fernando:2023obn}. Lattice QCD calculations are providing complementary first-principles information on TMD structure and evolution~\cite{Bollweg:2025ecn,LatticePartonCollaborationLPC:2022myp,Avkhadiev:2023poz,Yoon:2017qzo}. In the longitudinal-polarization sector, two independent phenomenological extractions of helicity TMDs from SIDIS data have recently been reported~\cite{Yang:2024drd,Bacchetta:2024yzl}.

In our preceding study~\cite{Yang:2024drd}, we performed the first global extraction of TMD helicity distributions using the HERMES~\cite{HERMES:2018awh} and CLAS~\cite{CLAS:2017yrm} SIDIS measurements. The extracted polarization rates exhibited a characteristic transverse-momentum dependence: the magnitude of quark polarization decreased with increasing $k_T$ in the valence-dominated region, qualitatively consistent with the Wigner-rotation expectation, whereas a contrasting trend was favored at smaller $x$. The latter behavior motivates further investigation of the dynamics involving sea quarks and gluons, whose contributions remain poorly constrained by existing measurements~\cite{HERMES:2018awh,COMPASS:2016klq,CLAS:2017yrm}. These findings call for a refined determination of what the present data establish, comparisons with complementary observables, and an assessment of the improvements achievable with future measurements.

In this work, we reanalyze the existing world SIDIS $A_{LL}$ data within an improved phenomenological framework. We incorporate kinematic corrections associated with the target and detected-hadron masses, upgrade the small-$b_T$ operator-product-expansion (OPE) matching for $f_1$, $g_{1L}$, and $D_1$ to next-to-next-to-leading order (NNLO), and retain next-to-next-to-leading logarithmic (NNLL) TMD evolution. We also refine the treatment of correlated systematic uncertainties by combining a penalty-term $\chi^2$ with the $T_0$ prescription to mitigate bias from multiplicative uncertainties. These refinements provide the basis for the updated extraction and the subsequent predictions and impact study. The present data constrain the up- and down-quark helicity TMDs, whereas the sea-quark and gluon distributions remain weakly constrained. The up- and down-quark distributions integrated over transverse momenta up to the hard scale agree with the corresponding collinear helicity distributions within uncertainties.

We examine the predictive implications of the extraction through complementary spin observables. Using the extracted helicity TMDs, we calculate the $\cos\phi_h$ azimuthal modulation of the longitudinal double-spin asymmetry in SIDIS within the Wandzura-Wilczek (WW)-type approximation~\cite{Avakian:2007mv,Accardi:2009au,Kanazawa:2015ajw,Scimemi:2018mmi}. The resulting predictions agree within uncertainties with the HERMES~\cite{HERMES:2018awh} and CLAS~\cite{CLAS:2017yrm} measurements that were not included in the fit, providing a consistency check of the description within this approximation. We also predict the azimuthally averaged SIDIS asymmetry $A_{LL}$ in charged-pion production at CLAS12. Its large-$x$ spin-structure program~\cite{Burkert:2008rj} and multidimensional SIDIS measurements with a $10.6\,\mathrm{GeV}$ electron beam~\cite{CLAS:2021opg} provide opportunities to examine the extracted transverse-momentum dependence in the valence region.

A complementary process is $W^\pm$ production in longitudinally polarized proton-proton collisions. The STAR measurements at the Relativistic Heavy Ion Collider (RHIC), performed at $\sqrt{s}=500$ and $510\,\mathrm{GeV}$, provide flavor-sensitive access to quark and antiquark helicities through leptonic $W$ decays~\cite{STAR:2010xwx,STAR:2018fty}. We use the SIDIS extraction to predict the charged-lepton-level single-spin asymmetries as functions of the $W$-boson transverse momentum at STAR kinematics, including the decay kinematics and lepton acceptance. These predictions extend the application of the extracted helicity TMDs to a different process at the electroweak scale and are compared with preliminary STAR measurements~\cite{Wang:2026nav} that are not included in the fit.

More precise flavor separation, particularly in the sea-quark sector, requires measurements with higher statistics and broader kinematic coverage. The proposed Electron-Ion Collider in China (EicC)~\cite{Anderle:2021wcy} is designed to collide polarized electron beams with polarized proton and $^3\mathrm{He}$ beams at center-of-mass energies of $15 \sim 20\,\mathrm{GeV}$ and high luminosity~\cite{Xiao:2026tbs}, with longitudinal polarization available for both beams~\cite{Du:2024sjt}. Its coverage of intermediate $x$, extending down to $x\simeq0.005$, complements the large-$x$ program at Jefferson Lab and the small-$x$, high-$Q^2$ reach of the future Electron-Ion Collider (EIC) at Brookhaven National Laboratory~\cite{Xiao:2026tbs,Huang:2025wdq}. High luminosity and polarization, together with measurements of identified hadrons on different polarized targets, make the EicC well suited to precision TMD studies~\cite{AbdulKhalek:2021gbh,Anderle:2021dpv,Yang:2024bfz,Zeng:2022lbo,Zeng:2023nnb}. We quantify its potential impact by fitting the existing world data together with projected EicC SIDIS pseudodata, using a more flexible flavor parametrization than in the world-data fit. Within the adopted framework, the projected data reduce the uncertainties of the quark and antiquark components, with substantial potential to improve constraints on the presently weakly determined sea-quark distributions.

The remainder of the paper is organized as follows.
Section~\ref{sec:framework} presents the theoretical framework, and Sec.~\ref{sec:numerical_study} describes the updated global analysis.
Section~\ref{sec:predictions} presents the WW-type comparison for the $\cos\phi_h$ modulation and the predictions for CLAS12 SIDIS and STAR $W^\pm$ production.
Section~\ref{sec:eicc_impact} assesses the projected impact of EicC measurements.
A summary is given in Sec.~\ref{sec:summary}.

\section{Theoretical framework}
\label{sec:framework}
\subsection{SIDIS Kinematics and Cross Section}

We consider the SIDIS process
\begin{align}
\ell(l) + N(P) \to \ell(l') + h(P_h) + X(P_X),
\end{align}
where the scattered lepton and a final-state hadron $h$ are detected, while $X$ denotes the unobserved final state. The symbols in parentheses specify the particle four-momenta. With $q=l-l'$, the kinematic variables are
\begin{align}
Q^2 &\equiv -q^2 = -(l-l')^2, \\
x &\equiv \frac{Q^2}{2P\cdot q}, \qquad
y \equiv \frac{P\cdot q}{P\cdot l}, \qquad
z \equiv \frac{P\cdot P_h}{P\cdot q}, \\
\gamma &\equiv \frac{2xM}{Q} = \frac{MQ}{P\cdot q},
\end{align}
where $M$ is the nucleon mass.

We work in the one-photon-exchange approximation and neglect the lepton mass. For a longitudinally polarized lepton with helicity $\lambda_e$ scattering off longitudinally polarized targets, one can express the differential cross section as~\cite{Bacchetta:2006tn}
\begin{align}
    \frac{\dd{\sigma}}{\dd{x} \dd{y} \dd{z} \dd{\phi_h} \dd{P_{h\perp}^2}}
    &=
    \frac{2\pi\alpha^2}{x y Q^2}
    \frac{y^2}{2(1-\varepsilon)}
    \left(1+\frac{\gamma^2}{2x}\right)
    \notag\\
    &\times
    \Big[
    F_{UU}
    + \lambda_e S_L \sqrt{1-\varepsilon^2}\,F_{LL}
    \notag\\
    &
    + \lambda_e S_L \sqrt{2\varepsilon(1-\varepsilon)}
      \cos\phi_h\,F_{LL}^{\cos\phi_h}
    \Big].
    \label{eq:CSLT}
\end{align}
Here $S_L$ denotes the longitudinal target-spin component; in the target rest frame, positive $S_L$ corresponds to spin antiparallel to the virtual-photon momentum. The labels $U$ and $L$ denote unpolarized and
longitudinally polarized states, respectively, with the first label referring to the lepton and the second to the target. An integration over the overall azimuth of the lepton scattering plane about the beam axis is performed. We only retain the azimuthally independent unpolarized contribution $F_{UU}$ and the double-longitudinal terms $F_{LL}$ and $F_{LL}^{\cos\phi_h}$ that are relevant to this study. The electromagnetic fine-structure constant is denoted by $\alpha$. The virtual-photon polarization parameter, defined as the ratio of longitudinal to transverse photon fluxes, is
\begin{align}
\varepsilon
&=
\frac{1-y-\frac{1}{4}\gamma^2 y^2}
     {1-y+\frac{1}{2}y^2+\frac{1}{4}\gamma^2 y^2}.
\end{align}

\begin{figure}
\centering
\includegraphics[width=0.45\textwidth]{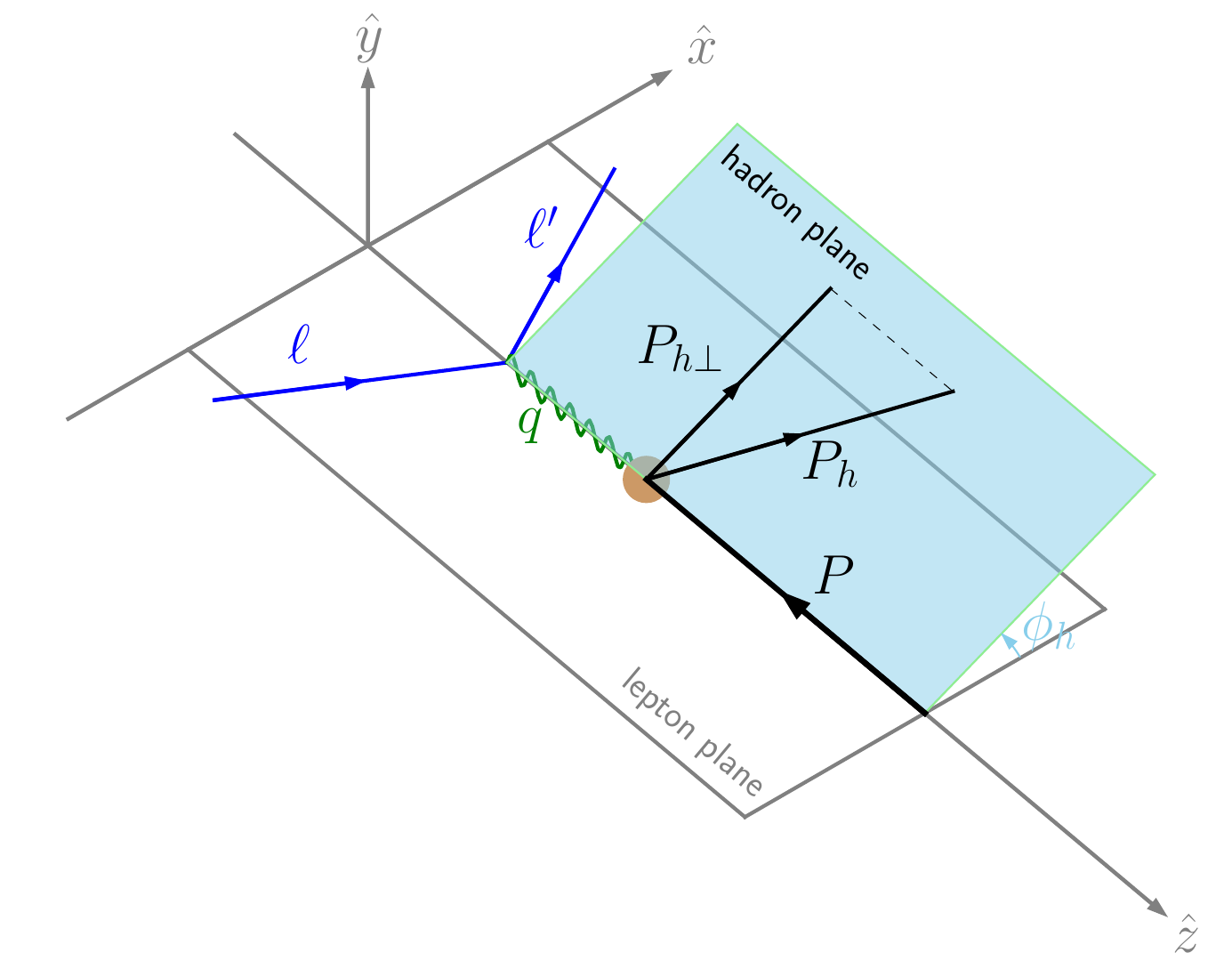}
\caption{SIDIS kinematics in the virtual-photon-nucleon frame, with azimuthal angles defined according to the Trento conventions. The virtual-photon direction defines the longitudinal axis, and the hadron transverse momentum $\boldsymbol P_{h\perp}$ lies in the transverse plane. The angles $\phi_h$ and $\phi_S$ specify the azimuths of the hadron momentum and transverse target-spin component relative to the lepton scattering plane.}
\label{Fig:Theoretical framework::trentoframe}
\end{figure}

Following the Trento conventions~\cite{Bacchetta:2004jz}, we define the measured hadron transverse momentum and azimuthal angle relative to the virtual-photon and target directions, as illustrated in Fig.~\ref{Fig:Theoretical framework::trentoframe}.
In a frame where the spatial momenta of $P$ and $q$ are defined along the longitudinal axis, $P_{h\perp}$ is the magnitude of the hadron transverse momentum. The corresponding covariant definitions are
\begin{align}
P_{h\perp}
&= \sqrt{-g^{\mu\nu}_\perp P_{h\mu}P_{h\nu}}, \\
\cos\phi_h
&= -\frac{l_\mu P_{h\nu}g^{\mu\nu}_\perp}
          {l_\perp P_{h\perp}}, \qquad
\sin\phi_h
= -\frac{l_\mu P_{h\nu}\epsilon^{\mu\nu}_\perp}
         {l_\perp P_{h\perp}},
\end{align}
where $l_\perp=\sqrt{-g^{\mu\nu}_\perp l_\mu l_\nu}$. The transverse metric and antisymmetric tensor are
\begin{align}
g^{\mu\nu}_\perp
&= g^{\mu\nu}
 - \frac{q^\mu P^\nu+P^\mu q^\nu}
        {P\cdot q(1+\gamma^2)}
 \notag\\
&\quad
 + \frac{\gamma^2}{1+\gamma^2}
   \left(
   \frac{q^\mu q^\nu}{Q^2}
   - \frac{P^\mu P^\nu}{M^2}
   \right), \\
\epsilon^{\mu\nu}_\perp
&= \epsilon^{\mu\nu\rho\sigma}
   \frac{P_\rho q_\sigma}
        {P\cdot q\sqrt{1+\gamma^2}}.
\end{align}

For TMD factorization, it is convenient to define the longitudinal axis by the target and detected-hadron momenta, with the target moving along the positive direction and the detected hadron along the negative direction. We introduce light-cone components $v^\pm=(v^0\pm v^3)/\sqrt{2}$ with respect to this axis. The TMD factorization applies in the low-transverse-momentum region,
$\delta\equiv q_T/Q\ll1$, where the quantity $q_T$ is the magnitude of the virtual-photon momentum transverse to these hadronic directions. 
Following Ref.~\cite{Scimemi:2019cmh}, the kinematic target- and detected-hadron-mass corrections are implemented through the relations:
\begin{align}
    x_S
    &\equiv -\frac{q^+}{P^+}
    = -\frac{2x}{\gamma^2}
      \bqty{
      1-\sqrt{1+\gamma^2\left(1-\frac{q_T^2}{Q^2}\right)}
      },
    \label{eq:xS} \\
    z_S
    &\equiv \frac{P_h^-}{q^-}
    = z\frac{x_S}{x}
      \frac{1+\sqrt{1-\varsigma^2}}
           {2\left(1-q_T^2/Q^2\right)},
    \label{eq:zS}
\end{align}
where
\begin{align}
    \varsigma
    &= \frac{\gamma M_h}{zQ}, \qquad
    q_T^2
    = \frac{P_{h\perp}^2}{z^2}
      \frac{1+\gamma^2}{1-\varsigma^2}.
    \label{eq:qT_mass_correction}
\end{align}
Here $M_h$ is the detected-hadron mass. These relations connect the measured variables to the momentum fractions and transverse momentum entering the TMD expressions.

Within TMD factorization~\cite{Bacchetta:2006tn,Boussarie:2023izj}, one can define the $b_T$-space SIDIS convolution by
\begin{align}
&\mathcal B_{\nu}[f^{(m)},D^{(n)}]
\equiv
\frac{x_S}{2\pi}
\sum_q e_q^2
\int_0^\infty \dd{b_T} b_T^{m+n+1} J_{\nu}(b_Tq_T)
\notag\\
&\qquad\times
f^{q(m)}(x_S,b_T;Q,Q^2)\,
D^{q(n)}(z_S,b_T;Q,Q^2).
\label{eq:tmd_b_convolution}
\end{align}
The sum includes quarks and antiquarks, and $e_q$ is their electric charge in units of the positron charge. The target dependence of $f^q$ and the detected-hadron dependence of $D^q$ are suppressed. The superscripts $(m)$ and $(n)$ denote the normalized $b_T$ derivatives of the TMD~PDF and TMD~FF, respectively, defined in Appendix~\ref{App:FTT}.
The function $J_\nu$ is the Bessel function of the first kind of order $\nu$.

Within the leading-power description, the structure functions are
\begin{align}
    F_{UU}
    &= |C_V(Q^2,Q^2)|^2\mathcal B_0[f_1,D_1],
    \label{eq::struct::FUU::bTspace} \\
    F_{LL}
    &= |C_V(Q^2,Q^2)|^2\mathcal B_0[g_{1L},D_1],
    \label{eq::struct::FLL::bTspace}
\end{align}
where $C_V$ is the hard matching coefficient. These expressions involve the zeroth derivatives, $m=n=0$, with the TMDs evaluated at the renormalization and rapidity scales $(\mu,\zeta)=(Q,Q^2)$. The functions $f_1$, $g_{1L}$, and $D_1$ are the unpolarized TMD~PDF, helicity TMD~PDF, and unpolarized TMD~FF, respectively. The variable $b_T$ is Fourier conjugate to the parton transverse momentum $k_T$ for PDFs and to the parent-parton transverse momentum $p_T$ for FFs. Our Fourier-transform conventions are given in Appendix~\ref{App:FTT}.

\subsection{Double spin asymmetries}

Following the polarization convention of Eq.~\eqref{eq:CSLT}, one can express the longitudinal double-spin asymmetry at fixed
$x$, $y$, $z$, and $P_{h\perp}^2$ as
\begin{align}
    A^{\mathrm{tot}}_{LL}(\phi_h)
    &\equiv \frac{1}{|\lambda_e||S_L|}
       \frac{\mathrm d\sigma_{LL}(+,+)-\mathrm d\sigma_{LL}(-,+)}
            {\mathrm d\sigma_{LL}(+,+)+\mathrm d\sigma_{LL}(-,+)}
    \notag\\
    &= \sqrt{1-\varepsilon^2}\,
       \frac{F_{LL}}{F_{UU}}
       +\sqrt{2\varepsilon(1-\varepsilon)}\,
        \cos\phi_h\,
        \frac{F_{LL}^{\cos\phi_h}}{F_{UU}}.
    \label{eq:helicity:crosssection}
\end{align}
In $\mathrm d\sigma_{LL}$, the first argument labels the lepton helicity and the second the target polarization. For the azimuthal angular moments below, we suppress the common differential measure and use $\sigma_{UU}$ for half the sum of $(+,+)$ and $(+,-)$ cross sections and $\sigma_{LL}$ for half their difference divided by $|\lambda_e||S_L|$, {\it i.e.}, the polarization-normalized spin-dependent contribution.

Integration over $\phi_h$ removes the $\cos\phi_h$ term, and the azimuthally averaged cross section asymmetry is
\begin{align}
    A_{LL}
    &\equiv
    \frac{\langle\sigma_{LL}\rangle}
         {\langle\sigma_{UU}\rangle}
    = \sqrt{1-\varepsilon^2}\,
      \frac{F_{LL}}{F_{UU}}.
    \label{eq:ALL_CSA}
\end{align}
This is the observable used to constrain the helicity TMDs in our analysis. Substituting the factorized expressions for $F_{UU}$ and $F_{LL}$, the common hard factor cancels and yields
\begin{align}
    A_{LL}
    &= \sqrt{1-\varepsilon^2}\,
       \frac{\mathcal B_0[g_{1L},D_1]}
            {\mathcal B_0[f_1,D_1]}.
\end{align}
It probes the helicity TMDs relative to the unpolarized TMDs through flavor-weighted convolutions with the unpolarized TMD FFs, and is therefore directly sensitive to the ratio $g_{1L}/f_1$ characterizing quark polarization rate.

The $\cos\phi_h$ modulation of the longitudinal double spin asymmetry is defined as
\begin{align}
    A_{LL}^{\cos\phi_h}
    &\equiv
    \frac{\langle 2\cos\phi_h\,\sigma_{LL}\rangle}
         {\langle\sigma_{UU}\rangle}
    \notag\\
    &= \sqrt{2\varepsilon(1-\varepsilon)}\,
       \frac{F_{LL}^{\cos\phi_h}}{F_{UU}},
    \label{eq:A_LL_cosphi_CSA}
\end{align}
where the factor of two compensates for the angular average. The structure function $F_{LL}^{\cos\phi_h}$ enters at the subleading power in the low-transverse-momentum expansion. While the measurements of this azimuthal modulation are not included in our main fit, we evaluate it in Sec.~\ref{sec:asymmetry} using the extracted helicity TMDs within the Wandzura-Wilczek-type approximation and compare with existing data. This comparison provides a consistency check of the description under that approximation.

\subsection{Structure of TMD~PDFs and TMD~FFs}

The TMDs used in this analysis combine perturbative matching onto collinear distributions at small $b_T$, a parametrization of their nonperturbative transverse dependence, and evolution to the renormalization and rapidity scales $(\mu,\zeta)$ relevant for the physical process. For a fixed parton flavor, we denote a TMD~PDF by $f(x,b_T)$ and a TMD~FF by $D(z,b_T)$, respectively, suppressing the scale arguments until the evolution factor is explicitly introduced. Here $x$ and $z$ denote the longitudinal momentum fraction arguments of the TMD PDF and FF.

At small $b_T$, one can match the TMDs onto collinear distributions through the operator product expansion (OPE):
\begin{align}
    f(x,b_T)
    &\simeq \sum_{f'} \int_x^1 \frac{\dd{y}}{y}
       C_{f\leftarrow f'}\pqty{\frac{x}{y},
          \mathbf{L}_{\mu_{\mathrm{OPE}}},
          a_s(\mu_{\mathrm{OPE}})} \notag\\
    &\qquad\times f'_{\mathrm{col}}(y,\mu_{\mathrm{OPE}})\, ,
    \label{eq:OPE_PDF} \\[6pt]
    D(z,b_T)
    &\simeq \sum_{f'} \int_z^1 \frac{\dd{y}}{y}
       \mathbb{C}_{f\to f'}\pqty{\frac{z}{y},
          \mathbf{L}_{\mu_{\mathrm{OPE}}},
          a_s(\mu_{\mathrm{OPE}})} \notag\\
    &\qquad\times
       \frac{d'_{\mathrm{col}}(y,\mu_{\mathrm{OPE}})}{y^2}\, ,
    \label{eq:OPE_FF}
\end{align}
where $\simeq$ indicates that power-suppressed terms in the small-$b_T$ expansion are neglected. The logarithm is $\mathbf{L}_{\mu}=\ln[b_T^2\mu^2/(4e^{-2\gamma_E})]$, with $\gamma_E$ the Euler-Mascheroni constant, and $a_s(\mu)=\alpha_s(\mu)/(4\pi)=g_s^2(\mu)/(4\pi)^2$, with $g_s$ the QCD gauge coupling. The scale $\mu_{\mathrm{OPE}}$ is the collinear matching scale; its choices for PDFs and FFs are given in Sec.~\ref{sec:unpolarized_TMDs}.
The summation of $f'$ runs over all parton flavors.
The functions $f'_{\mathrm{col}}$ and $d'_{\mathrm{col}}$ are the corresponding collinear PDFs and FFs. The Wilson coefficients $C_{f\leftarrow f'}$ and $\mathbb{C}_{f\to f'}$ are calculable in perturbation theory. Their dependence on the TMD renormalization and rapidity scales is suppressed. The additional factor $1/y^2$ in Eq.~\eqref{eq:OPE_FF} reflects the adopted normalization of the TMD~FF relative to the collinear FF.

To extend the small-$b_T$ expressions into the nonperturbative region, we adopt the multiplicative parametrization,
\begin{align}
    f(x,b_T)
    &= \Big[\,\sum_{f'} C_{f\leftarrow f'}
          \otimes f'_{\mathrm{col}}\,\Big]
       \times f_{\mathrm{NP}}(x,b_T)\, ,
    \label{eq:full_PDF} \\[4pt]
    D(z,b_T)
    &= \Big[\,\sum_{f'} \mathbb{C}_{f\to f'}
          \otimes d'_{\mathrm{col}}\,\Big]
       \times D_{\mathrm{NP}}(z,b_T),
    \label{eq:full_FF}
\end{align}
where $\otimes$ denotes the longitudinal momentum fraction convolutions in Eqs.~\eqref{eq:OPE_PDF} and \eqref{eq:OPE_FF}. The nonperturbative factors multiply the matching sum for each parton flavor. Their parametrizations approach unity as $b_T\to0$, with $f_{\mathrm{NP}}(x,0)=D_{\mathrm{NP}}(z,0)=1$, thereby preserving the leading small-$b_T$ matching behavior.
For the helicity TMD $g_{1L}$, the PDF matching coefficients and collinear distributions are replaced by their polarized counterparts, and the nonperturbative factor is denoted by $g_{1L,\mathrm{NP}}$. The explicit parametrizations of $f_{\mathrm{NP}}$, $D_{\mathrm{NP}}$, and $g_{1L,\mathrm{NP}}$ are given in Secs.~\ref{sec:unpolarized_TMDs}~and~\ref{sec:parametrizations}.

The evolution of TMD PDF and FF from the initial scales $(\mu_i,\zeta_i)$ to some other scales $(\mu,\zeta)$ is described by the factor $R$:
\begin{align}
    f(x,b_T;\mu,\zeta)
    &= \Big[\,\sum_{f'} C_{f\leftarrow f'}
          \otimes f'_{\mathrm{col}}\,\Big]
       \times f_{\mathrm{NP}}(x,b_T) \notag\\
    &\qquad\times
       R \big[b_T;(\mu_i,\zeta_i)\to(\mu,\zeta)\big]\, ,
    \label{eq:evolved_PDF} \\[6pt]
    D(z,b_T;\mu,\zeta)
    &= \Big[\,\sum_{f'} \mathbb{C}_{f\to f'}
          \otimes d'_{\mathrm{col}}\,\Big]
       \times D_{\mathrm{NP}}(z,b_T) \notag\\
    &\qquad\times
       R \big[b_T;(\mu_i,\zeta_i)\to(\mu,\zeta)\big].
    \label{eq:evolved_FF}
\end{align}
Here the matching expressions and nonperturbative factors define the initial TMDs at $(\mu_i,\zeta_i)$, with their initial-scale dependence suppressed. The scale $\mu_{\mathrm{OPE}}$ specifies the collinear matching, while $(\mu_i,\zeta_i)$ indicates the initial point of TMD evolution. The choice of initial scales and the explicit form of $R$ are discussed in the following subsection. The same evolution factor applies to the quark and antiquark TMDs $f_1$, $g_{1L}$, and $D_1$, while their matching coefficients and nonperturbative input functions differ. The evolution resums the large logarithms of $Qb_T$ associated with the separation between the hard scale $Q$ and the transverse scale $1/b_T$.

\subsection{Evolution of TMD~PDFs and TMD~FFs}

The dependence of the TMDs on the renormalization scale $\mu$ and rapidity scale $\zeta$ is governed by the coupled evolution equations
\begin{align}
    \mu^2 \frac{\partial}{\partial \mu^2}
    F(x,b_T;\mu,\zeta)
    &= \frac{\gamma_F(\mu,\zeta)}{2}
       F(x,b_T;\mu,\zeta),
    \label{Equ:mu_evolution} \\
    \zeta \frac{\partial}{\partial \zeta}
    F(x,b_T;\mu,\zeta)
    &= -\mathcal{D}(\mu,b_T)
       F(x,b_T;\mu,\zeta),
    \label{Equ:zeta_evolution}
\end{align}
where $\gamma_F$ is the ultraviolet anomalous dimension and $\mathcal{D}$ is the Collins-Soper kernel, also called the rapidity anomalous dimension. Here $F$ collectively denotes the TMDs $f_1$, $g_{1L}$, and $D_1$;
the longitudinal argument $x$ is replaced by $z$ for a TMD~FF.

The general solution takes the form
\begin{align}
    F(x,b_T;\mu,\zeta)
    &= R\big[b_T;(\mu_i,\zeta_i)\to(\mu,\zeta)\big]
       \notag\\
    &\qquad\times F(x,b_T;\mu_i,\zeta_i),
\end{align}
with
\begin{align}
    R
    &= \exp[
       \int_{\mathcal{P}}
       \pqty{
       \frac{\gamma_F(\bar\mu,\bar\zeta)}{\bar\mu}
       \,\dd{\bar\mu}
       -\frac{\mathcal{D}(\bar\mu,b_T)}{\bar\zeta}
       \,\dd{\bar\zeta}
       }].
\end{align}
The path $\mathcal{P}$ connects the initial scales $(\mu_i,\zeta_i)$ to the final scales $(\mu,\zeta)$ at fixed $b_T$. For the physical process, we choose $(\mu,\zeta)=(Q,Q^2)$.

The integrability condition~\cite{Collins:1981va},
\begin{align}
    \zeta\frac{\partial\gamma_F(\mu,\zeta)}{\partial\zeta}
    &= -\mu\frac{\partial\mathcal{D}(\mu,b_T)}{\partial\mu}
     = -\Gamma_{\rm cusp}(\mu),
    \label{Equ:integrability}
\end{align}
ensures path independence in the all-order formulation. Finite-order approximations can violate this condition and introduce residual path dependence.

In the formal integrable formulation, a scalar potential $\mathcal{F}(\mu,\zeta;b_T)$ can be defined whose gradient in $(\ln\mu^2,\ln\zeta)$ is $(\gamma_F/2,-\mathcal{D})$. Its equipotential curves are null-evolution curves: the TMD remains unchanged along each curve, and the evolution factor between two points on the same curve equals unity.

Following~\cite{Scimemi:2019cmh}, we adopt the $\zeta$ prescription to specify the initial TMDs. In the formal construction, the special null-evolution curve defining the so-called optimal TMD passes through the saddle point $(\mu_0,\zeta_0)$, determined by
\begin{align}
    \mathcal{D}(\mu_0,b_T)&=0,
    &
    \gamma_F(\mu_0,\zeta_0)&=0.
\end{align}
Both saddle-point scales depend on $b_T$. We denote the rapidity scale along this special curve by $\zeta_\mu(\mu,b_T)$. The optimal TMD is its common value,
\begin{align}
    F^{\mathrm{opt}}(x,b_T)
    &\equiv F(x,b_T;\mu_0,\zeta_0)
       \notag\\
    &=F\bigl(x,b_T;\mu,\zeta_\mu(\mu,b_T)\bigr),
\end{align}
which is independent of the point chosen along the curve.

The evolution from the saddle point to $(Q,Q^2)$ can therefore be represented by moving along the special null-evolution curve to $(Q,\zeta_\mu(Q,b_T))$, followed by rapidity evolution at fixed $\mu=Q$. The first step contributes unity, while the second gives
\begin{align}
    R\big[b_T;(\mu_0,\zeta_0)\to(Q,Q^2)\big]
    &=
    \left(
        \frac{Q^2}{\zeta_\mu(Q,b_T)}
    \right)^{-\mathcal{D}(Q,b_T)}.
\end{align}

For the numerical analysis, we implement this construction at NNLL accuracy using the kernel and rapidity-scale profile specified in Appendix~\ref{App:Evolution}, including their nonperturbative completion. This finite-order realization is combined with the NNLO small-$b_T$ matching described below.

\section{Numerical analysis}
\label{sec:numerical_study}

Using the theoretical framework described above, we perform an updated global fit to the world SIDIS $A_{LL}$ data. Compared with our previous study~\cite{Yang:2024drd}, the present analysis incorporates three refinements. First, we include kinematic corrections associated with the target and detected-hadron masses. Second, we upgrade the small-$b_T$ OPE matching for $f_1$, $g_{1L}$, and $D_1$ from NLO to NNLO. Third, we treat correlated experimental uncertainties using a penalty-term $\chi^2$ with the $T_0$ prescription~\cite{Ball:2009qv} to mitigate the D'Agostini bias~\cite{DAgostini:1993arp}.

\subsection{Unpolarized TMDs} 
\label{sec:unpolarized_TMDs}
\label{subsec: OPT Unpol PDF FF}

For the unpolarized TMD~PDFs $f_1(x,b_T)$ and TMD~FFs $D_1(z,b_T)$, we adopt the SV19 framework~\cite{Scimemi:2019cmh}. The OPE matching scales are $\mu_{\mathrm{OPE}}^{\mathrm{PDF}} = 2e^{-\gamma_E}/b_T + 2\,\mathrm{GeV}$ for the PDF and $\mu_{\mathrm{OPE}}^{\mathrm{FF}} = 2e^{-\gamma_E}z/b_T + 2\,\mathrm{GeV}$ for the FF. The additive $2\,\mathrm{GeV}$ term keeps both matching scales away from the Landau pole at large $b_T$. For the collinear inputs, we use NNPDF3.1 PDFs~\cite{NNPDF:2017mvq} and the DSS FFs~\cite{deFlorian:2014xna,deFlorian:2017lwf,deFlorian:2007ekg}.

In the adopted $\zeta$ prescription, the space-like matching coefficient for the unpolarized TMD~PDF has the NNLO expansion
\begin{widetext}
\begin{align}
&C_{f \leftarrow f'}(x, \mathbf{L}_{\mu_{\mathrm{OPE}}}, a_s) = \delta(\bar{x})\delta_{ff'} + a_s(\mu_{\mathrm{OPE}}) \left( -\mathbf{L}_{\mu_{\mathrm{OPE}}} P_{f \leftarrow f'}^{(1)} + C_{f \leftarrow f'}^{(1,0)} \right) \notag\\
&\quad+ a_s^2(\mu_{\mathrm{OPE}}) \Bigg[ \frac{P_{f \leftarrow k}^{(1)} \otimes P_{k \leftarrow f'}^{(1)} - \beta_0 P_{f \leftarrow f'}^{(1)}}{2} \mathbf{L}_{\mu_{\mathrm{OPE}}}^2 \notag\\
&\qquad- \mathbf{L}_{\mu_{\mathrm{OPE}}} \left( P_{f \leftarrow f'}^{(2)} + C_{f \leftarrow k}^{(1,0)} \otimes P_{k \leftarrow f'}^{(1)} - \beta_0 C_{f \leftarrow f'}^{(1,0)} \right) 
+ C_{f \leftarrow f'}^{(2,0)} + \frac{d^{(2,0)}\gamma_1}{\Gamma_0} \delta(\bar{x})\delta_{ff'} \Bigg]\notag\\
&\quad + \mathcal{O}(a_s^3).
\label{eq:unpolarized_pdf_matching}
\end{align}
\end{widetext}
Here $\bar{x}\equiv1-x$, $P_{f\leftarrow f'}^{(i)}$ are the $i$-loop space-like Dokshitzer-Gribov-Lipatov-Altarelli-Parisi (DGLAP) splitting kernels, $C_{f\leftarrow f'}^{(i,0)}$ denote the $\mathbf{L}_{\mu_{\mathrm{OPE}}}$-independent finite matching coefficients, and the repeated index $k$ is summed over parton flavors. The QCD coefficients $\beta_0$, $\gamma_1$, and $\Gamma_0$ follow the conventions and explicit expressions collected in Appendix~\ref{App:Evolution}. The final diagonal term in Eq.~\eqref{eq:unpolarized_pdf_matching} follows from the boundary condition of the $\zeta$ prescription and involves the two-loop Collins-Soper constant
\begin{equation}
    d^{(2,0)}
    = C_F C_A\left(\frac{404}{27}-14\zeta_3\right)
      -\frac{112}{27}C_F T_F n_f,
    \label{eq:d20}
\end{equation}
where $C_F=4/3$, $C_A=3$, and $T_F=1/2$ are the usual $SU(3)$ color factors, $n_f$ is the number of active quark flavors, and $\zeta_3$ is Ap\'ery's constant. For the TMD~FF, the analogous perturbative expansion is written for the rescaled coefficient $z^2\mathbb{C}_{f\to f'}$. The space-like kernels and finite coefficients are replaced by the time-like splitting kernels and the rescaled finite coefficients $z^2\mathbb{C}_{f\to f'}^{(i,0)}$, respectively, with $x$ replaced by $z$. The unscaled coefficient $\mathbb{C}_{f\to f'}$ enters Eq.~\eqref{eq:OPE_FF} with the factor $1/y^2$. The one-loop splitting kernels and finite matching coefficients for the unpolarized PDF and FF are collected in Appendix~\ref{App:matching_kernels}.

For the optimal unpolarized TMDs, the nonperturbative factors entering Eqs.~(\ref{eq:full_PDF})~and~(\ref{eq:full_FF}) are parametrized as
\begin{align}
 f_{\mathrm{NP}}(x,b_{T})
    = \exp[
        -\frac{\lambda_{1}^{f}(1-x)+\lambda_{2}^{f}x+x(1-x)\lambda_{5}^{f}}
               {\sqrt{1+\lambda_{3}^{f}x^{\lambda_{4}^{f}}b_{T}^{2}}}\,
        b_{T}^{2}
      ] , 
\end{align}
\begin{align}
 D_{\mathrm{NP}}(z,b_{T})
    &= \exp[
        -\frac{\eta_{1}z+\eta_{2}(1-z)}{\sqrt{1+\eta_{3}(b_{T}/z)^{2}}}\,
        \frac{b_{T}^{2}}{z^{2}}
      ]
      \notag\\
    &\quad\times\bigl(1+\eta_{4}\,b_{T}^{2}/z^{2}\bigr).
\end{align}
The numerical values of the parameters are given in Table~\ref{tab:SV19 para}, as determined in the SV19 global fit~\cite{Scimemi:2019cmh}.

\begin{table}
\caption{Parameters of the nonperturbative functions for the optimal unpolarized TMD~PDF and TMD~FF. All parameters have units of $\mathrm{GeV}^2$, except $\lambda_4^{f}$, which is dimensionless. 
 }
        \label{tab:SV19 para}
\begin{tabular}{ccccccccc}
\hline\hline
$\lambda_1^{f}$ & $\quad$ & $\lambda_2^{f}$ & $\quad$ & $\lambda_3^{f}$ & $\quad$ & $\lambda_4^{f}$ & $\quad$ & $\lambda_5^{f}$ \\
$ 0.198 $   & $\quad$ & $ 9.3$      &  $\quad$ &  $ 431$     &  $\quad$ & $ 2.12 $ & $\quad$ & $ -4.44$    \\\hline\hline
$\eta_1$    & $\quad$ & $\eta_2$    & $\quad$ & $\eta_3$    & $\quad$ & $\eta_4$     \\
$ 0.260$   & $\quad$ & $ 0.476$      & $\quad$ & $ 0.478$     & $\quad$ & $ 0.483$       \\\hline\hline
\end{tabular}

\end{table}

\subsection{TMD helicity distribution parametrization}
\label{sec:parametrizations}

The optimal TMD helicity distribution at the saddle point is parametrized as
\begin{equation}
    g_{1L}^{f\leftarrow P}(x,b_T)
    = \bigl[\,\Delta C_{f\leftarrow f'}\otimes g_{1L}^{f'\leftarrow P}(x)\,\bigr]\,
      g_{1L,\mathrm{NP}}(x,b_T)\, ,
    \label{eq:g1L_struct}
\end{equation}
where $\Delta C_{f\leftarrow f'}$ is the helicity matching coefficient. Its perturbative expansion follows the same structure as the unpolarized PDF coefficient in Eq.~(\ref{eq:unpolarized_pdf_matching}), with $P^{(i)}$ and $C^{(i,0)}$ replaced by $\Delta P^{(i)}$ and $\Delta C^{(i,0)}$, respectively. The one-loop polarized splitting kernels and finite matching coefficients are collected in Appendix~\ref{App:matching_kernels}.

We parametrize the longitudinal input to the matching convolution as a possible $x$-dependent deformation of a reference collinear helicity PDF:
\begin{equation}
\begin{split}
    g_{1L}^{f\leftarrow P}(x)
    &= N_f\,
      \frac{(1-x)^{\alpha_f}x^{\beta_f}(1+\varepsilon_f x)}
           {n(\alpha_f,\beta_f,\varepsilon_f)}\,\\
    &\quad\times g_{1}^{f\leftarrow P}\bigl(x,
         Q=\mu_{\mathrm{OPE}}^{\mathrm{PDF}}\bigr)\, .
    \label{eq:g1L_coll}
\end{split}
\end{equation}
The reference distribution $g_{1}^{f\leftarrow P}$ is taken from NNPDFpol1.1~\cite{Nocera:2014gqa} and evaluated at the matching scale $\mu_{\mathrm{OPE}}^{\mathrm{PDF}} = 2e^{-\gamma_E}/b_T + 2\,\mathrm{GeV}$. The scale argument of $g_{1L}^{f\leftarrow P}(x)$ is suppressed; its matching-scale dependence enters through the reference distribution.

The nonperturbative transverse dependence is parametrized by
\begin{equation}
    g_{1L,\mathrm{NP}}(x,b_T)
    = \exp\!\Bigl[-\bigl(\lambda_{1}(1-x)+\lambda_{2}x
                       +x(1-x)\lambda_{5}\bigr)\,b_T^{2}\Bigr]\, ,
    \label{eq:g1L_NP}
\end{equation}
where $\lambda_1$, $\lambda_2$, and $\lambda_5$ have dimensions of $\mathrm{GeV}^2$. This factor approaches unity as $b_T\to0$. A common transverse factor is used for all flavors in the world-data fit of Sec.~\ref{sec:world_data_fit}; this restriction is relaxed for the EicC projection in Sec.~\ref{sec:eicc_impact}.

The auxiliary normalization factor is defined by
\begin{equation}
\begin{aligned}
    n(\alpha,\beta,\varepsilon)
    &= \mathrm{B}(\alpha+\alpha_{\mathrm{ref}},\beta+\beta_{\mathrm{ref}})\times \\
    &\frac{\bigl[(\alpha+\alpha_{\mathrm{ref}}+\beta+\beta_{\mathrm{ref}})
                   +\varepsilon(\beta+\beta_{\mathrm{ref}})\bigr]}
                  {\alpha+\alpha_{\mathrm{ref}}+\beta+\beta_{\mathrm{ref}}}\, ,
\end{aligned}
    \label{eq:n_factor}
\end{equation}
where $\mathrm{B}$ is the Euler Beta function, with $\alpha_{\mathrm{ref}}=4$ and $\beta_{\mathrm{ref}}=2$. This factor is introduced to reduce correlations between the normalization parameter $N_f$ and the shape parameters $\alpha_f$, $\beta_f$, and $\varepsilon_f$. Together with the transverse parameters $\lambda_{1,2,5}$, these parameters define the fit ansatz, subject to the restrictions specified below.

\subsection{Isospin and charge-conjugation symmetries}

Isospin symmetry is applied in this analysis. It relates the neutron TMD PDFs to proton TMD PDFs by interchanging the up- and down-quark flavors and their antiquarks, while leaving the strange-quark and gluon distributions unchanged. For a generic TMD~PDF $f\in\{f_{1},\,g_{1L}\}$,
\begin{equation}
\begin{aligned}
f^{u \leftarrow n}(x,b_T) &= f^{d \leftarrow p}(x,b_T), \\
f^{\bar{u} \leftarrow n}(x,b_T) &= f^{\bar{d} \leftarrow p}(x,b_T), \\
f^{d \leftarrow n}(x,b_T) &= f^{u \leftarrow p}(x,b_T), \\
f^{\bar{d} \leftarrow n}(x,b_T) &= f^{\bar{u} \leftarrow p}(x,b_T), \\
f^{s \leftarrow n}(x,b_T) &= f^{s \leftarrow p}(x,b_T), \\
f^{\bar{s} \leftarrow n}(x,b_T) &= f^{\bar{s} \leftarrow p}(x,b_T), \\
f^{g \leftarrow n}(x,b_T) &= f^{g \leftarrow p}(x,b_T).
\end{aligned}
\end{equation}

For deuteron targets, we define the TMD PDFs normalized per nucleon. The D-wave component of the deuteron wave function leads to a reduction of the nucleon polarization~\cite{Wiringa:1994wb}. In the effective-polarization approximation, the helicity distribution
is modeled as
\begin{equation}
g_{1L}^{f\leftarrow d}(x,b_T)
= \frac{P_{d}^{n}\,g_{1L}^{f\leftarrow n}(x,b_T)
   + P_{d}^{p}\,g_{1L}^{f\leftarrow p}(x,b_T)}{2},
\end{equation}
where we adopt $P_{d}^{n}=P_{d}^{p}=0.925$ for the effective neutron and proton polarizations relative to the deuteron polarization. The unpolarized distribution is approximated by
\begin{equation}
f_1^{f\leftarrow d}(x,b_T)
= \frac{f_1^{f\leftarrow p}(x,b_T)+f_1^{f\leftarrow n}(x,b_T)}{2}.
\end{equation}
This treatment neglects nucleon momentum smearing and other dynamical nuclear modifications.

For charged pions and kaons, charge-conjugation symmetry relates the
TMD~FFs according to
\begin{equation}
D_1^{q\to h^-}(z,b_T)=D_1^{\bar q\to h^+}(z,b_T),
\qquad h\in\{\pi,K\},
\end{equation}
with the analogous relation obtained by interchanging $q\leftrightarrow\bar q$. The gluon FFs satisfy $D_1^{g\to h^-}(z,b_T)=D_1^{g\to h^+}(z,b_T)$.

For neutral pions, we adopt the approximation used in the DSS FF parametrization~\cite{deFlorian:2014xna}
\begin{equation}
D_{1}^{f\to\pi^{0}}(z,b_T)
= \frac{D_{1}^{f\to\pi^{+}}(z,b_T)
       +D_{1}^{f\to\pi^{-}}(z,b_T)}{2},
\end{equation}
for each quark, antiquark, and gluon flavor $f$. In the adopted TMD parametrization, this relation follows from the corresponding collinear relation, the linearity of the matching convolution, and the use of the same nonperturbative factor $D_{\mathrm{NP}}(z,b_T)$ for all three pion charge states.

\subsection{World data fit}
\label{sec:world_data_fit}

Building on Ref.~\cite{Yang:2024drd}, we perform an updated global analysis of the longitudinal double spin asymmetry $A_{LL}$ in SIDIS to extract TMD helicity distributions.

We use the HERMES~\cite{HERMES:2018awh} and CLAS~\cite{CLAS:2017yrm} measurements of $A_{LL}$. HERMES provides 477 $A_{LL}$ data points and 180 $A_{LL}^{\cos\phi_h}$ data points for pion and kaon production on proton and deuteron targets, and CLAS provides 21 data points for each asymmetry in neutral-pion production on a proton target.
The corresponding data sets are summarized in Table~\ref{tab:EXP:HERMES_CLAS}. 
\footnote{COMPASS~\cite{COMPASS:2016klq} also reports azimuthal asymmetry amplitudes for positive and negative charged hadrons in muon-deuteron SIDIS, but the published information is insufficient for the present analysis, and thus these data are not included in the fit.}

For the determination of TMD helicity distributions, we fit only the $A_{LL}$ data.
The $A_{LL}^{\cos\phi_h}$ measurements receive higher-twist contributions and are only used for a consistency check under the WW-type approximation in Sec.~\ref{sec:asymmetry}.
We apply the cut $\delta<0.5$ to select the low-transverse-momentum region for the TMD analysis~\cite{Collins:1989gx,Collins:2011zzd}, leaving 246 $A_{LL}$ data points in the fit.

\begin{table*}
\caption{SIDIS longitudinal double spin asymmetry data sets considered in this analysis. Numbers outside (inside) parentheses represent the numbers of $A_{LL}$ ($A_{LL}^{\cos\phi_h}$) data points.
The fifth column gives the number of points remaining after the $\delta<0.5$ cut.}
\label{tab:EXP:HERMES_CLAS}
	\begin{ruledtabular}
\begin{tabular}{llcccll}
Data set & Target & Lepton beam & Points & Points after cut & Process & Measurement \\
\hline
HERMES~\cite{HERMES:2018awh}& H$_2$     & $27.6\,\mathrm{GeV}\, e^{\pm}$    &80(30) &42(10)& $e^{\pm}p\rightarrow e^{\pm} \pi^{+}X$   & $A_{LL}(A_{LL}^{\cos\phi_h})$  \\
HERMES~\cite{HERMES:2018awh}& H$_2$     & $27.6\,\mathrm{GeV}\, e^{\pm}$    &80(30) &42(11)& $e^{\pm}p\rightarrow e^{\pm} \pi^{-}X$   & $A_{LL}(A_{LL}^{\cos\phi_h})$  \\
HERMES~\cite{HERMES:2018awh}& D$_2$     & $27.6\,\mathrm{GeV}\, e^{\pm}$    &80(30) &39(10)& $e^{\pm}d\rightarrow e^{\pm} \pi^{+}X$   & $A_{LL}(A_{LL}^{\cos\phi_h})$  \\
HERMES~\cite{HERMES:2018awh}& D$_2$     & $27.6\,\mathrm{GeV}\, e^{\pm}$    &80(30) &37(9)& $e^{\pm}d\rightarrow e^{\pm} \pi^{-}X$   & $A_{LL}(A_{LL}^{\cos\phi_h})$  \\
HERMES~\cite{HERMES:2018awh}& D$_2$     & $27.6\,\mathrm{GeV}\, e^{\pm}$    &79(30) &40(9)& $e^{\pm}d\rightarrow e^{\pm} K^{+}X$   & $A_{LL}(A_{LL}^{\cos\phi_h})$  \\
HERMES~\cite{HERMES:2018awh}& D$_2$     & $27.6\,\mathrm{GeV}\, e^{\pm}$    &78(30) &38(9)& $e^{\pm}d\rightarrow e^{\pm} K^{-}X$   & $A_{LL}(A_{LL}^{\cos\phi_h})$  \\
\hline
CLAS~\cite{CLAS:2017yrm} & $^{14}$NH$_{3}$ &$6\,\mathrm{GeV}\, e^{-}$ & 21(21) & 8(8)& $e^{-}p\rightarrow e^{-} \pi^{0}X$ & $A_{LL}(A_{LL}^{\cos\phi_h})$ \\
\hline
Total       &            &            & 498(201) & 246(66) & & \\
\end{tabular}
\end{ruledtabular}
\end{table*}

To quantify the agreement between theory and data, we construct a $\chi^2$ function with nuisance parameters for correlated normalization uncertainties. We denote the uncorrelated uncertainty of point $i$ by $E_{i,\mathrm{uncor}}$ and the relative normalization uncertainty of group $s$ by $E_{s,\mathrm{cor}}^R$. Each group has a nuisance parameter $r_s$, with a penalty $r_s^2$ for its deviation from zero. The index $s$ runs over the three groups $\mathrm{CLAS}$, $\mathrm{HERMES}\;p$, and $\mathrm{HERMES}\;d$, while $i$ runs over the data points within each group. For the initial iteration, we use
\begin{equation}
    \chi^2_{r0} = \sum_s\sum_{i\in s}\left(
        \frac{D_i-T_i-r_s\,E_{s,\mathrm{cor}}^R\,D_i}
             {E_{i,\mathrm{uncor}}}
    \right)^{\!2}+\sum_s r_s^2\, ,
    \label{eq:chi2_r0}
\end{equation}
where $D_i$ is the measured central value and $T_i$ is the corresponding theory prediction. The nuisance parameters are varied together with the parameters of the TMD parametrization.

In Eq.~(\ref{eq:chi2_r0}), the correlated shift is scaled by the measured central value $D_i$.
To avoid the D'Agostini bias~\cite{DAgostini:1993arp}, we use the $T_0$ prescription~\cite{Ball:2009qv} in subsequent iterations:
\begin{equation}
    \chi^2 = \sum_s\sum_{i\in s}\pqty{
        \frac{D_i-T_i-r_s\,E_{s,\mathrm{cor}}^R\,T_i^{(0)}}
             {E_{i,\mathrm{uncor}}}
    }^2+\sum_s r_s^2 ,
    \label{eq:chi2_r1}
\end{equation}
where $T_i^{(0)}$ is the theory prediction from the preceding fit. These reference predictions are held fixed during each iteration and updated between iterations until the fitted parameters converge. The final fit statistic is denoted by $\chi^2$.

\begin{table}
\caption{Fit quality after convergence of the $T_0$ iterations for the $A_{LL}$ data points within the $\delta<0.5$ cut.}
\label{tab:chi2_results}
\begin{ruledtabular}
\begin{tabular}{lcc}
Data set & $N$ & $\chi^2/N$ \\
\hline
CLAS & 8 & 0.531 \\
HERMES $p$ & 84 & 0.952 \\
HERMES $d$ & 154 & 1.132 \\
\hline
Total & 246 & 1.051 \\
\end{tabular}
\end{ruledtabular}
\end{table}

The limited precision of the existing world data motivates a reduced parameter set. We impose $N_s=N_{\bar{s}}$. For the up- and down-quark distributions, we use common parameters for the $x$-dependent prefactor, $\alpha_u=\alpha_d\equiv\alpha$, $\beta_u=\beta_d\equiv\beta$, and $\varepsilon_u=\varepsilon_d\equiv\varepsilon$. The sea-quark and gluon collinear inputs receive no additional $x$-dependent modification, i.e., $\alpha_{\bar{u}}=\alpha_{\bar{d}}=\alpha_s=\alpha_{\bar{s}}=\alpha_g=0$, and similarly for the $\beta$ and $\varepsilon$ parameters. A common set of nonperturbative parameters is used for all flavors, $\lambda_i^f=\lambda_i$. These restrictions leave 12 physics parameters, supplemented by the three nuisance parameters.

To estimate the uncertainties, we generate $100$ Monte Carlo replicas of the original data, taking into account the experimental uncertainties and correlations included in the analysis.
Each replica is fitted by minimizing the objective function in Eq.~(\ref{eq:chi2_r1}), with the reference predictions $T_i^{(0)}$ held fixed during each iteration.
The resulting fits form an ensemble.
For the extracted quantities, the central values are the replica means, and the uncertainties are represented by 68\% intervals.
The overall fit yields $\chi^2/N=1.051$, as summarized in Table~\ref{tab:chi2_results}.
The central values and uncertainties of the fitted parameters are collected in Table~\ref{tab:parameter results}.

\begin{figure*}
    \centering
    \includegraphics[width=0.52\linewidth]{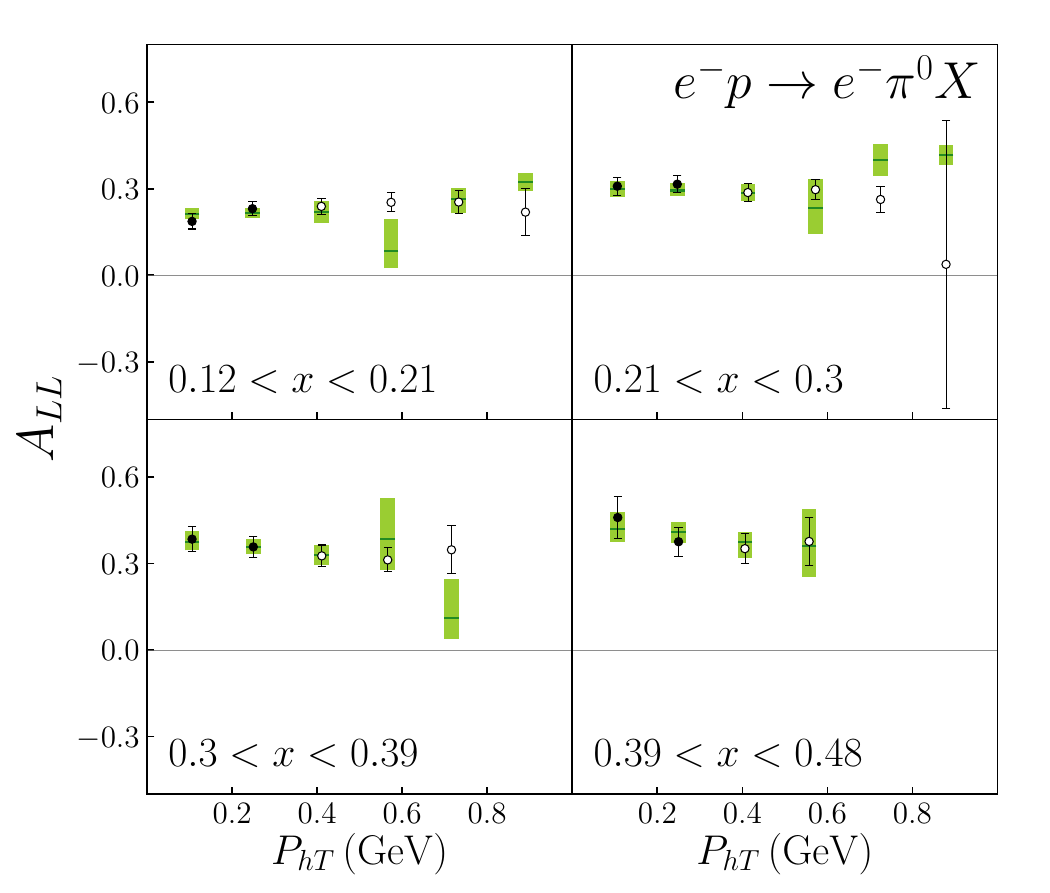}
    \caption{Comparison of fit results and the CLAS $A_{LL}$ data~\cite{CLAS:2017yrm} for the process $e^{-}p\to e^{-}\pi^{0}X$ in four bins of $x$. Filled points satisfy the $\delta<0.5$ cut and enter the fit; open points lie outside this region. The green bands represent 68\% confidence intervals.}
    \label{fig:world:clas}
\end{figure*}

\begin{figure*}
    \centering
    \includegraphics[width=0.495\linewidth]{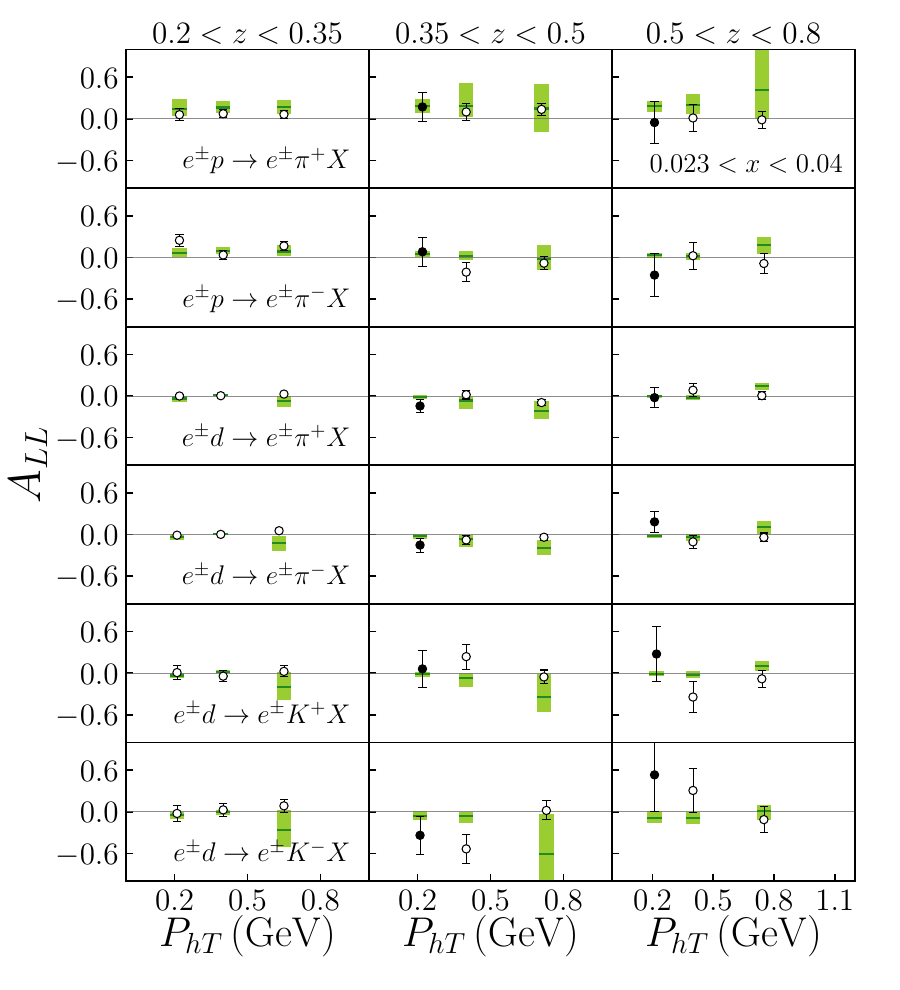}
    \includegraphics[width=0.495\linewidth]{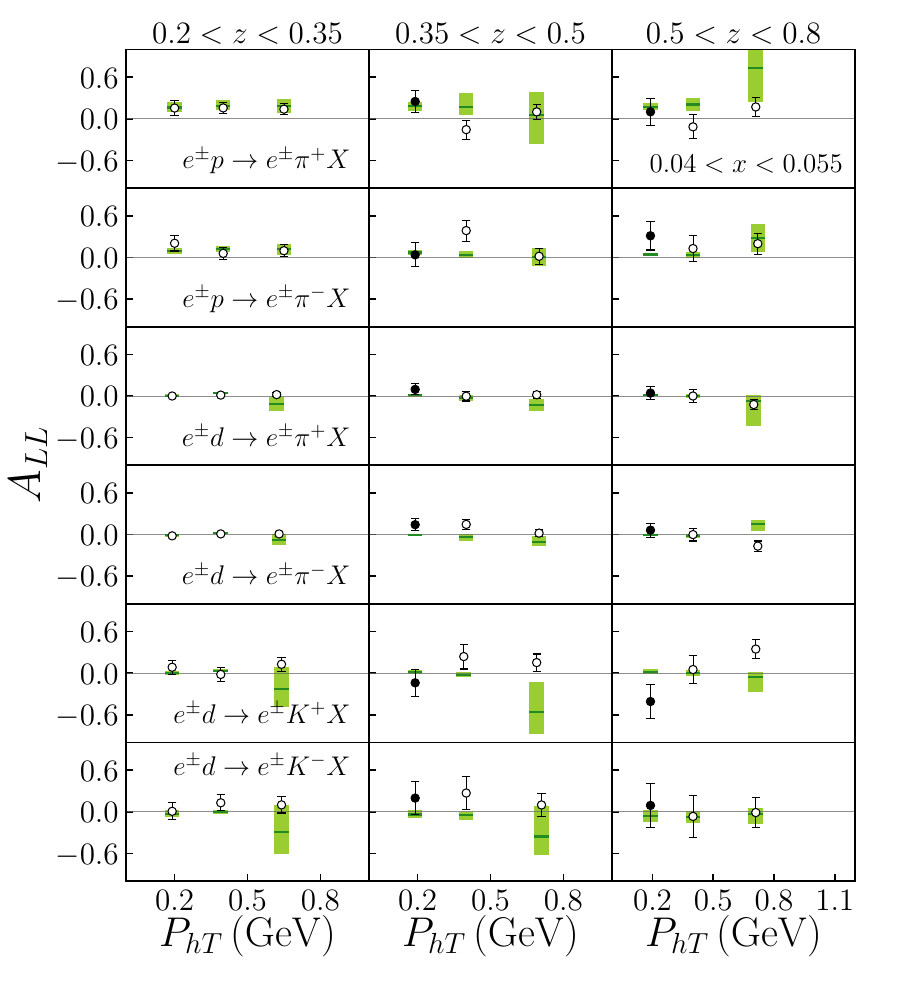}
    \includegraphics[width=0.495\linewidth]{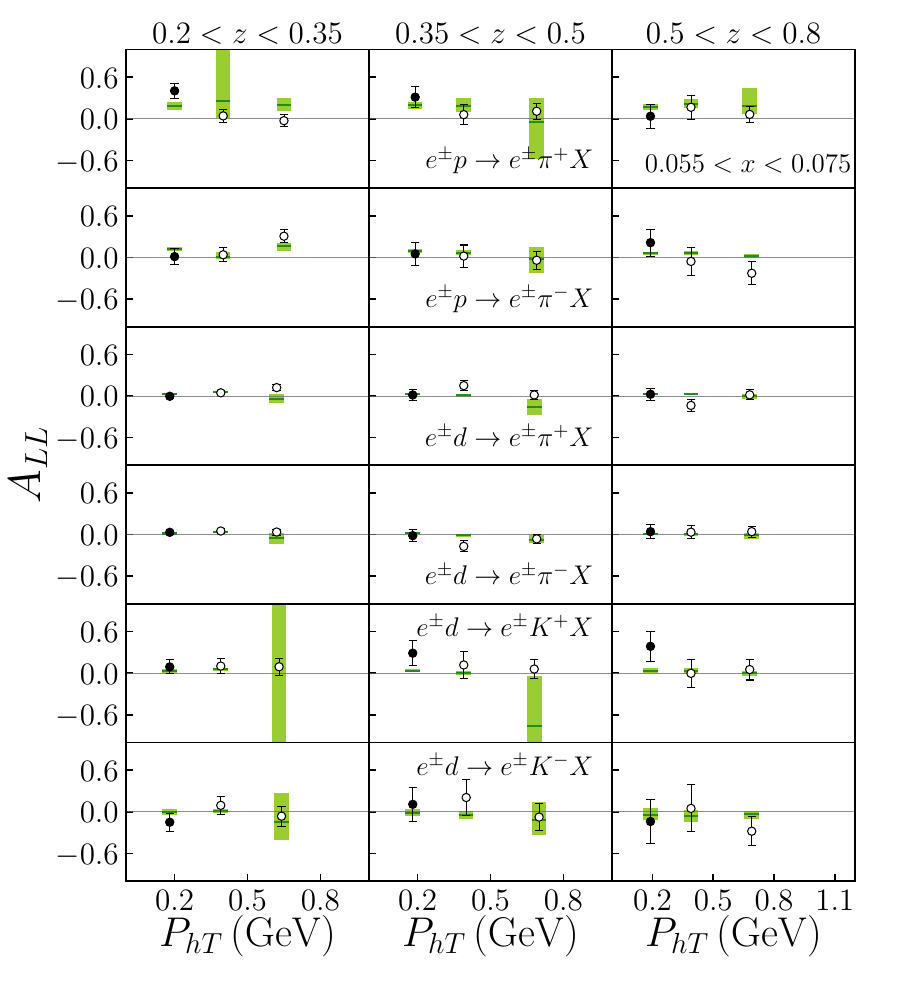}
    \includegraphics[width=0.495\linewidth]{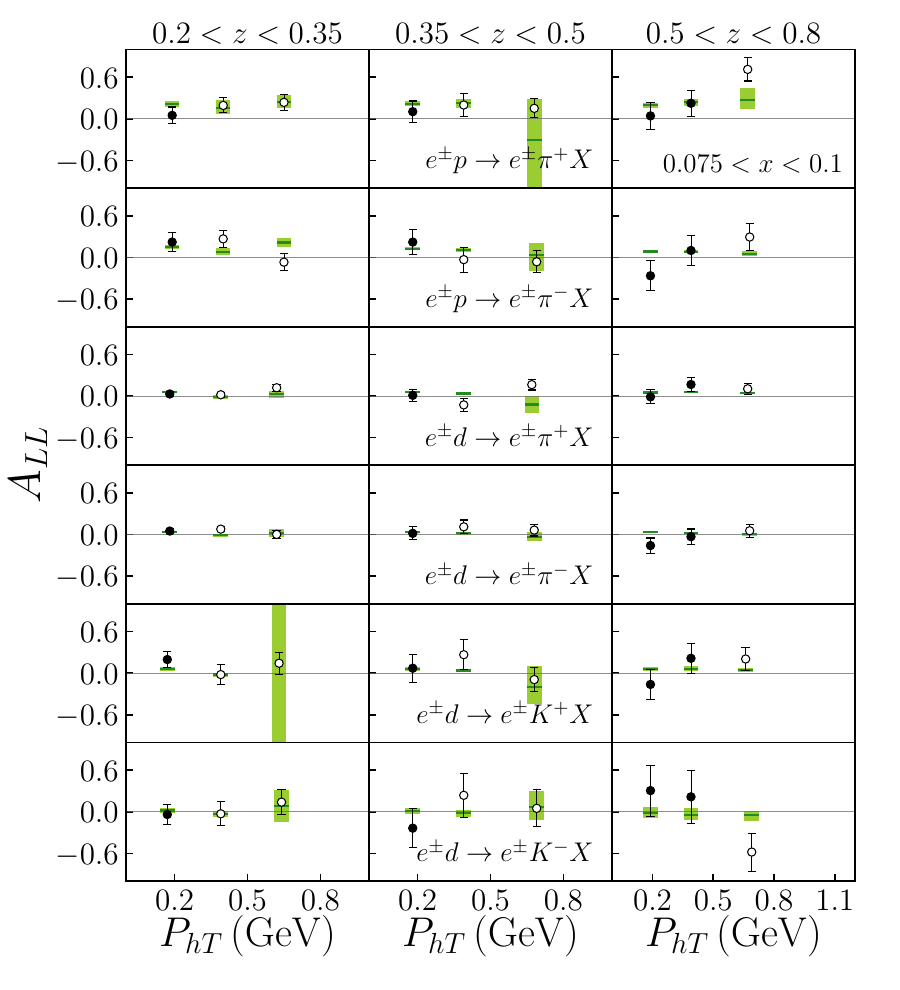}
    \caption{Comparison of fit results and the HERMES $A_{LL}$ data~\cite{HERMES:2018awh}. The four panels correspond to $x$ intervals: $0.023<x<0.04$, $0.04<x<0.055$, $0.055<x<0.075$, and $0.075<x<0.1$. Within each panel, the six rows correspond to different SIDIS processes, and the three columns correspond to the $z$ bins. Filled points satisfy the $\delta<0.5$ cut and enter the fit; open points lie outside this region. The green bands represent 68\% confidence intervals.}
    \label{fig:hermes:low_x}
\end{figure*}

\begin{figure*}
    \centering
    \includegraphics[width=0.495\linewidth]{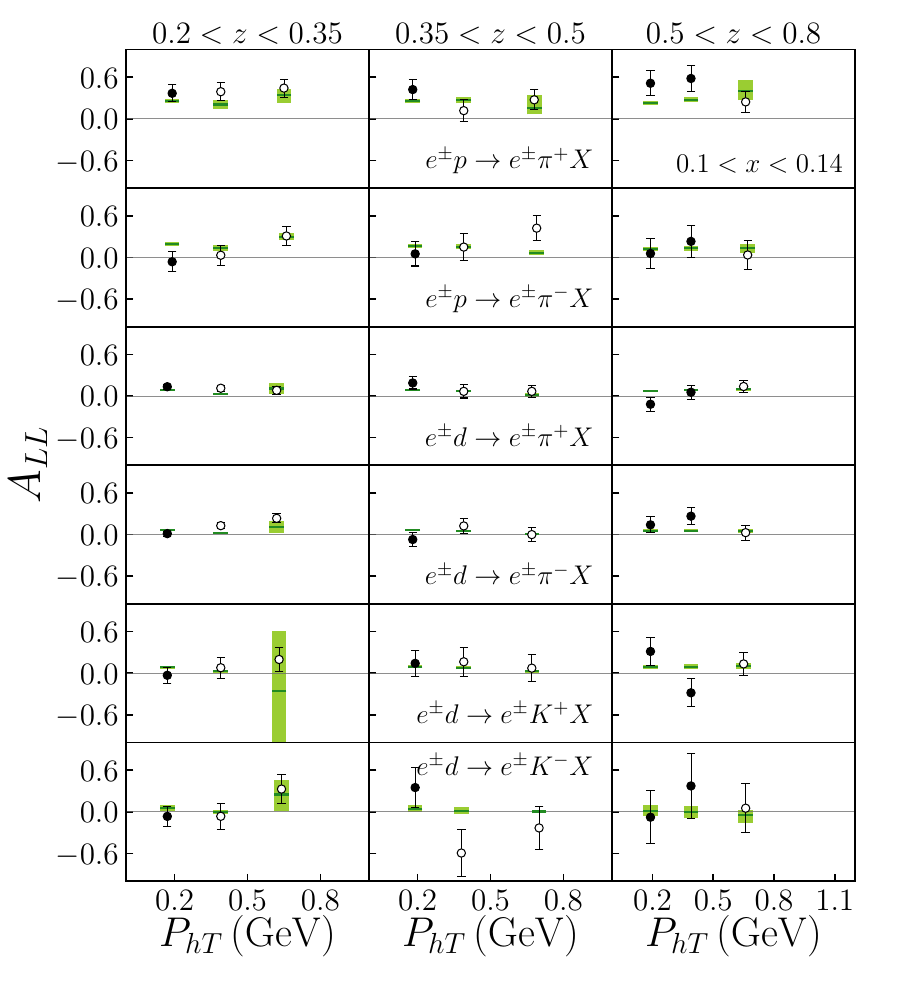}
    \includegraphics[width=0.495\linewidth]{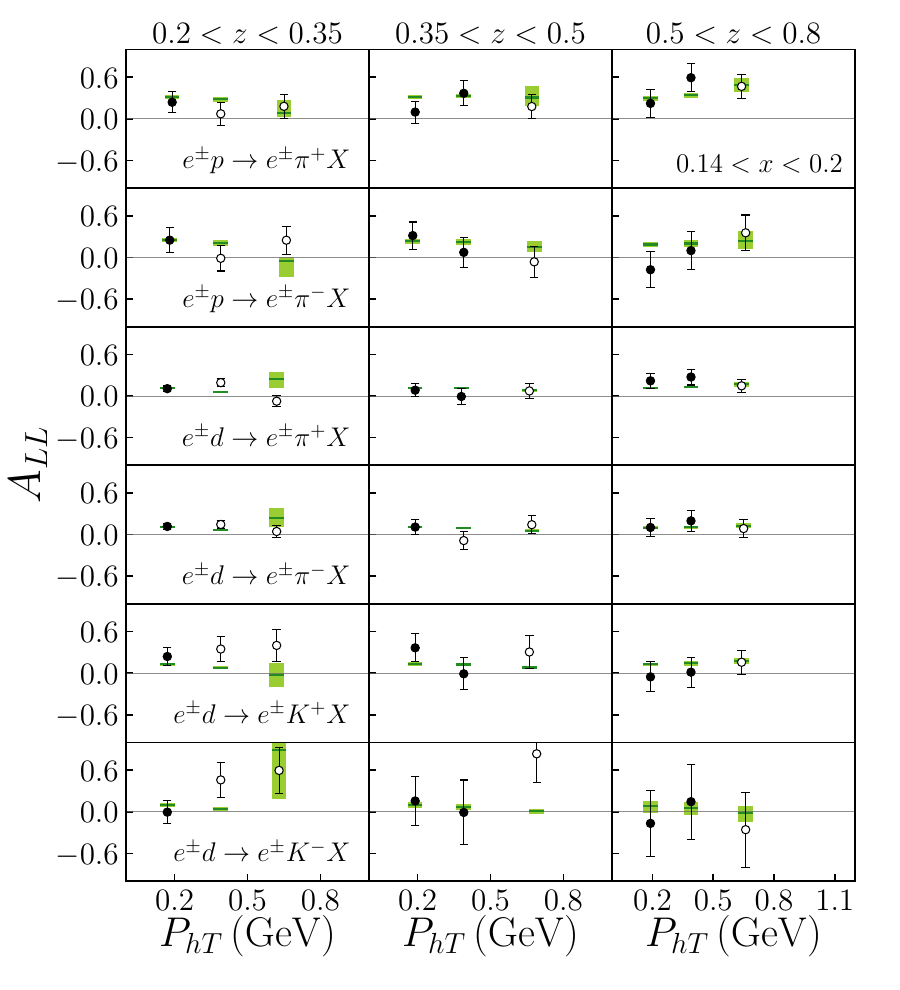}
    \includegraphics[width=0.495\linewidth]{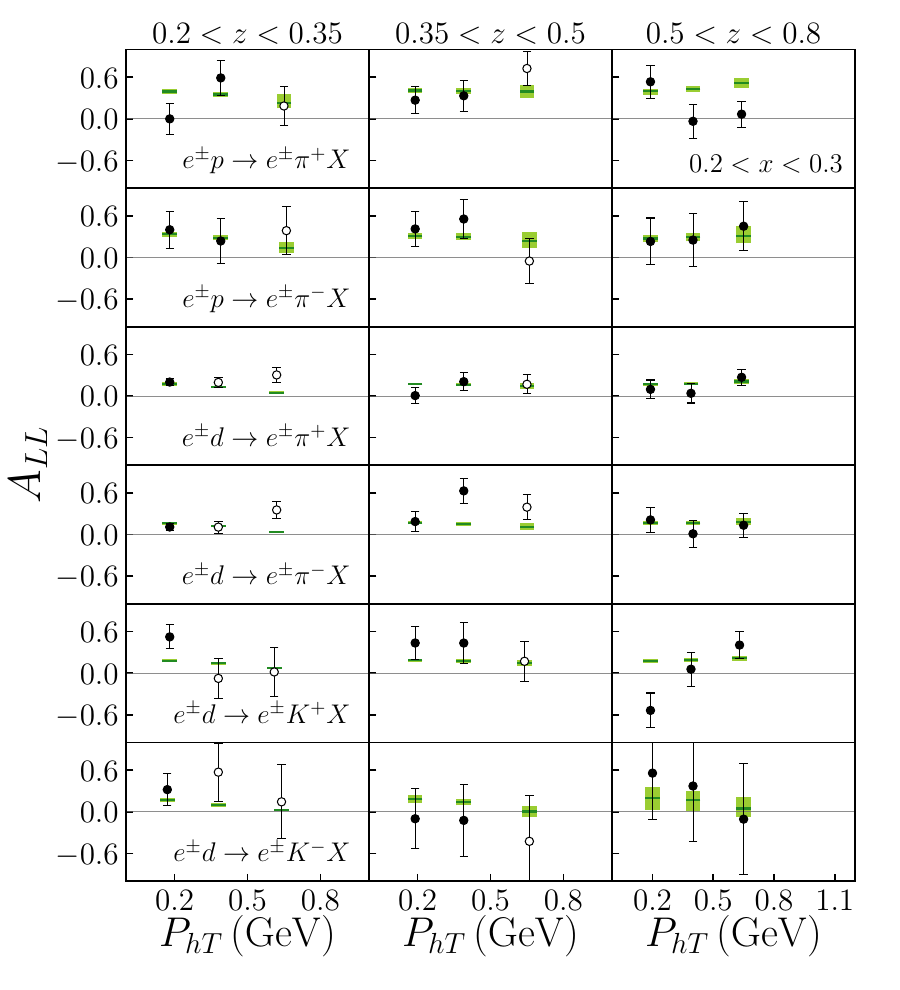}
    \includegraphics[width=0.495\linewidth]{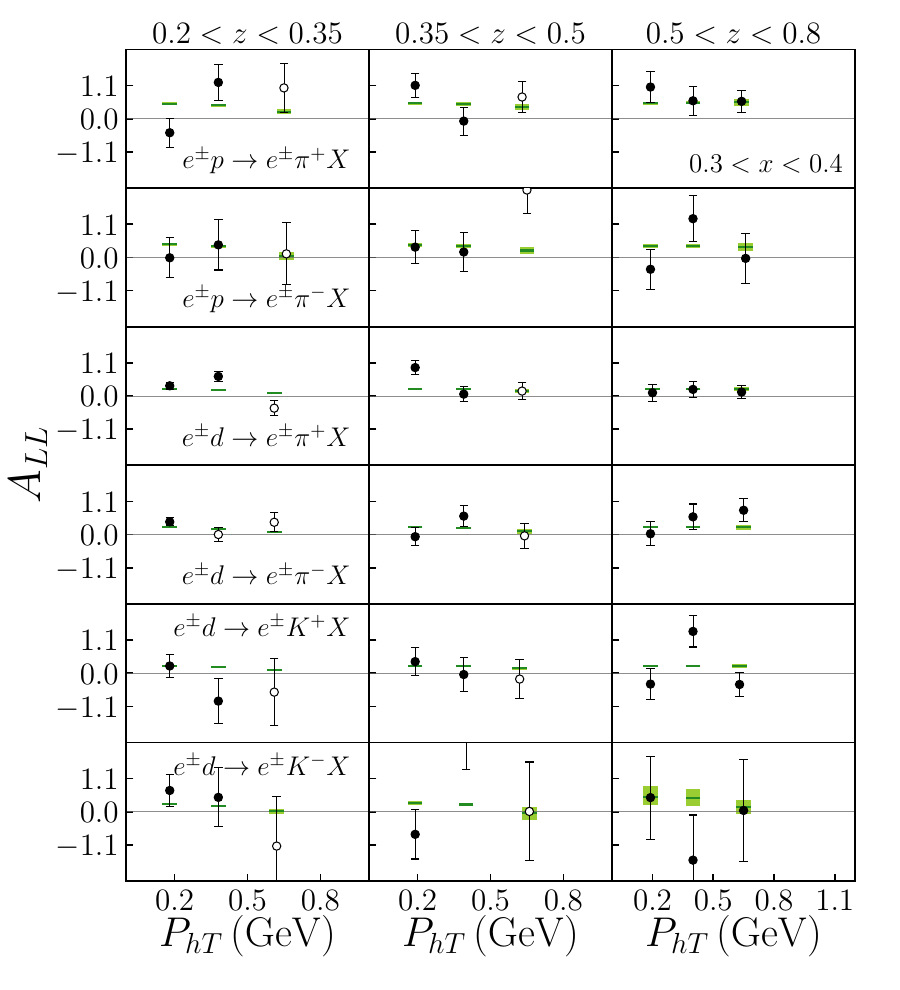}
    \caption{The same as Fig.~\ref{fig:hermes:low_x}, but the panels correspond to $x$ intervals: $0.1<x<0.14$, $0.14<x<0.2$, $0.2<x<0.3$, and $0.3<x<0.4$.}
    \label{fig:hermes:mid_x}
\end{figure*}

\begin{figure*}
    \centering
    \includegraphics[width=0.495\linewidth]{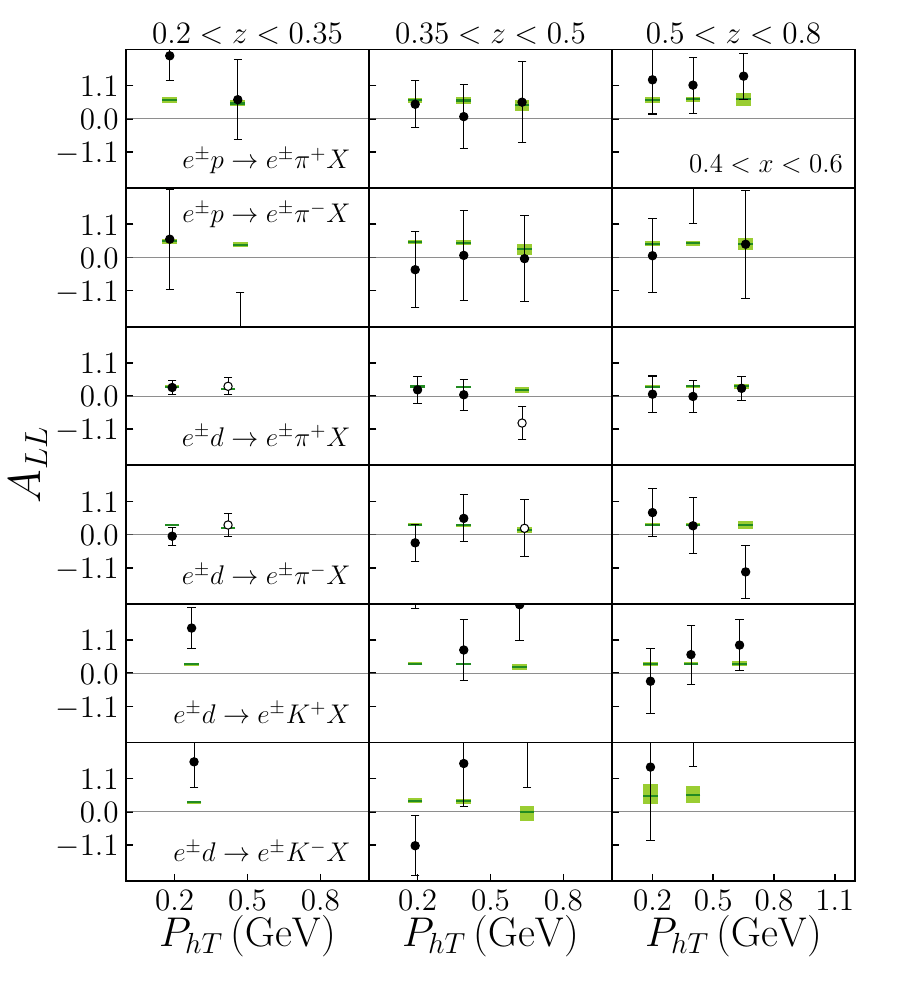}
    \caption{The same as Fig.~\ref{fig:hermes:low_x}, but for the $x$ interval $0.4<x<0.6$.}
    \label{fig:hermes:high_x}
\end{figure*}

The comparisons between the fit results and the experimental data are shown in Figs.~\ref{fig:world:clas}--\ref{fig:hermes:high_x}.
For data points outside the fit region, the mass-corrected light-cone fraction $x_S$ in Eq.~(\ref{eq:xS}) can become negative for $q_T>Q$.
When $x_S<0$, we just evaluate the displayed prediction with $x_S$ and $z_S$ replaced by $x$ and $z$, respectively. We note that predictions outside the $\delta<0.5$ cut are extrapolations and do not enter the fit.

\begin{table*}
\caption{Central values and $68\%$ confidence intervals of the parameters from the world-data fit.}
\label{tab:parameter results}
\begin{tabular}{p{4cm}p{4cm}p{4cm}p{4cm}}
\hline\hline
Parameter       &  Value                                            & Parameter         & Value \\[2pt]\hline
$N_{u}$         &  $0.0540^{+0.0353}_{-0.0152}$                    & $N_{\bar{u}}$     & $0.0752^{+0.181}_{-0.124}$   \\[5pt]
$N_{d}$         &  $0.0561^{+0.0462}_{-0.0243}$                     & $N_{\bar{d}}$     & $0.0389^{+0.0805}_{-0.0632}$      \\[5pt]
$N_{s}$         &  $-0.00547^{+0.0507}_{-0.0423}$                   & $N_{g}$           & $-0.250^{+0.207}_{-0.334}$     \\[5pt]
$\alpha$        &  $-1.18^{+1.68}_{-1.71}$                           & $\beta$           & $-0.782^{+0.535}_{-0.489}$  \\[5pt]
$\varepsilon$   &  $3.84^{+6.38}_{-3.55}$                            & $\lambda_{1}$     & $0.215^{+0.113}_{-0.0837}$  \\[5pt]
$\lambda_{2}$   &  $0.602^{+0.734}_{-0.516}$                         & $\lambda_{5}$     & $-1.20^{+0.953}_{-1.42}$  \\[5pt]
$r_{\mathrm{CLAS}}$   &  $0.0960^{+0.458}_{-0.424}$              & $r_{\mathrm{HERMES}\,p}$   & $-0.338^{+0.396}_{-0.406}$  \\[5pt]
$r_{\mathrm{HERMES}\,d}$   &  $0.218^{+0.163}_{-0.165}$    & & \\[5pt]
\hline\hline
\end{tabular}
\end{table*}

\begin{figure*}
    \centering
    \includegraphics[width=0.30\linewidth]{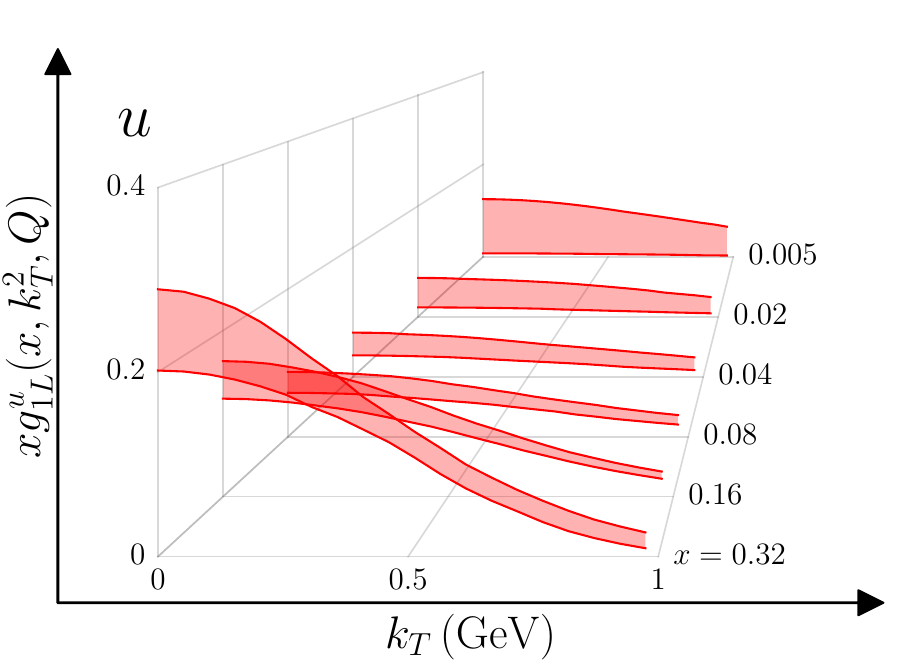}\hfill
    \includegraphics[width=0.30\linewidth]{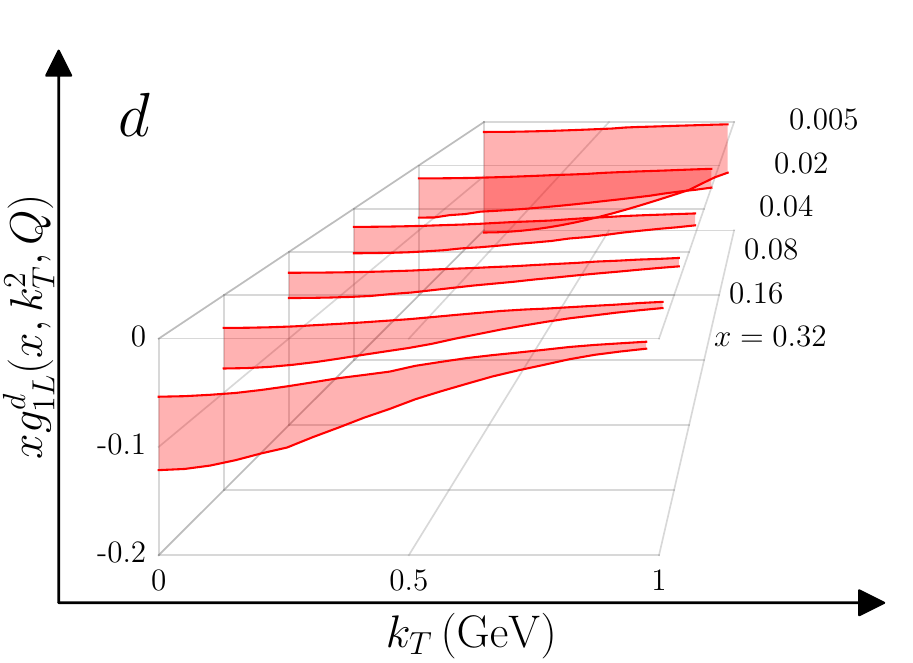}\hfill
    \includegraphics[width=0.30\linewidth]{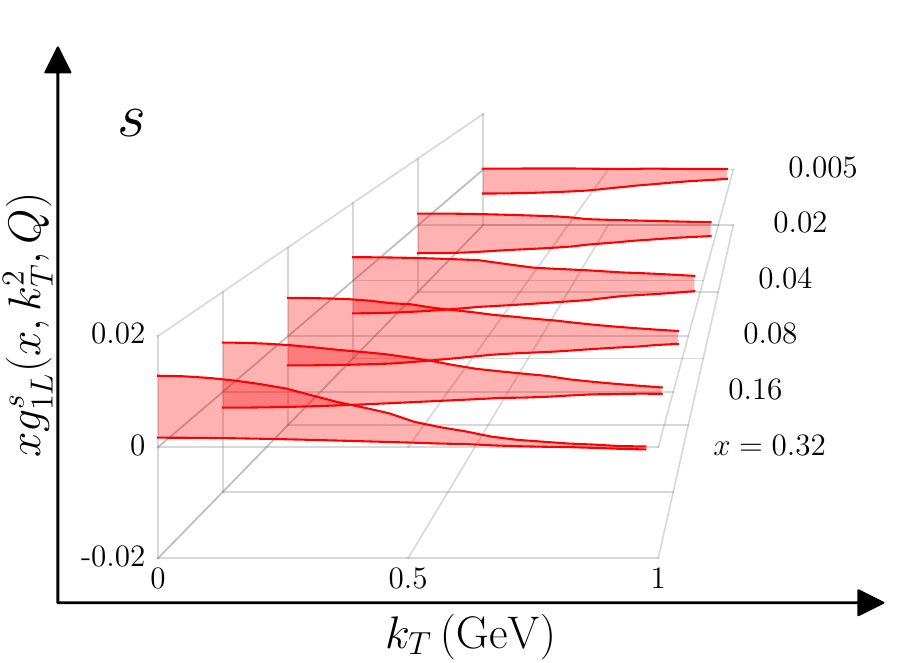}\\
    \includegraphics[width=0.30\linewidth]{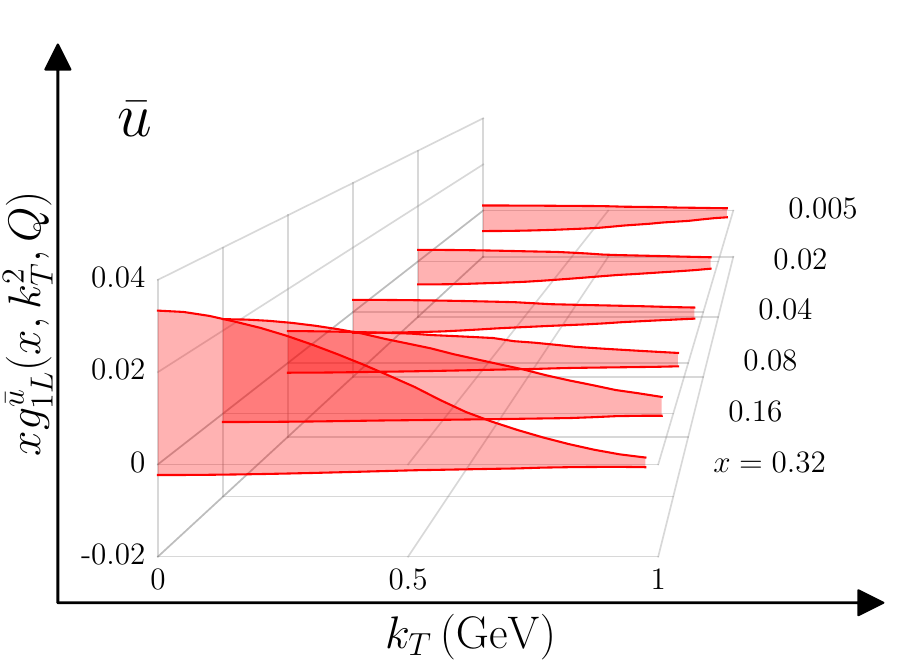}\hfill
    \includegraphics[width=0.30\linewidth]{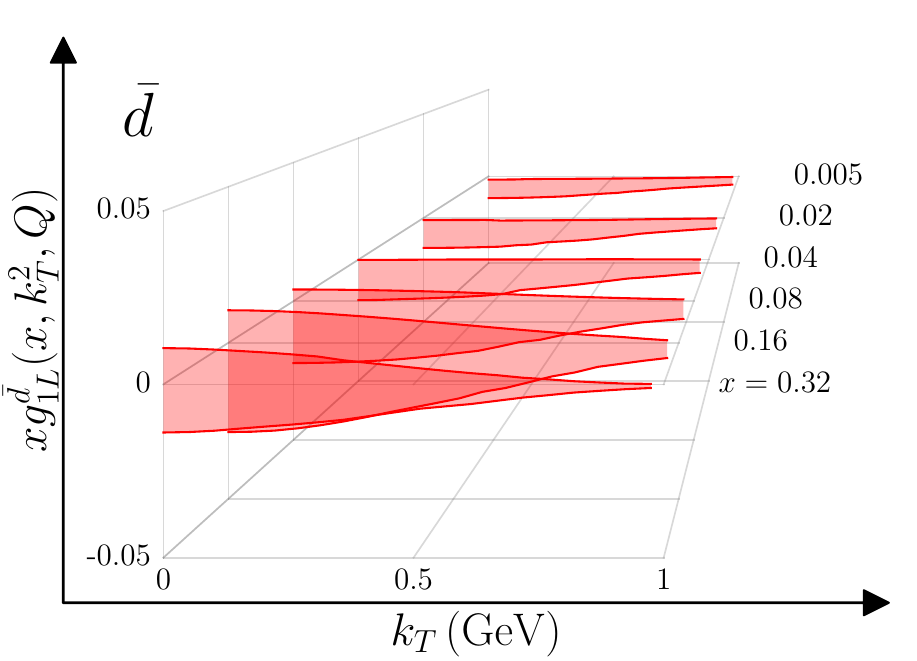}\hfill
    \includegraphics[width=0.30\linewidth]{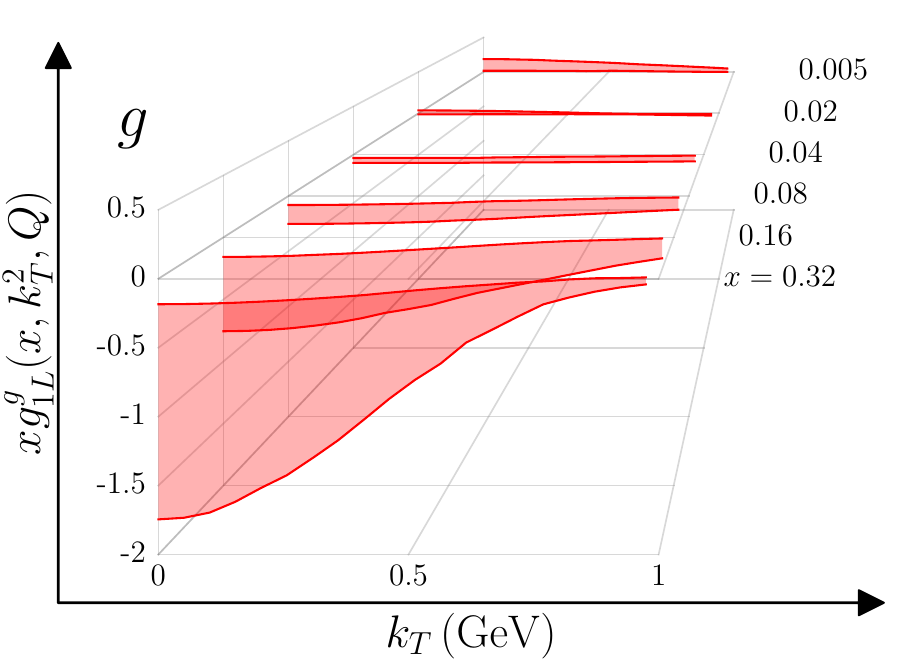}
    \caption{The extracted helicity distributions $xg_{1L}(x,k_T,Q)$ at $Q=2\,\mathrm{GeV}$ for the $u$, $d$, and $s$ quarks (upper row) and the $\bar{u}$, $\bar{d}$ quarks and the gluon (bottom row). The bands represent the $68\%$ confidence intervals estimated from the replicas at fixed values of $x$.}
    \label{fig:xg1LxkT}
\end{figure*}

The extracted helicity distributions $xg_{1L}(x,k_T,Q)$ are shown in Fig.~\ref{fig:xg1LxkT} for the $u$, $d$, and $s$ quarks, as well as the $\bar{u}$, $\bar{d}$ and the gluon.
Among these components, the up- and down-quark distributions are the best determined: both are clearly away from zero over the probed $k_T$ range.
The sea-quark and gluon distributions remain compatible with zero within uncertainties, and the present data are not sufficient to determine their magnitudes or signs.
Improved constraints on these components may be provided by higher-precision measurements with broader kinematic coverage at future facilities.
The projected impact of the EicC is investigated in Sec.~\ref{sec:eicc_impact}.
Despite the different degrees of constraint, the magnitudes of the distributions generally decrease with increasing $k_T$.

To compare with the collinear helicity distributions, we integrate the extracted TMD helicity distributions over transverse momenta up to $k_T^{\mathrm{Max}}=Q$:
\begin{equation}
    g_{1L}(x,Q) = 2\pi\int_{0}^{k_T^{\mathrm{Max}}}
    \dd{k_T} k_T\,g_{1L}(x,k_T,Q).
    \label{eq:g1L_kTintegrated}
\end{equation}
Here $g_{1L}(x,Q)$ denotes the finite-cutoff integral.
With the TMD evaluated at the hard-scattering scales $\mu=Q$ and $\zeta=Q^2$ introduced in Sec.~\ref{sec:framework}, this cutoff corresponds to the natural scale choice $\mu=\sqrt{\zeta}=k_T^{\mathrm{Max}}$.
For $Q\gg\Lambda_{\mathrm{QCD}}$, the relation between such cumulative integrals and collinear PDFs involves perturbative matching and power-suppressed corrections~\cite{Ebert:2022cku}.

\begin{figure*}
    \centering
    \subfloat[]{\includegraphics[width=0.44\linewidth]{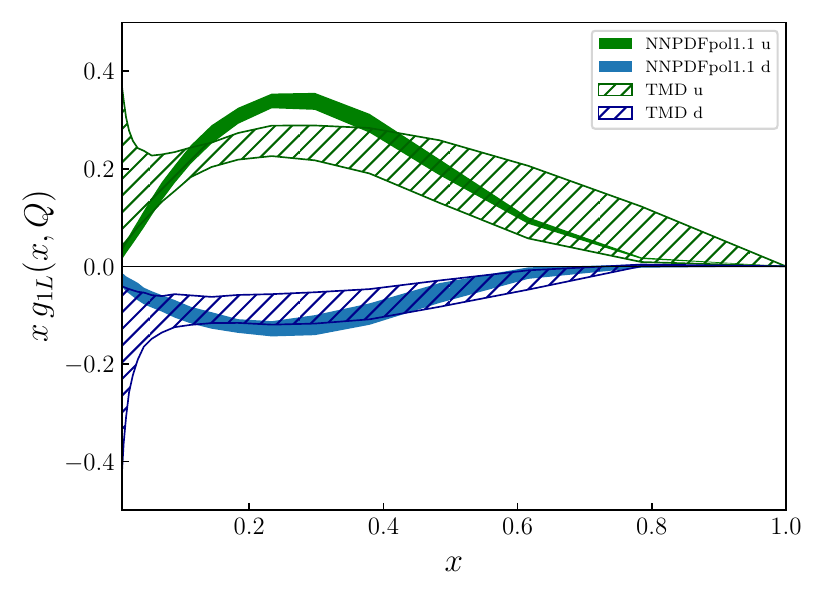}}\hfill
    \subfloat[]{\includegraphics[width=0.44\linewidth]{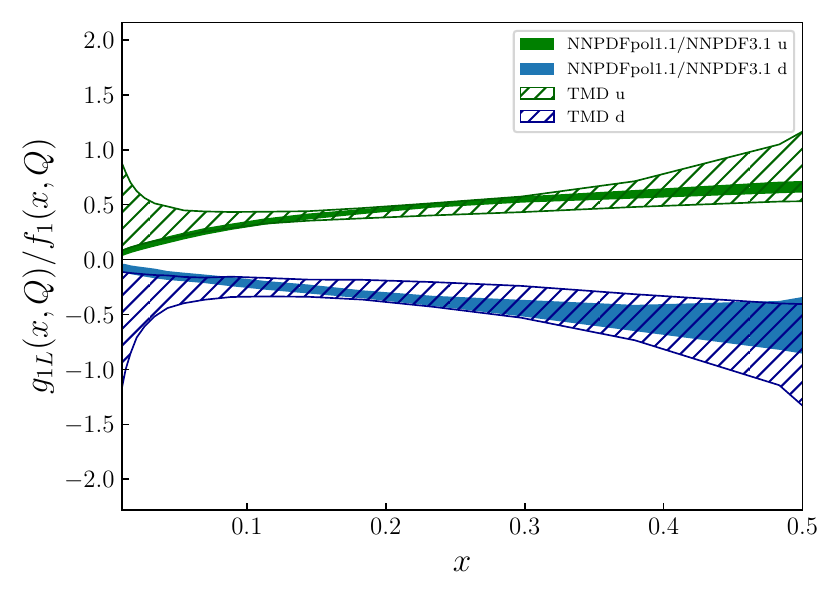}}
    \caption{Left panel: the comparison between the transverse-momentum-integrated TMD helicity distributions $xg_{1L}(x)$ and the collinear helicity distributions. Right panel: comparison of the polarization rates $g_{1L}(x)/f_1(x)$ evaluated from the transverse-momentum-integrated TMD distributions and from collinear distributions. The distributions are evaluated at $Q=2\,\mathrm{GeV}$. The hatched bands are from present TMD extractions, and the filled bands are from collinear distributions with the collinear helicity distributions of NNPDFpol1.1~\cite{Nocera:2014gqa} and the collinear unpolarized distributions of NNPDF3.1~\cite{NNPDF:2017mvq}.}
    \label{fig:g1Lx_comp}
\end{figure*}
The transverse-momentum-integrated distributions are compared with the collinear helicity distributions of NNPDFpol1.1~\cite{Nocera:2014gqa} in Fig.~\ref{fig:g1Lx_comp}, for $xg_{1L}(x)$ and for the polarization rates $g_{1L}(x)/f_1(x)$.
For the up and down quarks, the extracted distributions agree with their collinear counterparts within uncertainties.

\section{Predictions}
\label{sec:predictions}

\subsection{\texorpdfstring{$\cos\phi_h$}{cos phi-h} modulation of longitudinal double spin asymmetry}
\label{sec:asymmetry}

The $\cos\phi_h$ modulation of the longitudinal double spin asymmetry, defined in Eq.~(\ref{eq:A_LL_cosphi_CSA}), involves the structure function $F_{LL}^{\cos\phi_h}$. At the first subleading power in $1/Q$, this structure function reads~\cite{Bacchetta:2006tn}
\begin{equation}
\begin{aligned}
F_{LL}^{\cos\phi_h}
&=
\frac{2M}{Q}
\Bigg\{
-z^2 M_h\mathcal B_1[xe_L,H_1^{\perp(1)}]
\\
&\quad+\frac{z^2 M_h^2}{M}\mathcal B_1\!\left[g_{1L},\frac{\tilde D^{\perp(1)}}{z}\right]
-M\mathcal B_1[x g_L^{\perp(1)},D_1]
\\
&\quad-M_h\mathcal B_1\!\left[h_{1L}^{\perp(1)},\frac{\tilde E}{z}\right]
\Bigg\},
\label{eq:FLL_cosphi_full}
\end{aligned}
\end{equation}
where $e_L$ and $g_L^{\perp}$ are twist-3 TMD distributions, while $h_{1L}^{\perp}$ is a twist-2 TMD distribution. $H_1^{\perp}$ is the Collins FF, and $\tilde D^{\perp}$ and $\tilde E$ are genuine twist-3 TMD~FFs. Here $M_h$ is the detected-hadron mass, and $\mathcal B_\nu$ is the convolution defined in Eq.~(\ref{eq:tmd_b_convolution}).

The QCD equations of motion relate the twist-3 distribution $g_L^{\perp}$ to the helicity distribution $g_{1L}$~\cite{Bacchetta:2006tn},
\begin{equation}
    x g_L^{\perp}
    = x \tilde g_L^{\perp} + g_{1L} + \frac{m}{M}\,h_{1L}^{\perp} ,
    \label{eq:eom_gLperp}
\end{equation}
where $\tilde g_L^{\perp}$ denotes the genuine twist-3 contribution and $m$ the quark mass. In the WW-type approximation, in which the genuine twist-three contributions are dropped, and neglecting the quark-mass term, the structure function reduces to
\begin{equation}
    F_{LL}^{\cos\phi_h}
    \;\approx\;
    -\frac{2M^2}{Q}\,
    \mathcal B_1[g_{1L}^{(1)},D_1] \, ,
    \label{eq:FLL_cosphi_WW}
\end{equation}
where $g_{1L}^{(1)}(x,b_T;Q,Q^2)$ denotes the normalized first derivative of the helicity TMD with respect to $b_T^2$, defined in Appendix~\ref{App:FTT} following the convention of Ref.~\cite{Boussarie:2023izj}. We evaluate this expression using the evolved leading-twist TMDs, obtaining a phenomenological WW-type estimate of the $\cos\phi_h$ modulation.

\begin{figure*}
    \centering
    \includegraphics[width=0.44\linewidth]{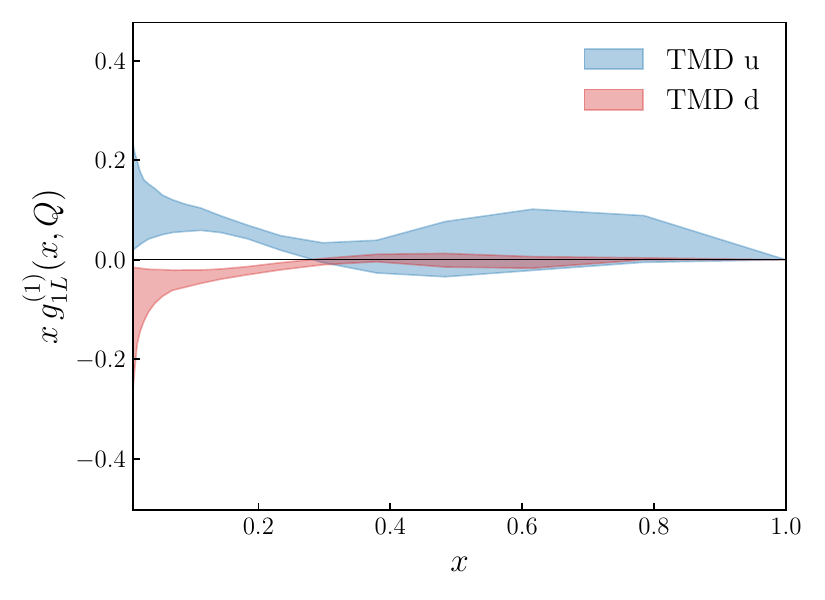}\hfill
    \includegraphics[width=0.44\linewidth]{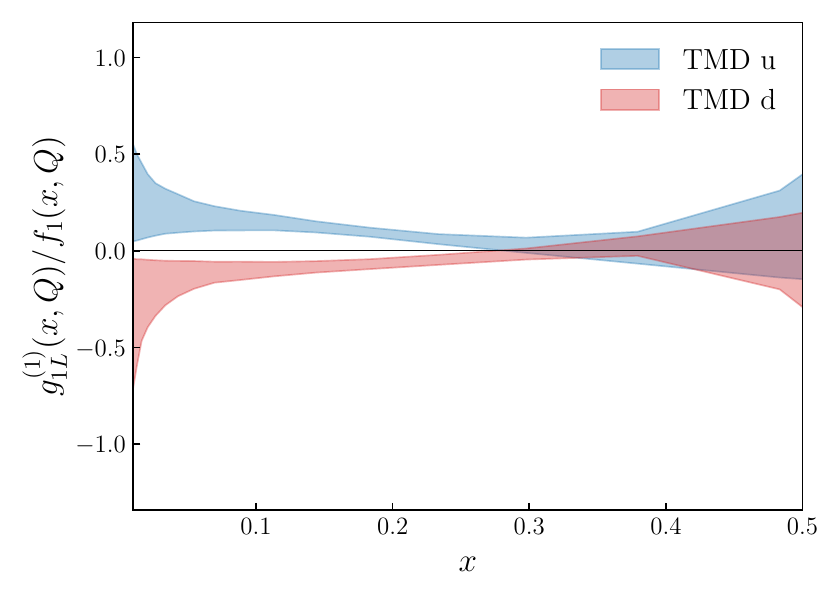}
    \caption{The first-order transverse-momentum moment of the helicity TMD, evaluated with $k_T^{\mathrm{Max}}=Q$ at $Q=2\,\mathrm{GeV}$ for the up (blue) and down (red) quarks, shown as the distributions $xg_{1L}^{(1)}(x,Q)$ (left panel) and the ratios $g_{1L}^{(1)}(x,Q)/f_1(x,Q)$ (right panel). The bands represent the $68\%$ confidence intervals obtained from the extracted TMD helicity distributions.}
    \label{fig:g1L1x_comp}
\end{figure*}

To characterize the transverse-momentum dependence of the extracted helicity TMD, we also consider its first transverse-momentum moment
with the finite cutoff $k_T^{\mathrm{Max}}=Q$,
\begin{equation}
    g_{1L}^{(1)}(x,Q)
    \equiv
    \int_{k_T\le Q}\!\dd[2]{\boldsymbol{k}_T}
    \frac{k_T^2}{2M^2}\,g_{1L}(x,k_T,Q).
    \label{eq:g1L_first_kT_moment}
\end{equation}
This truncated moment is distinct from the finite-$b_T$ derivative entering Eq.~(\ref{eq:FLL_cosphi_WW}).
The relation between the untruncated moment and the derivative at $b_T=0$, discussed in Appendix~\ref{App:FTT}, is formal and requires an ultraviolet prescription when perturbative transverse-momentum tails are included.
Figure~\ref{fig:g1L1x_comp} shows $xg_{1L}^{(1)}(x,Q)$ and the ratio $g_{1L}^{(1)}(x,Q)/f_1(x,Q)$ for up and down quarks.
Here $f_1(x,Q)$ is the unpolarized TMD integrated over transverse momenta up to the same cutoff $k_T=Q$.
These quantities illustrate the flavor dependence of the transverse momentum moments.

\begin{figure*}
    \centering
    \includegraphics[width=0.52\linewidth]{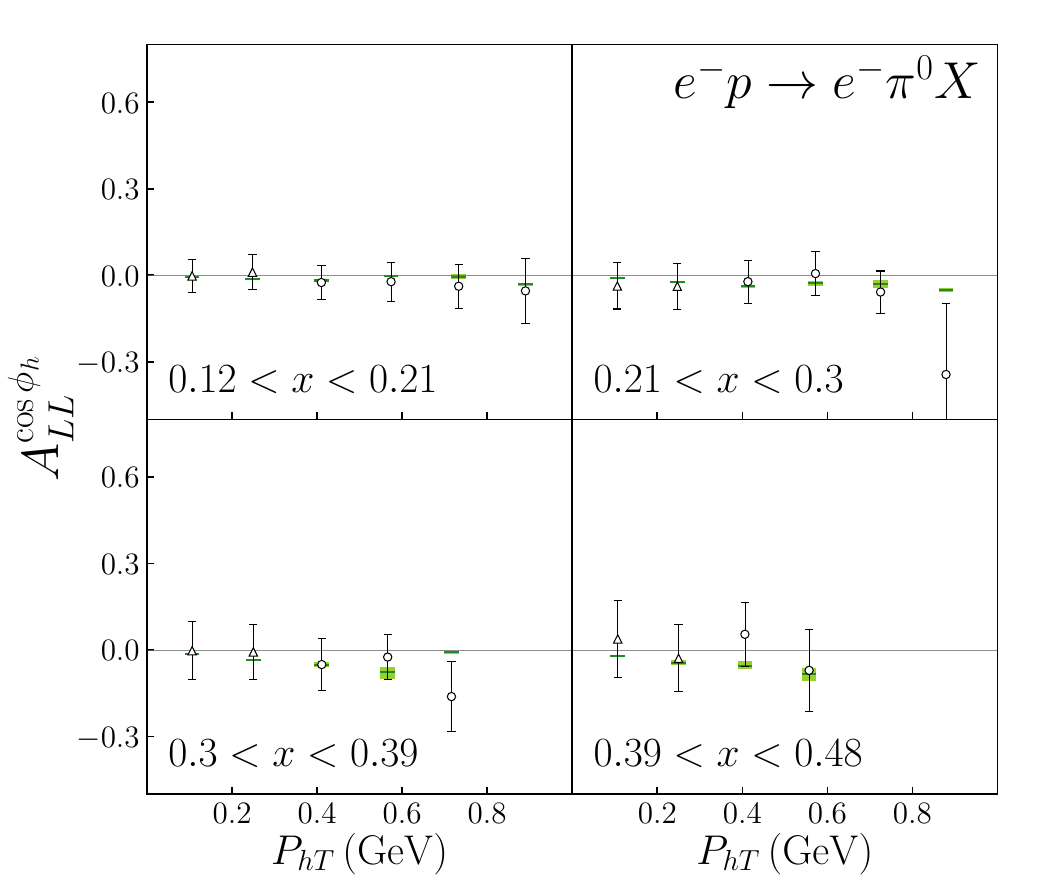}
    \caption{Comparison of the theory prediction (bands) with the CLAS
    $A_{LL}^{\cos\phi_h}$ data~\cite{CLAS:2017yrm} in four bins of $x$, as a function of $P_{h\perp}$. The bands represent the $68\%$ confidence intervals from the extracted TMD helicity distributions. All experimental data shown were excluded from the fit. Triangles indicate $\delta<0.5$, while circles indicate $\delta\geq0.5$, where the predictions are extrapolations.}
    \label{fig:cosphi_clas}
\end{figure*}

\begin{figure*}
    \centering
    \includegraphics[width=0.495\linewidth]{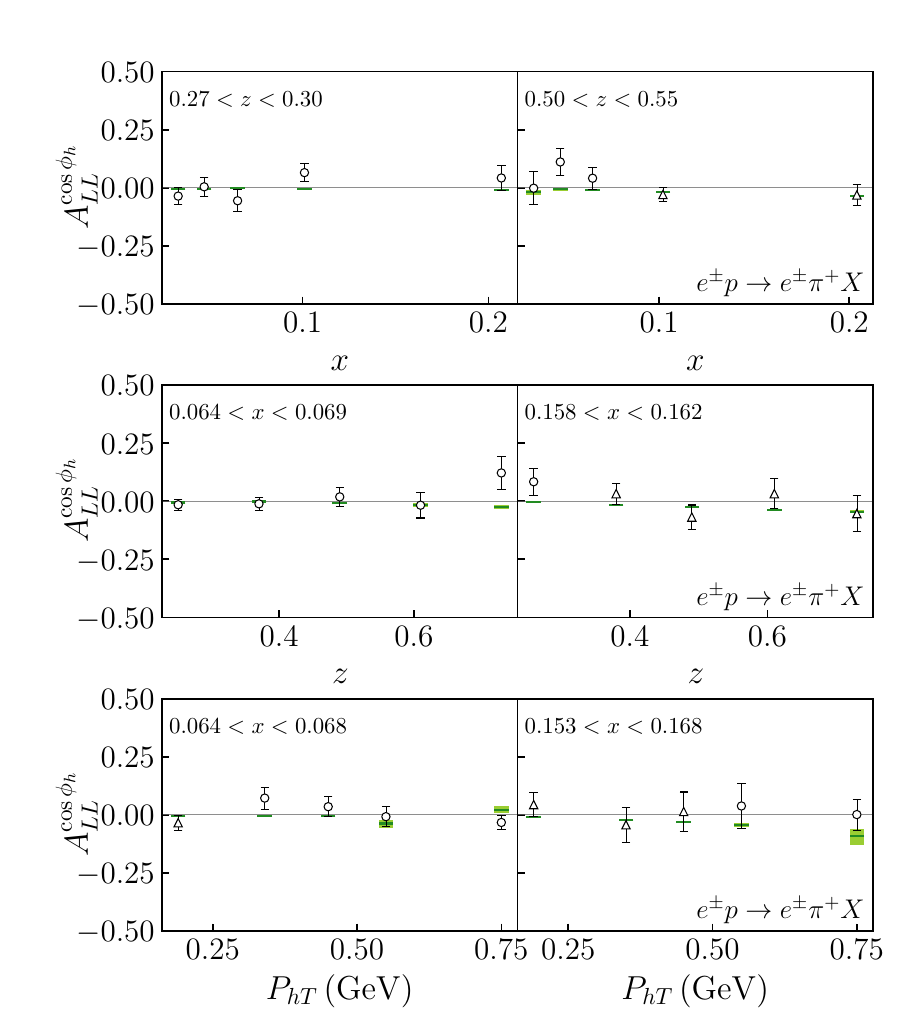}
    \includegraphics[width=0.495\linewidth]{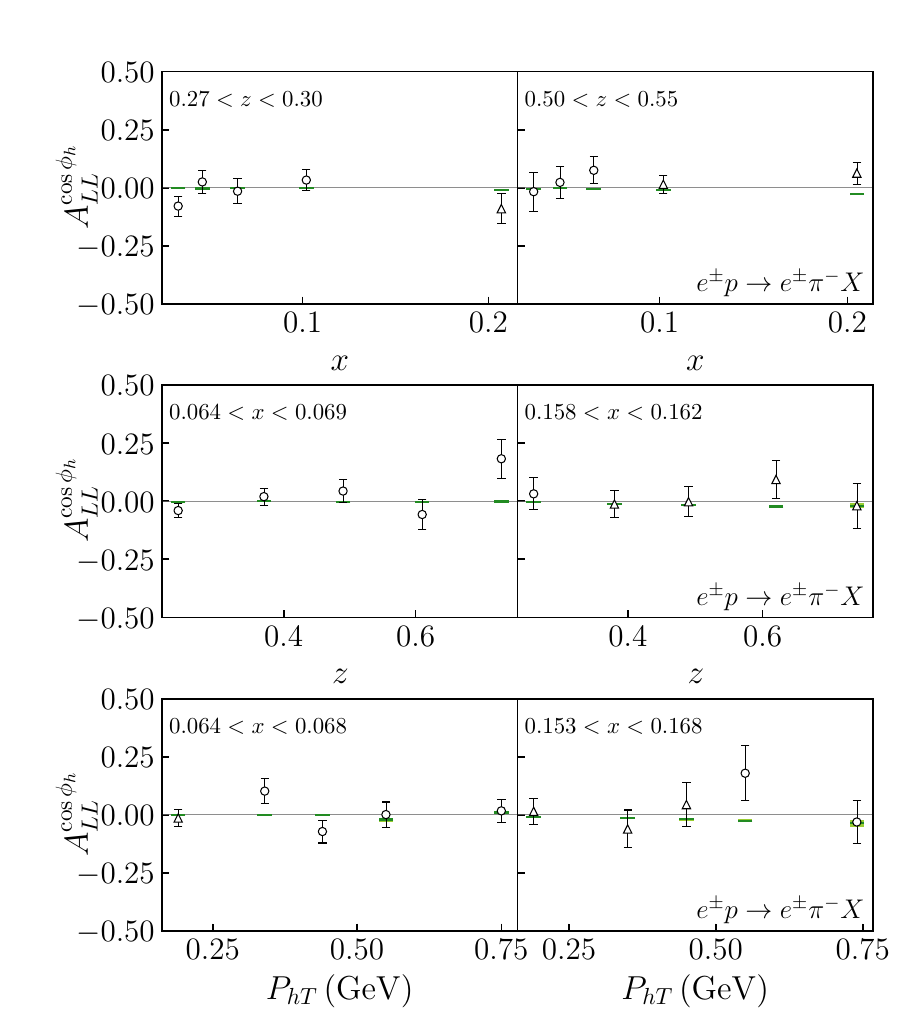}
    \includegraphics[width=0.495\linewidth]{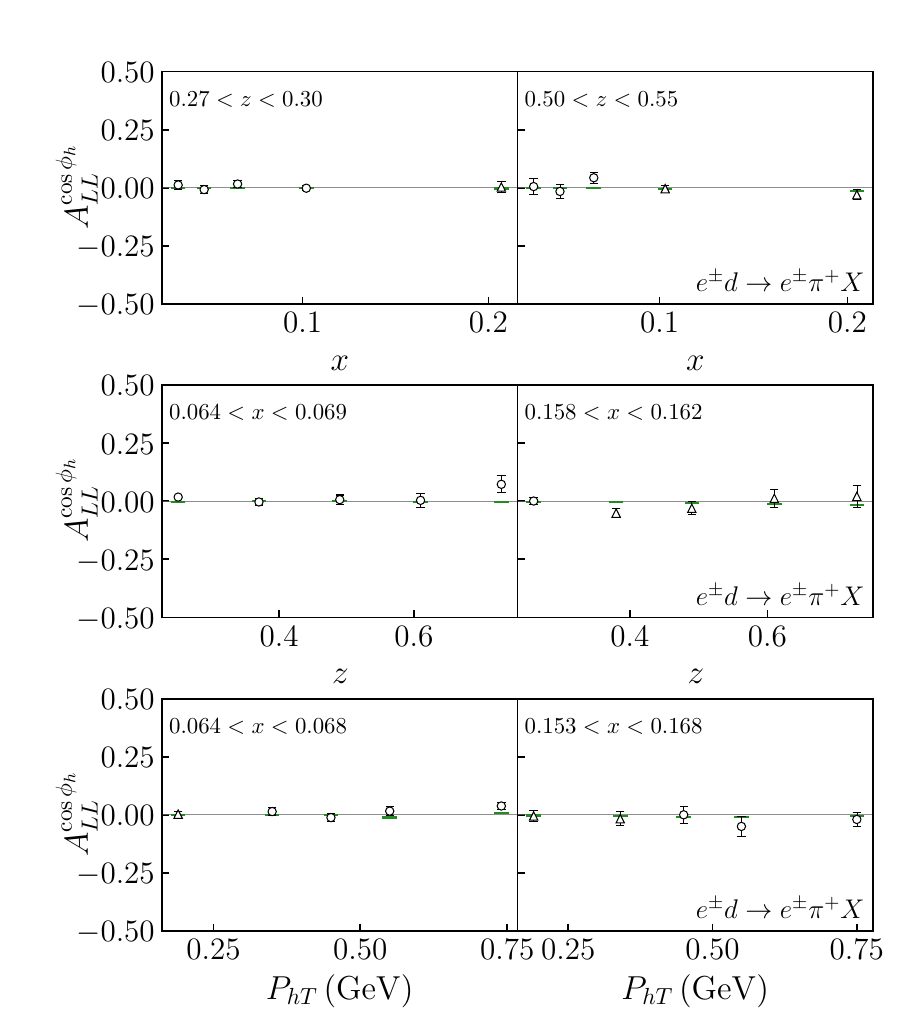}
    \includegraphics[width=0.495\linewidth]{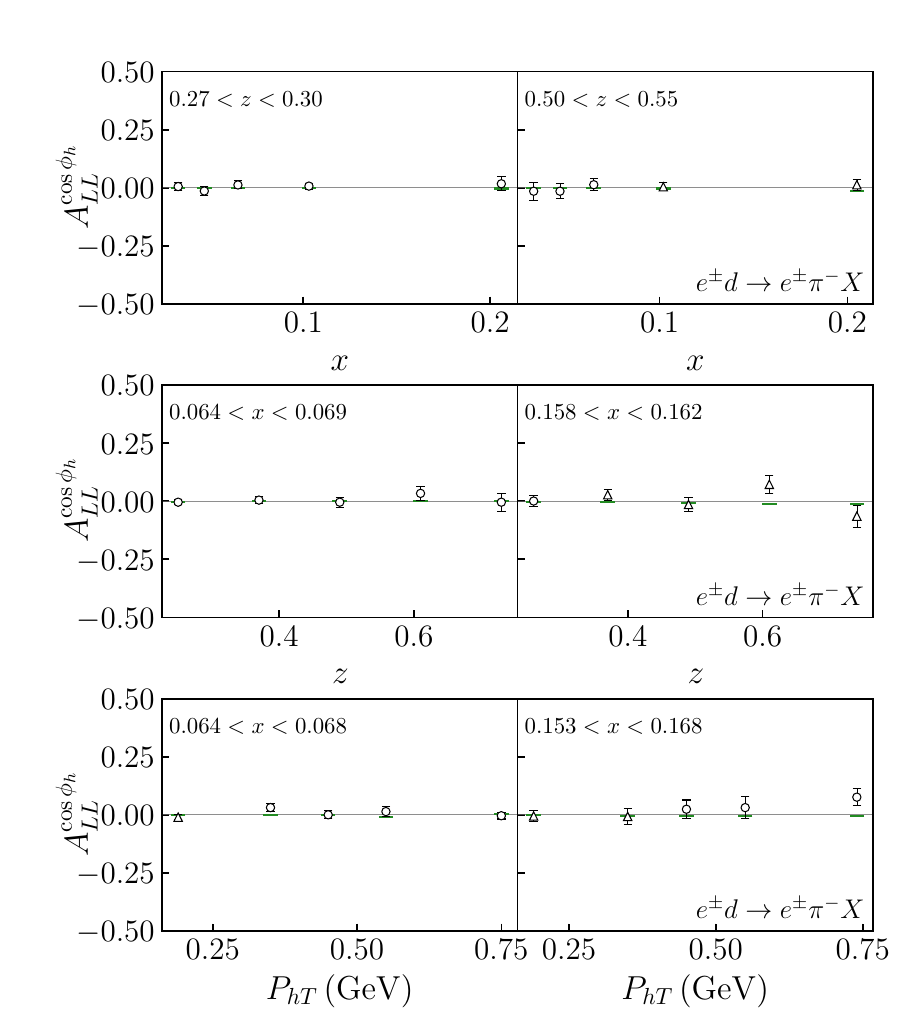}
    \caption{Comparison of the theory prediction (bands) with the HERMES
    $A_{LL}^{\cos\phi_h}$ data~\cite{HERMES:2018awh}. The panels correspond to different targets and produced hadrons. Within each panel, the three rows show the dependence on $x$, $z$, and $P_{h\perp}$, respectively, for the kinematic selections indicated. The bands represent the $68\%$ confidence intervals from the extracted TMD helicity distributions. All experimental data shown were excluded from the fit. Triangles indicate $\delta<0.5$, while circles indicate $\delta\geq0.5$, where the predictions are extrapolations.}
    \label{fig:cosphi_hermes_pion}
\end{figure*}

\begin{figure*}
    \centering
    \includegraphics[width=0.495\linewidth]{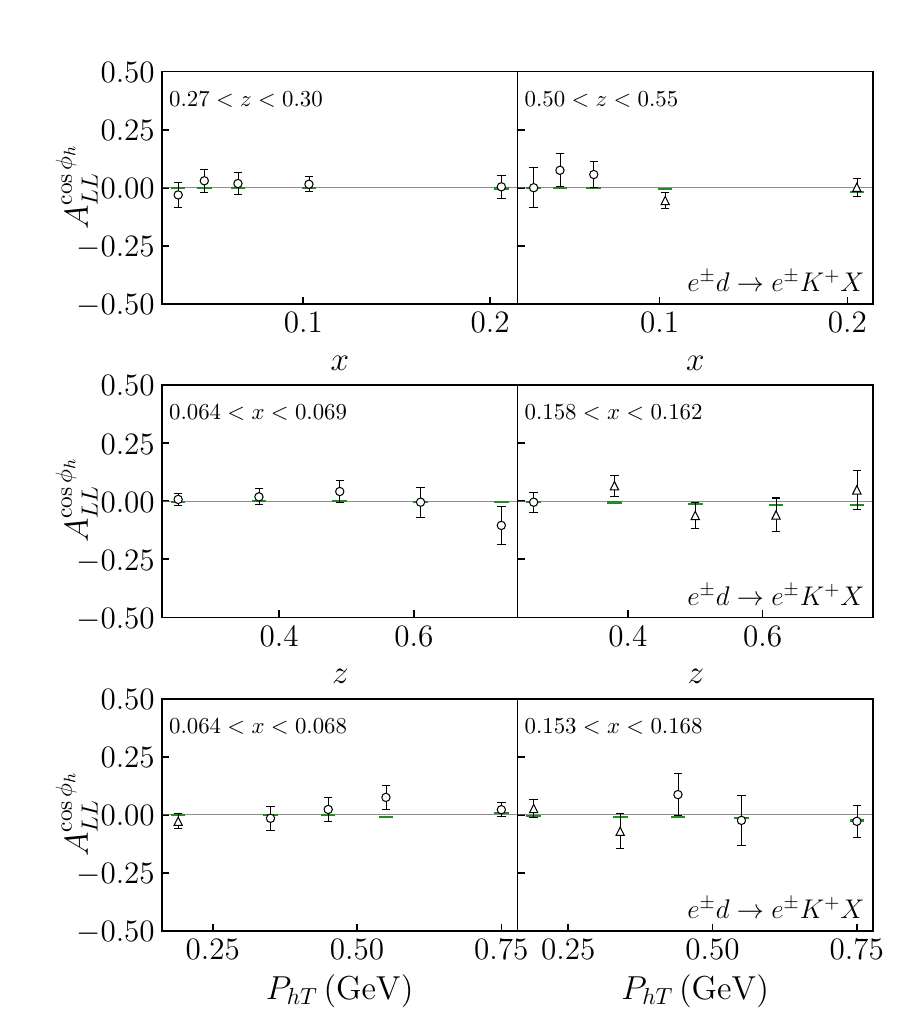}
    \includegraphics[width=0.495\linewidth]{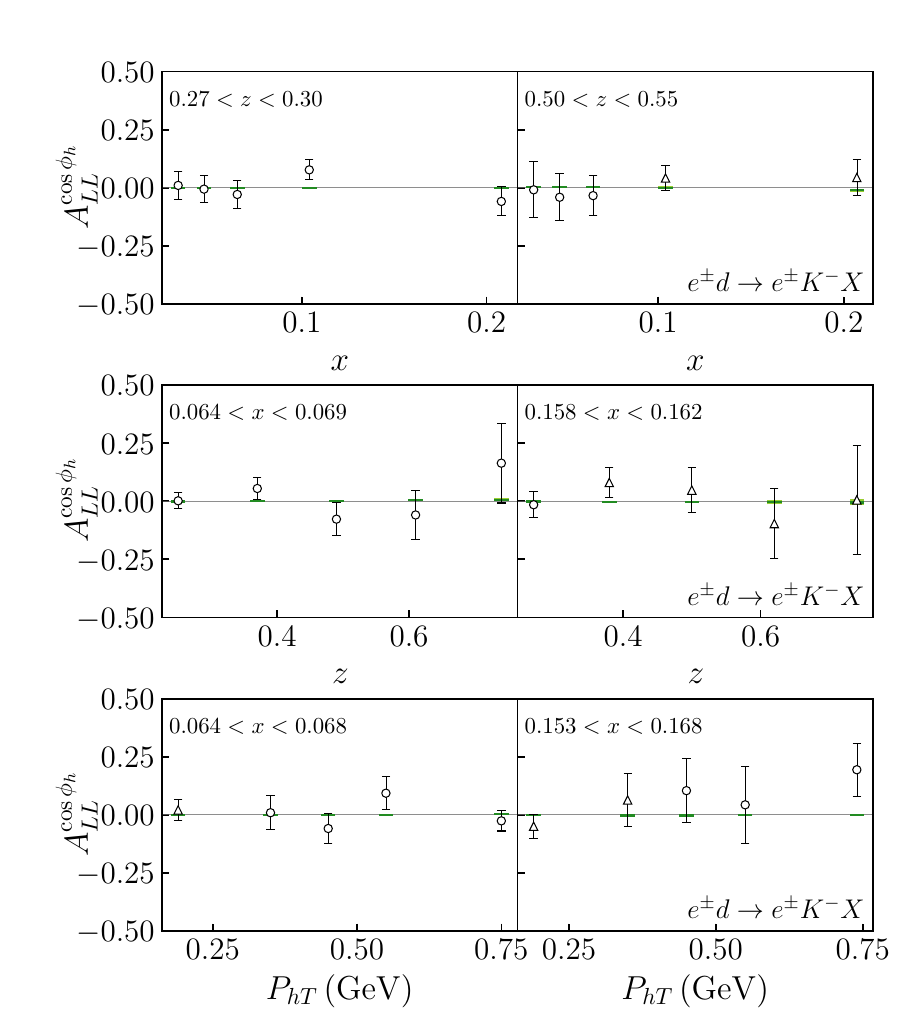}
    \caption{Same as Fig.~\ref{fig:cosphi_hermes_pion}, but for $e^{\pm}d\to e^{\pm}K^{+}X$ (left panel) and $e^{\pm}d\to e^{\pm}K^{-}X$ (right panel).}
    \label{fig:cosphi_hermes_kaon}
\end{figure*}

Using the extracted TMDs, we evaluate the $\cos\phi_h$ asymmetry of Eq.~(\ref{eq:A_LL_cosphi_CSA}) in the WW-type approximation and compare it with the measurements of HERMES~\cite{HERMES:2018awh} and CLAS~\cite{CLAS:2017yrm}, which were not included in the fit.
The comparisons are shown in Figs.~\ref{fig:cosphi_clas}--\ref{fig:cosphi_hermes_kaon}.
Considering only uncorrelated experimental uncertainties, we obtain $\chi^2_{\rm uncor}/N=0.710$ for the 66 data points satisfying $\delta<0.5$.
Including all 201 points gives $\chi^2_{\rm uncor}/N=0.852$; this comparison also includes extrapolations outside the low transverse-momentum region adopted in the fit.
The comparison indicates that the WW-type approximation provides a good description of the measured azimuthal modulation of the longitudinal double spin asymmetry within the present experimental uncertainties.

\subsection{CLAS12 SIDIS \texorpdfstring{$A_{LL}$}{A-LL} predictions}
\label{sec:clas12_prediction}

We next evaluate the azimuthally averaged longitudinal double spin asymmetry $A_{LL}$ defined in Eq.~(\ref{eq:ALL_CSA}) for charged-pion production at representative fixed-target kinematics relevant to CLAS12.
The calculation uses the factorized structure functions of Sec.~\ref{sec:framework} for a $10.6\,\mathrm{GeV}$ electron beam incident on a proton target, with $z=0.5$.
At fixed beam energy, the inelasticity is determined by $x$ and $Q^2$.

\begin{figure*}
    \centering
    \includegraphics[width=0.31\linewidth]{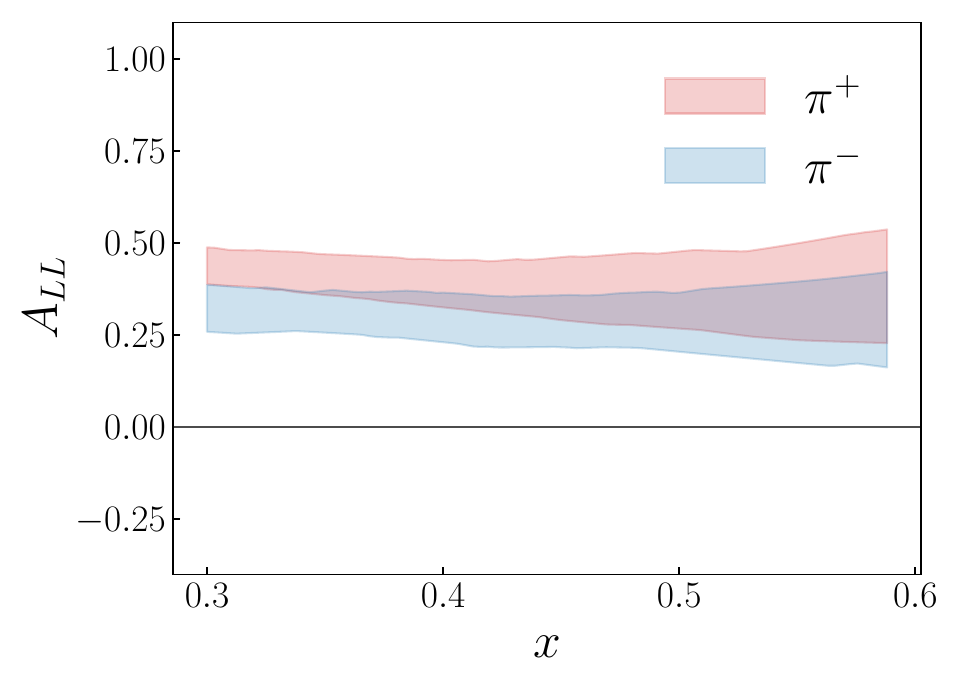}\hfill
    \includegraphics[width=0.31\linewidth]{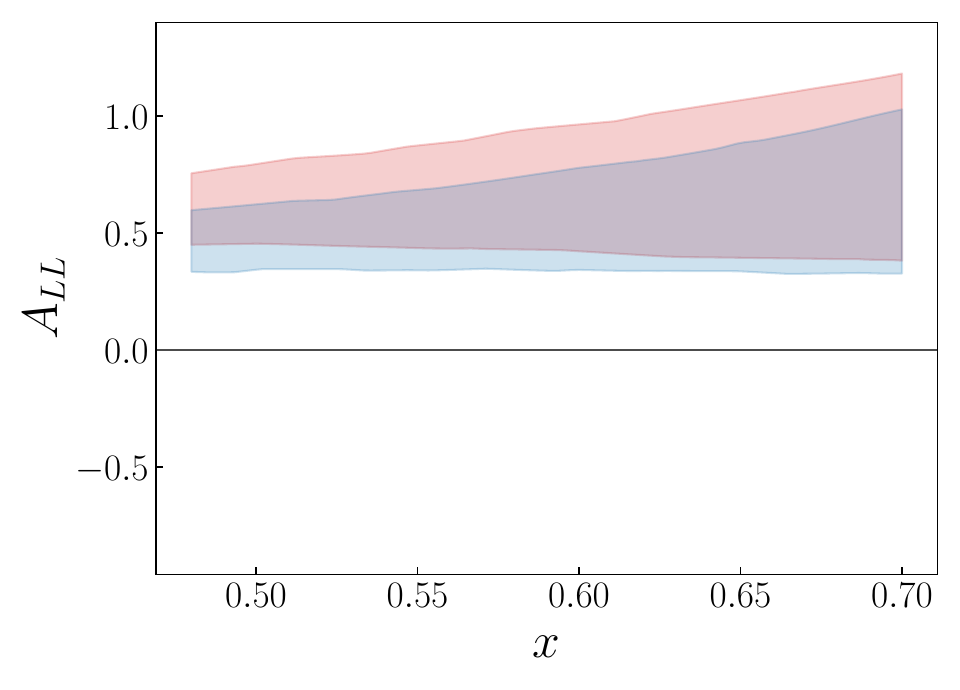}\hfill
    \includegraphics[width=0.31\linewidth]{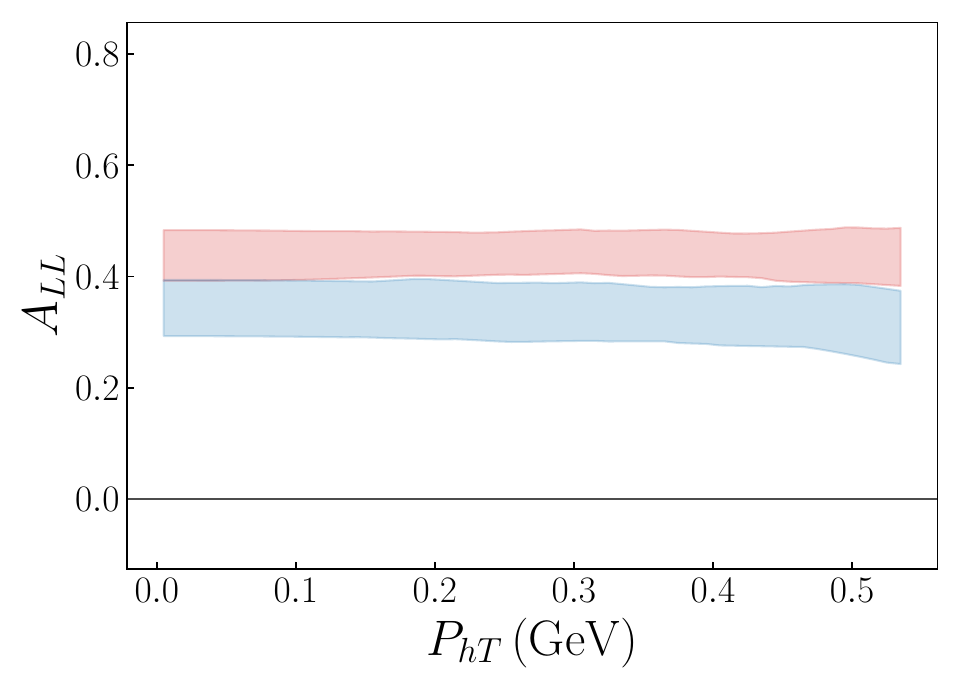}\\[3pt]
    \includegraphics[width=0.31\linewidth]{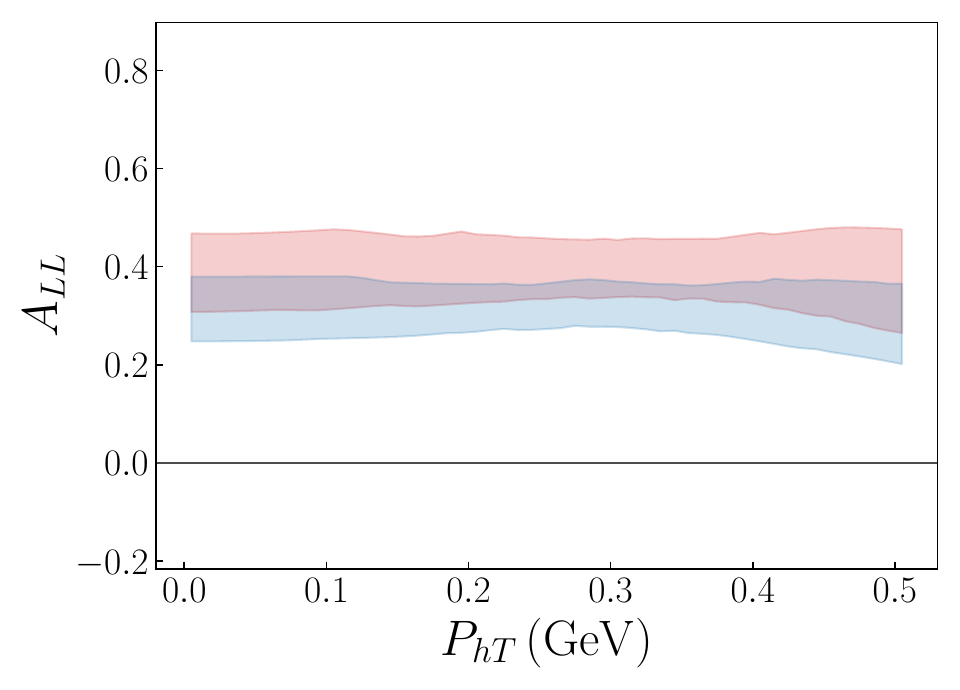}\hfill
    \includegraphics[width=0.31\linewidth]{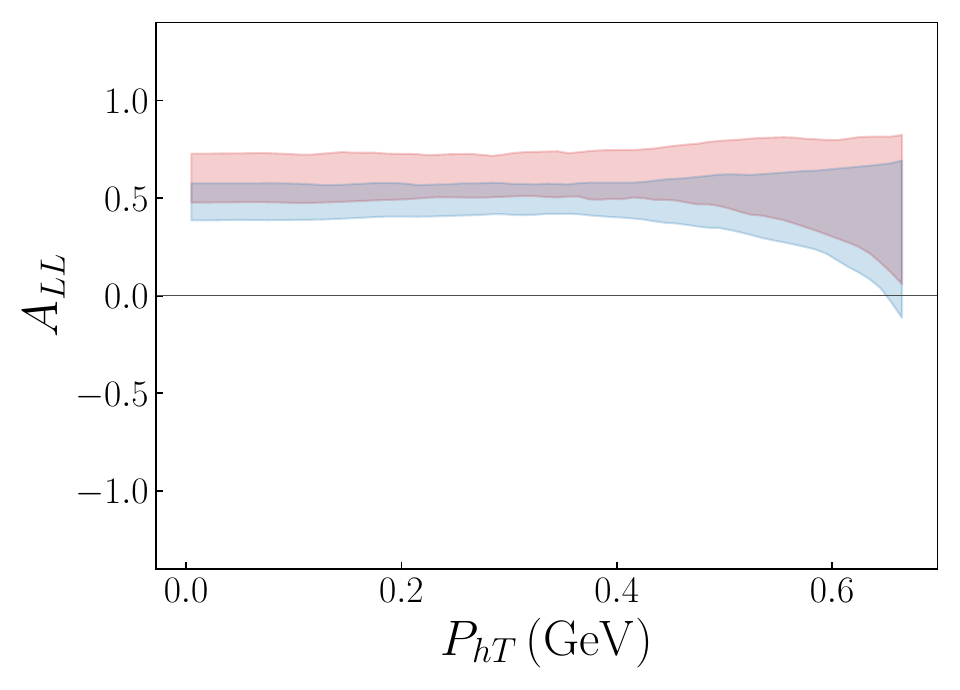}\hfill
    \includegraphics[width=0.31\linewidth]{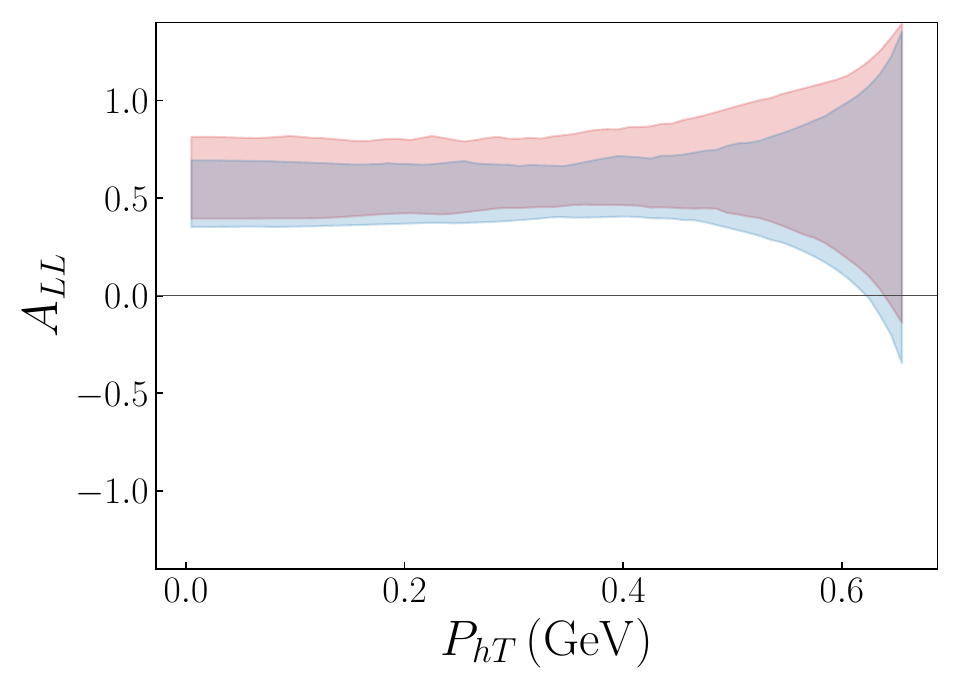}
    \caption{Predictions for the azimuthally averaged SIDIS longitudinal double spin asymmetry $A_{LL}$ for a $10.6\,\mathrm{GeV}$ electron beam on a proton target at $z=0.5$. The upper row shows the $x$ dependence at $Q^2=5$ and $8\,\mathrm{GeV}^2$ and the $P_{h\perp}$ dependence at $(x,Q^2)=(0.3,5\,\mathrm{GeV}^2)$.
The lower row shows the $P_{h\perp}$ dependence at $(x,Q^2)=(0.5,5\,\mathrm{GeV}^2)$, $(0.5,8\,\mathrm{GeV}^2)$, and $(0.6,8\,\mathrm{GeV}^2)$. Red and blue bands correspond to $\pi^+$ and $\pi^-$ production, respectively, and represent the $68\%$ confidence intervals from the extracted TMD helicity distributions. Only kinematics satisfying $\delta<0.5$ are shown.}
    \label{fig:clas12_all}
\end{figure*}

Figure~\ref{fig:clas12_all} shows the predictions for $\pi^+$ and $\pi^-$ production as functions of $x$ at $Q^2=5$ and $8\,\mathrm{GeV}^2$, and as functions of $P_{h\perp}$ at the fixed $(x,Q^2)$ values. Only kinematics satisfying $\delta<0.5$, the cut adopted in the fit, are shown. The bands represent the $68\%$ confidence intervals from the extracted TMD helicity distributions.

\subsection{STAR \texorpdfstring{$W^\pm$}{W+/-} single-spin-asymmetry prediction}
\label{sec:star_prediction}

We use the helicity TMDs extracted from SIDIS to predict the parity-violating longitudinal single-spin asymmetry in $W^\pm$ production, with fiducial cuts on the charged decay lepton:
\begin{equation}
    \begin{aligned}
        p(P_A,S_L)+p(P_B)&\to W^\pm(q)+X\\
                          &\to
        \begin{cases}
            e^+(l)+\nu_e+X, & W^+,\\
            e^-(l)+\bar\nu_e+X, & W^-.
        \end{cases}
    \end{aligned}
\end{equation}
Beam $A$ moves along the positive laboratory $\hat{z}$ axis and is longitudinally polarized; the polarization of beam $B$ is averaged. Positive $S_L$ denotes proton spin aligned with the momentum of beam $A$. We use the positive-minus-negative helicity convention for the polarization-normalized asymmetry,
\begin{equation}
    A_{LU}^{W^\pm}(q_T)=
    \frac{F_{LU}^{W^\pm,\mathrm{acc}}(q_T)}
         {F_{UU}^{W^\pm,\mathrm{acc}}(q_T)} ,
\label{eq:star_alu-new}
\end{equation}
where the first (second) polarization label refers to beam $A$ ($B$), and the superscript ``acc'' denotes the lepton-level fiducial acceptance.

We consider proton-proton collisions at $\sqrt{s}=510\,\mathrm{GeV}$ and treat the $W$ boson as on shell, with hard scale $Q=M_W$. Here $q_T$ and $y_W$ denote the transverse momentum and rapidity of the produced boson. In the leading-power TMD description, valid for $q_T\ll M_W$, the longitudinal partonic momentum fractions are
\begin{equation}
    x_1=\frac{M_W}{\sqrt{s}}e^{y_W},\qquad
    x_2=\frac{M_W}{\sqrt{s}}e^{-y_W}.
\label{eq:star_x12-new}
\end{equation}
Requiring $0<x_1,x_2\leq1$ gives
\begin{equation}
    -\ln\frac{\sqrt{s}}{M_W}\leq y_W\leq
     \ln\frac{\sqrt{s}}{M_W}.
\end{equation}
The transverse mass,
\begin{equation}
    M_T=\sqrt{M_W^2+q_T^2},
\end{equation}
determines the physical boson four-momentum used in the decay kinematics,
\begin{equation}
    q^\mu=(M_T\cosh y_W,q_T,0,M_T\sinh y_W).
\label{eq:star-w-four-momentum}
\end{equation}
We retain the transverse recoil in the decay kinematics while using the leading-power momentum fractions and rapidity limits above.

For a flavor product $\Phi(b_T,M_W)$, define
\begin{equation}
\begin{split}
    \mathcal{C}_{ud}[\Phi]&\equiv
    |V_{ud}|^2H_W(M_W)\\
    &\quad\times\int_0^\infty\frac{b_T \dd{b_T}}{2\pi}\,
    J_0(b_Tq_T)\Phi(b_T,M_W),
\label{eq:star-convolution-operator}
\end{split}
\end{equation}
where $V_{ud}$ is the Cabibbo-Kobayashi-Maskawa (CKM) matrix element, and $H_W$ is the charged-current hard factor.
The common hard factor cancels in the asymmetry in Eq.~(\ref{eq:star_alu-new}); it is kept here to make the factorized structure explicit.
Throughout this subsection, $f_1(x,b_T,M_W)$ and $g_{1L}(x,b_T,M_W)$ denote the evolved TMDs evaluated at $(\mu,\zeta)=(M_W,M_W^2)$.

We include only the charged-current channels involving $u$ and $d$
quarks and antiquarks. Contributions involving strange and charm
quarks are neglected in this approximation. With the beam ordering
specified above, the $W^+$ contributions are~\cite{Huang:2015vpy}
\begin{align}
 \mathcal F_{UU}^{u_A\bar d_B}
 &=\mathcal C_{ud}\!\left[
 f_1^{u/A}(x_1,b_T,M_W)
 f_1^{\bar d/B}(x_2,b_T,M_W)\right], \notag\\
 \mathcal F_{LU}^{u_A\bar d_B}
 &=\mathcal C_{ud}\!\left[
 -g_{1L}^{u/A}(x_1,b_T,M_W)
 f_1^{\bar d/B}(x_2,b_T,M_W)\right], \notag\\
 \mathcal F_{UU}^{\bar d_Au_B}
 &=\mathcal C_{ud}\!\left[
 f_1^{\bar d/A}(x_1,b_T,M_W)
 f_1^{u/B}(x_2,b_T,M_W)\right], \notag\\
 \mathcal F_{LU}^{\bar d_Au_B}
 &=\mathcal C_{ud}\!\left[
 +g_{1L}^{\bar d/A}(x_1,b_T,M_W)
 f_1^{u/B}(x_2,b_T,M_W)\right].
\label{eq:star-wplus-structure-functions}
\end{align}
Similarly, the contributions to $W^-$ production are
\begin{align}
 \mathcal F_{UU}^{d_A\bar u_B}
 &=\mathcal C_{ud}\!\left[
 f_1^{d/A}(x_1,b_T,M_W)
 f_1^{\bar u/B}(x_2,b_T,M_W)\right], \notag\\
 \mathcal F_{LU}^{d_A\bar u_B}
 &=\mathcal C_{ud}\!\left[
 -g_{1L}^{d/A}(x_1,b_T,M_W)
 f_1^{\bar u/B}(x_2,b_T,M_W)\right], \notag\\
 \mathcal F_{UU}^{\bar u_Ad_B}
 &=\mathcal C_{ud}\!\left[
 f_1^{\bar u/A}(x_1,b_T,M_W)
 f_1^{d/B}(x_2,b_T,M_W)\right], \notag\\
 \mathcal F_{LU}^{\bar u_Ad_B}
 &=\mathcal C_{ud}\!\left[
 +g_{1L}^{\bar u/A}(x_1,b_T,M_W)
 f_1^{d/B}(x_2,b_T,M_W)\right].
\label{eq:star-wminus-structure-functions}
\end{align}
The relative signs of the two $F_{LU}$ beam orderings follow from the left-handed charged-current coupling and the quark-antiquark interchange rule in Ref.~\cite{Huang:2015vpy}.

To impose the lepton cuts, we construct the Collins-Soper basis in the $W$ rest frame~\cite{Collins:1977iv}. The laboratory frame coincides with the hadronic center-of-mass frame. Neglecting the proton masses, the beam momenta are
\begin{equation}
    P_A^\mu=\left(\frac{\sqrt{s}}{2},0,0,+\frac{\sqrt{s}}{2}\right),
    \qquad
    P_B^\mu=\left(\frac{\sqrt{s}}{2},0,0,-\frac{\sqrt{s}}{2}\right),
\end{equation}
The $W$ boson velocity in the laboratory frame is
\begin{equation}
    \boldsymbol\beta_W=\frac{\boldsymbol q}{q^0}.
\end{equation}
The boosted beam four-momenta and their unit spatial directions are defined by
\begin{equation}
    P_{i*}^\mu=\Lambda^\mu{}_{\nu}(-\boldsymbol\beta_W)P_i^\nu
    =\left(E_{i*},\boldsymbol p_{i*}\right),
    \qquad
    \widehat{\boldsymbol p}_{i*}=
    \frac{\boldsymbol p_{i*}}{|\boldsymbol p_{i*}|},
\end{equation}
where $i=A,B,$ and $\Lambda(-\boldsymbol\beta_W)$ is the Lorentz boost from the laboratory frame to the $W$ rest frame. The orthonormal basis is
\begin{align}
 \boldsymbol e_z&=
 \frac{\widehat{\boldsymbol p}_{A*}-\widehat{\boldsymbol p}_{B*}}
 {\left|\widehat{\boldsymbol p}_{A*}-\widehat{\boldsymbol p}_{B*}\right|},
 &
 \boldsymbol e_x&=
 \frac{\widehat{\boldsymbol p}_{A*}+\widehat{\boldsymbol p}_{B*}}
 {\left|\widehat{\boldsymbol p}_{A*}+\widehat{\boldsymbol p}_{B*}\right|},
 &
 \boldsymbol e_y&=\boldsymbol e_z\times\boldsymbol e_x .
\label{eq:star-collins-soper-basis}
\end{align}
At $q_T=0$, for which the numerator defining $\boldsymbol e_x$ vanishes, an arbitrary transverse unit vector may be used for $\boldsymbol e_x$; the azimuthal integration makes the result independent of this choice. In the limit $q_T\to0$, $\boldsymbol e_z$ approaches the direction of beam $A$.

Let $c\equiv\cos\theta$, where $\theta$ is the polar angle of the charged lepton relative to $\boldsymbol e_z$, and let $\phi$ be its azimuthal angle. The charged lepton is treated as massless, so that its momentum in the $W$ rest frame is
\begin{align}
 l_*^\mu&=
 \left(\frac{M_W}{2},\frac{M_W}{2}\boldsymbol n(c,\phi)\right).\label{eq:star-lepton-rest-frame-momentum}\\
 \boldsymbol n(c,\phi)
 &=\sqrt{1-c^2}\cos\phi\,\boldsymbol e_x+
   \sqrt{1-c^2}\sin\phi\,\boldsymbol e_y+
   c\,\boldsymbol e_z.
\end{align}
Boosting $l_*^\mu=(E_*,\boldsymbol p_*)$ to the laboratory frame with $\boldsymbol\beta_W$ gives
\begin{align}
 E_e&=\gamma_W\left(E_*+\boldsymbol\beta_W\cdot\boldsymbol p_*\right),\notag\\
 \boldsymbol p_e&=\boldsymbol p_*+
 \left[
 \frac{\gamma_W^2}{\gamma_W+1}
 \left(\boldsymbol\beta_W\cdot\boldsymbol p_*\right)
 +\gamma_WE_*
 \right]\boldsymbol\beta_W,\\
 \gamma_W&=(1-\beta_W^2)^{-1/2},
 \qquad \beta_W=|\boldsymbol\beta_W|.
\label{eq:star-lepton-laboratory-momentum}
\end{align}
For $l^\mu=(E_e,p_{e,x},p_{e,y},p_{e,z})$, the fiducial variables are
\begin{equation}
 E_T^e=p_{T,e}=\sqrt{p_{e,x}^2+p_{e,y}^2},
 \qquad
 \eta_e=\operatorname{asinh}\left(\frac{p_{e,z}}{p_{T,e}}\right).
\label{eq:star-lepton-variables}
\end{equation}
Here $E_T^e$ and $\eta_e$ are the transverse energy and pseudorapidity of the charged lepton.

For the decay angular distribution, we take the normalized
Born $V\!-\!A$ weights appropriate to the leading-power
small-$q_T/M_W$ limit~\cite{CDF:2003tlc}:
\begin{align}
 w_q^{W^+}(c)&=\frac{3}{16\pi}(1-c)^2,
 &
 w_{\bar q}^{W^+}(c)&=\frac{3}{16\pi}(1+c)^2, \notag\\
 w_q^{W^-}(c)&=\frac{3}{16\pi}(1+c)^2,
 &
 w_{\bar q}^{W^-}(c)&=\frac{3}{16\pi}(1-c)^2.
\label{eq:star-va-weights}
\end{align}
Here the subscripts $q$ and $\bar q$ indicate whether beam $A$ supplies the quark or antiquark, respectively. Each weight is normalized over the full solid angle:
\begin{equation}
    \int_{-1}^{1}\dd{c}
    \int_0^{2\pi}\dd{\phi}
    w_{q,\bar q}^{W^\pm}(c)=1.
\end{equation}
The finite transverse recoil is included through the lepton boost. Additional finite-$q_T$ production angular coefficients are not
included.

Define the lepton-level acceptance indicator
\begin{align}
 \mathcal I(q_T,y_W;c,\phi)
 &=\Theta\!\left(\eta_e+1\right)
   \Theta\!\left(1-\eta_e\right) \notag\\
 &\quad\times
   \Theta\!\left(E_T^e-25\,\mathrm{GeV}\right)
   \Theta\!\left(50\,\mathrm{GeV}-E_T^e\right).
\label{eq:star-acceptance-indicator}
\end{align}
Here $\Theta$ denotes the Heaviside step function. The two beam-order-dependent acceptance factors are therefore
\begin{align}
 \mathcal A_q^{W^\pm}(q_T,y_W)
 &=\int_{-1}^{1}\dd{c} \int_0^{2\pi}\dd{\phi}
   \mathcal I(q_T,y_W;c,\phi)\,w_q^{W^\pm}(c), \notag\\
 \mathcal A_{\bar q}^{W^\pm}(q_T,y_W)
 &=\int_{-1}^{1}\dd{c}\int_0^{2\pi}\dd{\phi}
   \mathcal I(q_T,y_W;c,\phi)\,w_{\bar q}^{W^\pm}(c).
\label{eq:star-acceptance}
\end{align}
They contain the parity-violating decay distribution and the kinematic lepton selection, and hence are not constant detector efficiencies.

For $X=UU,LU$, the accepted structure functions are
\begin{align}
 F_X^{W^+,\mathrm{acc}}(q_T)
 &=\int_{-\ln(\sqrt{s}/M_W)}^{+\ln(\sqrt{s}/M_W)}
 \dd{y_W}
 \notag\\
 &\times\big[
 \mathcal A_q^{W^+}\mathcal F_X^{u_A\bar d_B}
 +\mathcal A_{\bar q}^{W^+}\mathcal F_X^{\bar d_Au_B}
 \big], \\
 F_X^{W^-,\mathrm{acc}}(q_T)
 &=\int_{-\ln(\sqrt{s}/M_W)}^{+\ln(\sqrt{s}/M_W)}
 \dd{y_W}
 \notag\\
 &\times\big[
 \mathcal A_q^{W^-}\mathcal F_X^{d_A\bar u_B}
 +\mathcal A_{\bar q}^{W^-}\mathcal F_X^{\bar u_Ad_B}
 \big].
\label{eq:star-accepted-structure-functions}
\end{align}
The arguments $(q_T,y_W)$ of the acceptance factors and structure functions are implicit on the right-hand sides.

\begin{figure*}
    \centering
    \includegraphics[width=0.62\linewidth]{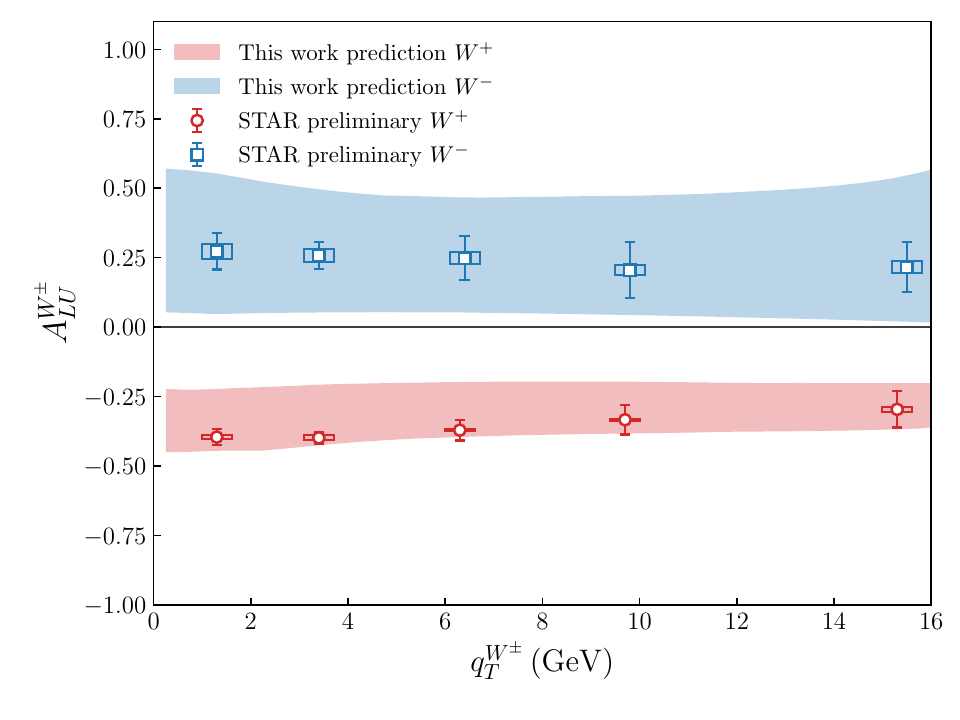}
    \caption{Predictions for the longitudinal single-spin asymmetries $A_{LU}^{W^+}$ and $A_{LU}^{W^-}$ in polarized proton-proton collisions at $\sqrt{s}=510\,\mathrm{GeV}$ with fiducial cuts on the charged decay lepton. The red ($W^+$) and blue ($W^-$) bands represent $68\%$ confidence intervals from the extracted helicity TMDs after applying $25<E_T^e<50\,\mathrm{GeV}$ and $-1<\eta_e<1$. The STAR preliminary~\cite{Wang:2026nav} $W^+$ and $W^-$ data are shown for qualitative comparison. Vertical bars denote statistical uncertainties, and open boxes denote point-to-point systematic uncertainties. The common relative-luminosity uncertainty and beam-polarization scale uncertainty are not shown.}
    \label{fig:star_w_alu}
\end{figure*}

Figure~\ref{fig:star_w_alu} shows the predicted asymmetries with the lepton cuts specified above. The bands represent the $68\%$ confidence intervals obtained from extracted helicity TMDs. The predicted asymmetries are negative for $W^+$ and positive for $W^-$ over the displayed range, with a broader replica band for $W^-$. The preliminary data of STAR~\cite{Wang:2026nav} are overlaid for qualitative comparison and were not included in the fit.

\section{Impact of the EicC projection}
\label{sec:eicc_impact}

The EicC is designed to collide polarized electron beams with polarized proton and $^3$He beams at center-of-mass energies of $15 \sim 20\,\mathrm{GeV}$. For electron-proton collisions, the design parameters include a peak luminosity of $4.25\times10^{33}\,\mathrm{cm}^{-2}\mathrm{s}^{-1}$ and an annual integrated luminosity of about $107\,\mathrm{fb}^{-1}$~\cite{Xiao:2026tbs}. The projected longitudinal polarizations of the electron and proton beams are $P_e=(80\pm1.6)\%$ and $P_p=(70\pm3.5)\%$, respectively~\cite{Du:2024sjt}. The EicC provides coverage of the intermediate-$x$ region $0.005<x<0.3$ over momentum transfers $1\,\mathrm{GeV}^{2}<Q^{2}<30\,\mathrm{GeV}^{2}$, offering opportunities for precision SIDIS studies~\cite{Xiao:2026tbs}.

To assess the impact of the EicC SIDIS program on the determination of $g_{1L}$, we construct pseudodata for the longitudinal double spin asymmetry $A_{LL}$ using simulations supplied by the EicC collaboration. The samples cover charged-pion and charged-kaon production on protons and effective neutrons, with the neutron asymmetries extracted from simulated $e^-\,^3$He collisions. The samples contain $13545$ points before the TMD selection and correspond to an assumed integrated luminosity of $50\,\mathrm{fb}^{-1}$ for each of the $e^-p$ and $e^-\,^3$He collision samples. Statistical uncertainties are estimated in four-dimensional bins of $(x,Q^2,z,P_{h\perp})$. Applying the cut $\delta<0.5$, as in the world-data fit, leaves $7595$ EicC pseudodata points. Their central values are generated from the world-data fit. We assign relative systematic uncertainties of $2\%$ to the electron-beam polarization, $5\%$ to the proton or ion beam polarization, and $5\%$ to the nuclear effects associated with the extraction of neutron asymmetries from $^3$He. The polarization uncertainties are treated as uncorrelated between the proton and neutron samples, which correspond to separate running periods.

For the projection fit, we relax the reduced parametrization used for the world-data fit of Sec.~\ref{sec:world_data_fit}. The constraint $N_s=N_{\bar{s}}$ is released, and the parameters $(N_f,\alpha_f,\beta_f,\varepsilon_f)$ are allowed to vary independently for $f=u,d,s,\bar{u},\bar{d},\bar{s},g$. Separate sets of the transverse parameters $(\lambda_1,\lambda_2,\lambda_5)$ are assigned to the $u$, $d$, sea-quark, and gluon sectors, with a common set for $s$, $\bar{s}$, $\bar{u}$, and $\bar{d}$. This parametrization contains $37$ physics parameters. The five normalization groups are CLAS, HERMES proton, HERMES deuteron, EicC proton, and EicC neutron, giving five nuisance parameters. We fit the world data and EicC pseudodata simultaneously using the penalty-term $\chi^2$ and estimate the resulting uncertainties with $100$ Monte Carlo replicas.

\begin{figure*}
    \centering
    \includegraphics[width=0.30\linewidth]{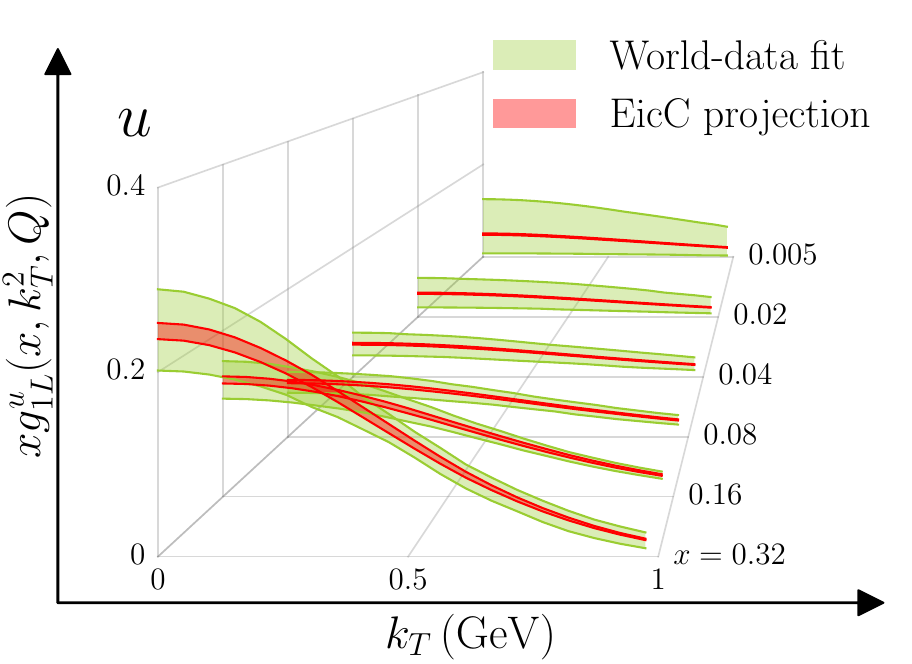}\hfill
    \includegraphics[width=0.30\linewidth]{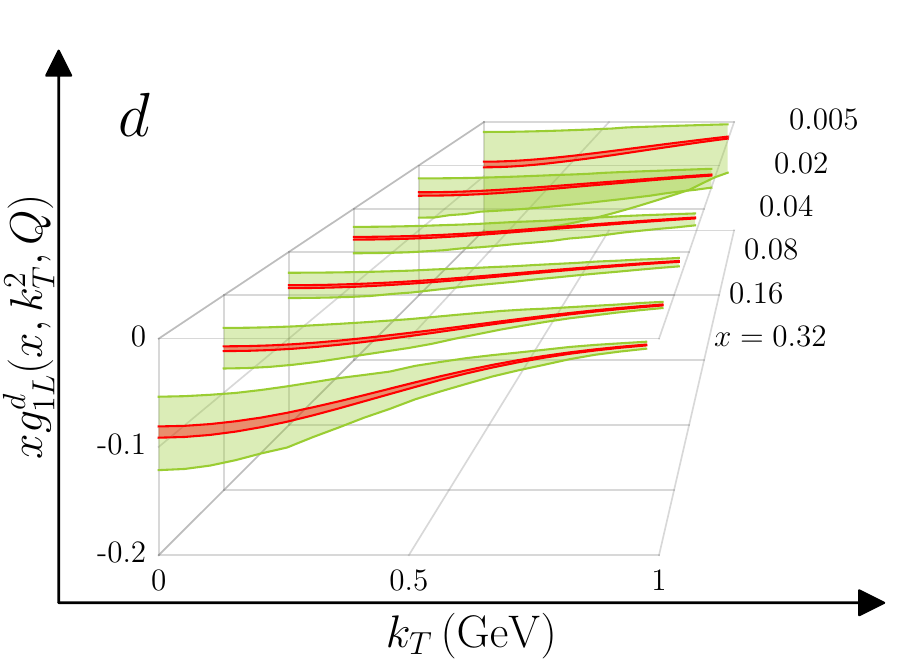}\hfill
    \includegraphics[width=0.30\linewidth]{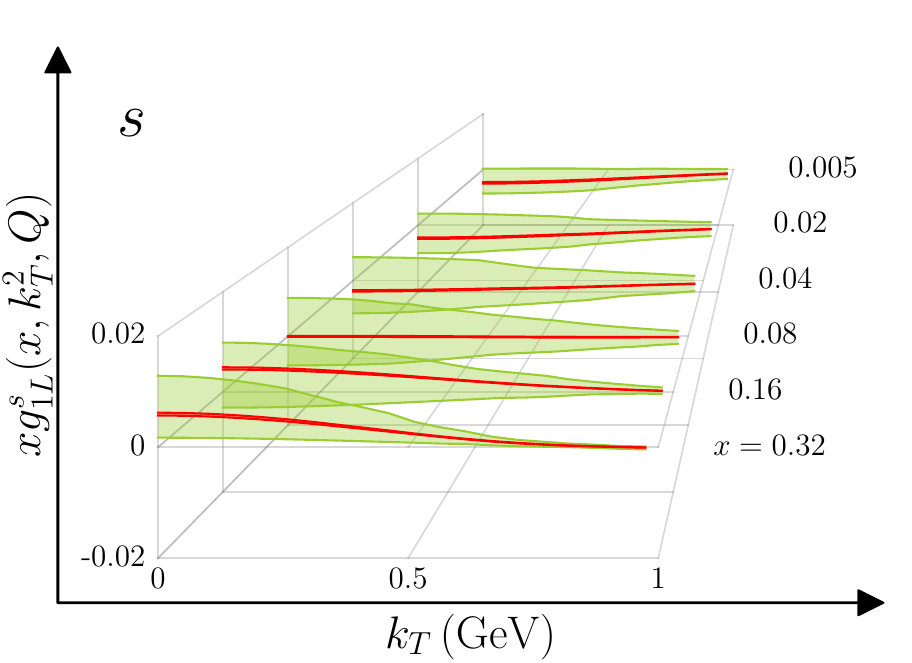}\\
    \includegraphics[width=0.30\linewidth]{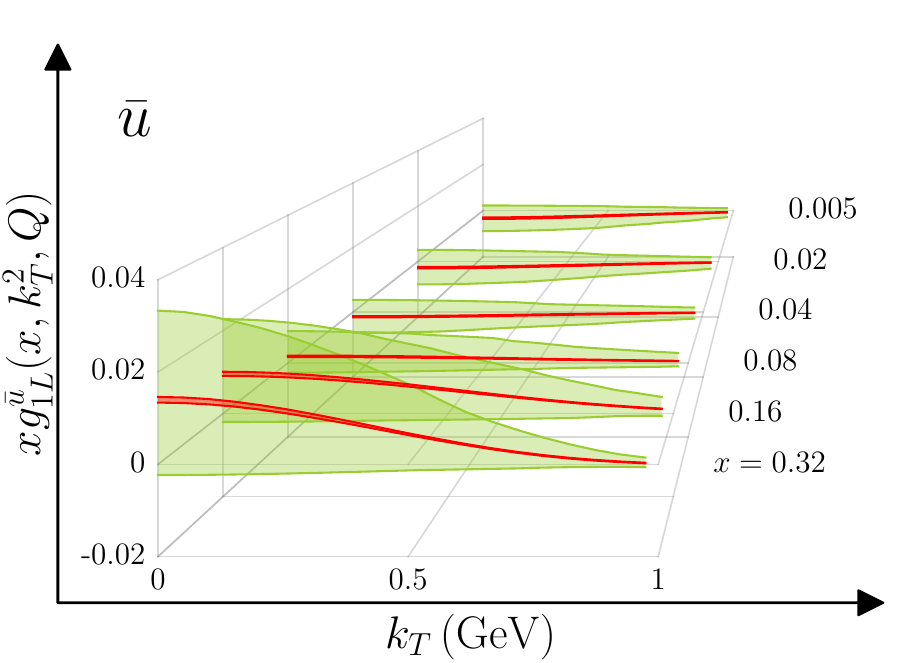}\hfill
    \includegraphics[width=0.30\linewidth]{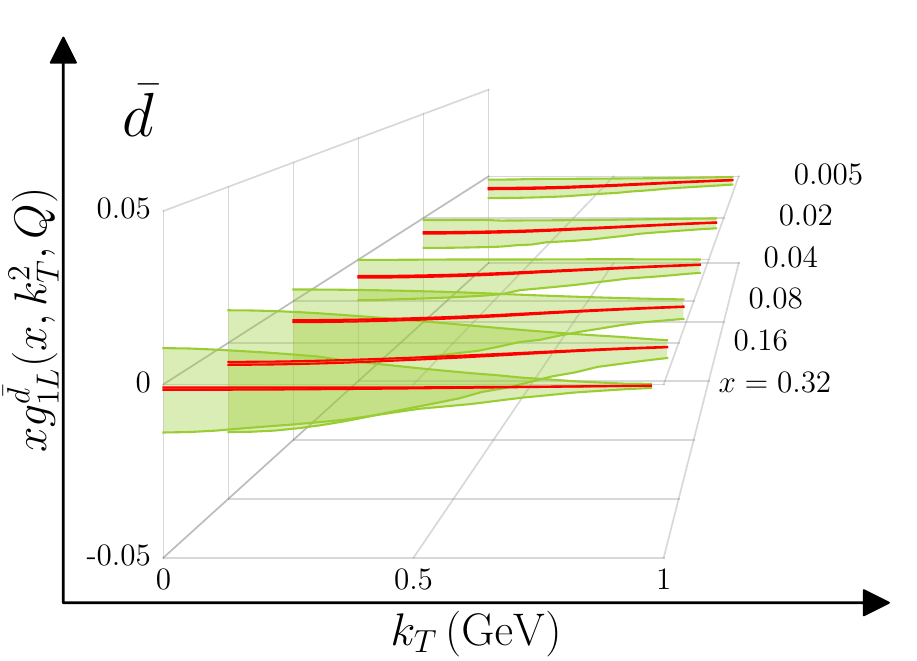}\hfill
    \includegraphics[width=0.30\linewidth]{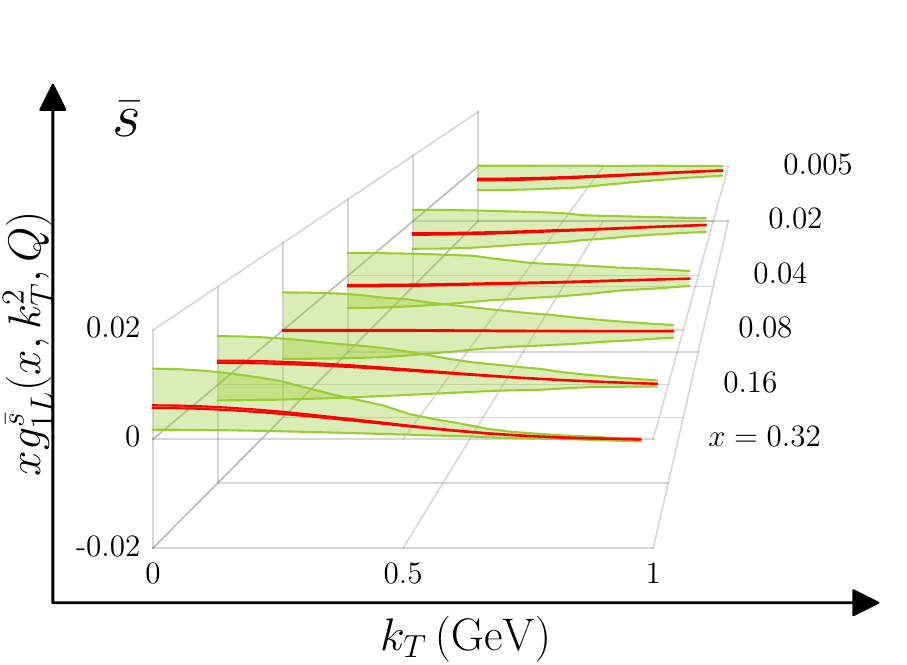}
    \caption{The extracted helicity TMDs $xg_{1L}(x,k_T,Q)$ at $Q=2\,\mathrm{GeV}$ for $u$, $d$, and $s$ (top row) and $\bar{u}$, $\bar{d}$, and $\bar{s}$ (bottom row). The green and red bands show the $68\%$ confidence intervals from the world-data fit and the combined fit to world data and EicC pseudodata, respectively.}
    \label{fig:eicc_g1LxkT}
\end{figure*}

Figure~\ref{fig:eicc_g1LxkT} compares the world-data fit results and EicC projections with $68\%$ uncertainty intervals for the helicity TMDs $xg_{1L}(x,k_T,Q)$ at $Q=2\,\mathrm{GeV}$, showing the components $u$, $d$, $s$, $\bar{u}$, $\bar{d}$, and $\bar{s}$. The projected EicC data are expected to significantly reduce the uncertainties for both quark and antiquark helicity TMD distributions. The relative uncertainties are approximately $5\% \sim 7\%$ for the EicC projected helicity TMDs.

\section{Summary}
\label{sec:summary}

We have presented an updated global analysis of the transverse-momentum-dependent helicity distributions of the nucleon using measurements of the longitudinal double spin asymmetry $A_{LL}$ in SIDIS. The analysis incorporates kinematic target- and detected-hadron-mass corrections, NNLO small-$b_T$ OPE matching for $f_1$, $D_1$, and $g_{1L}$, NNLL TMD evolution, and a penalty-term $\chi^2$ with the $T_0$ prescription for correlated normalization uncertainties.

The up- and down-quark helicity TMDs are constrained away from zero by the existing world data, whereas the sea-quark and gluon distributions remain weakly constrained and compatible with zero within uncertainties. The up- and down-quark distributions integrated over transverse momenta up to $k_T=Q$ agree with the corresponding collinear helicity distributions within uncertainties. Over the analyzed kinematic range, the magnitudes of the extracted helicity TMD distributions generally decrease with increasing $k_T$.

Under the Wandzura-Wilczek-type approximation, the predicted $\cos\phi_h$ modulation of the longitudinal double spin asymmetry agrees with the HERMES and CLAS measurements within the present experimental uncertainties. These measurements were not included in the fit and provide a consistency check of the description within this approximation.

We have also provided predictions for the azimuthally averaged SIDIS asymmetry $A_{LL}$ in $\pi^\pm$ production at representative CLAS12 kinematics satisfying $q_T/Q<0.5$, and for the longitudinal single-spin asymmetries $A_{LU}^{W^\pm}$ at STAR kinematics, with fiducial cuts on the charged decay lepton. The $W^\pm$ predictions are compared qualitatively with preliminary STAR measurements. Finalized STAR measurements could provide complementary constraints on helicity TMDs, including sea-quark distributions.

Using EicC proton SIDIS pseudodata and neutron asymmetries extracted from simulated $e^-\,^3$He collisions, we have assessed the potential impact of the projected measurements on the determination of helicity TMDs. From the combined fit to world data and EicC pseudodata, we find that the projected EicC measurements can substantially reduce the uncertainties of the quark and antiquark helicity TMDs in the intermediate-$x$ region, including the presently weakly constrained sea-quark distributions. The numerical projection yields relative uncertainties of approximately $5\% \sim 7\%$ for the extracted helicity TMDs.

\acknowledgments{
This work is supported in part by the National Natural Science Foundation of China under Grant No.~12547160, No.~12321005, No.~12675108, and No.~12335006, by the National Key R\&D Program of China No.~2024YFA1611004, and by the Shandong Provincial Postdoctoral Innovation Project No. SDCX-ZG-202602016.
P. Sun is supported by the Natural Science Foundation of China under Grant No.~11975127 and No.~12061131006.
Y. X. Zhao is supported by the CAS Project for Young Scientists in Basic Research under Grant No.~YSBR-117.
}

\appendix
\section{Fourier transforms of TMDs}
\label{App:FTT}

For azimuthally symmetric TMDs, the two-dimensional Fourier transforms between
transverse-momentum space and impact-parameter space reduce to Hankel
transforms.  With $b_T=|\boldsymbol b_T|$, $k_T=|\boldsymbol k_T|$, and
$p_T=|\boldsymbol p_T|$, we use the conventions
\begin{align}
    f(x,b_T;Q,Q^2)
    &= 2\pi\int_0^\infty \dd{k_T} k_T
       J_0(b_Tk_T)\,f(x,k_T,Q), \\
    f(x,k_T,Q)
    &= \frac{1}{2\pi}\int_0^\infty \dd{b_T} b_T \notag\\
    &\quad\times J_0(b_Tk_T)\,f(x,b_T;Q,Q^2),
    \label{eq:fourier_pdf}
\end{align}
for a generic TMD~PDF $f$, and
\begin{align}
    D(z,b_T;Q,Q^2)
    &= 2\pi\int_0^\infty \dd{p_T} p_T
       J_0\!\left(b_Tp_T\right)D(z,zp_T;Q), \\
    D(z,zp_T,Q)
    &= \frac{1}{2\pi}\int_0^\infty \dd{b_T} b_T \notag\\
    &\quad\times J_0\!\left(b_Tp_T\right)D(z,b_T;Q,Q^2),
    \label{eq:fourier_ff}
\end{align}
for a generic TMD~FF $D$, where $\boldsymbol p_T$ is the transverse momentum
of the parent quark. Thus $\boldsymbol b_T$ is Fourier conjugate to
$\boldsymbol k_T$ for PDFs and to $\boldsymbol p_T$ for FFs.
These normalizations lead to the factor $b_T J_0(b_Tq_T)/(2\pi)$ in the TMD structure functions in Sec.~\ref{sec:framework}.

For a TMD~PDF, we define the $n$th-order $b_T$ derivative by
\begin{equation}
    f^{(n)}(x,b_T,Q)
    = n!\left(-\frac{2}{M^2}\frac{\partial}{\partial b_T^2}\right)^n
      f(x,b_T;Q,Q^2),
    \label{eq:fourier_pdf_moment}
\end{equation}
where $M$ is the nucleon mass. Provided that the transverse-momentum
integral converges, the derivative at $b_T=0$ equals the corresponding
untruncated transverse-momentum moment:
\begin{equation}
    f^{(n)}(x,Q)
    = \int \dd[2]{\boldsymbol{k}_T}\,
      \left(\frac{k_T^2}{2M^2}\right)^n f(x,k_T,Q).
    \label{eq:fourier_pdf_moment_origin}
\end{equation}
For a TMD~FF, the corresponding derivative is defined by
\begin{equation}
    D^{(n)}(z,b_T,Q)
    = n!\left(-\frac{2}{z^2M_h^2}\frac{\partial}{\partial b_T^2}\right)^n
      D(z,b_T;Q,Q^2),
    \label{eq:fourier_ff_moment}
\end{equation}
where $M_h$ is the detected-hadron mass.
For TMDs with perturbative transverse-momentum tails, the correspondence
between derivatives at $b_T=0$ and untruncated transverse-momentum
moments is formal and requires a consistent ultraviolet prescription.
In particular, the finite-$b_T$ derivative $g_{1L}^{(1)}$ entering
Sec.~\ref{sec:asymmetry} is distinct from the finite-cutoff moment
defined in Eq.~(\ref{eq:g1L_first_kT_moment}).

\section{Evolution of TMD~PDFs and TMD~FFs}
\label{App:Evolution}

This appendix specifies the evolution factor used in the numerical analysis.
The general evolution equations and the optimal-TMD construction were introduced in Sec.~\ref{sec:framework}.
For an optimal TMD defined on the $\zeta$-prescription curve, the evolution to $(\mu,\mu^2)$ is
\begin{align}
    F(x,b_T;\mu,\mu^2)&=R(\mu,b_T)F^{\rm opt}(x,b_T),\\
    R(\mu,b_T)&=\left(\frac{\mu^2}{\zeta_\mu(\mu,b_T)}\right)^{-\mathcal{D}(\mu,b_T)}.
    \label{eq:evolution_factor}
\end{align}
Here $F$ denotes a quark or antiquark TMD~PDF or TMD~FF,
and $\mathcal{D}$ is the Collins-Soper kernel.  The same factor is used for the unpolarized TMDs $f_1$ and $D_1$ and
for the helicity TMD $g_{1L}$; their distinct nonperturbative input functions
are all defined at the optimal point.

The kernel in Eq.~\eqref{eq:evolution_factor} follows the SV19
implementation of the $\zeta$ prescription~\cite{Scimemi:2019cmh}.  Its
perturbative component is evaluated at the regulated impact parameter
\begin{align}
    b_T^*&=\frac{b_T}{\sqrt{1+b_T^2/B_{\rm NP}^2}}, \\
    \mathcal{D}(\mu,b_T)&=\mathcal{D}_{\rm resum}(\mu,b_T^*)+c_0b_Tb_T^*.
    \label{eq:evolution_kernel}
\end{align}
The $b_T^*$ map keeps the resummed perturbative kernel away from its
large-$b_T$ singular region, while the second term supplies the
nonperturbative large-$b_T$ contribution.  The parameters used here are
$B_{\rm NP}=1.93\,\mathrm{GeV}^{-1}$ and
$c_0=0.0391\,\mathrm{GeV}^{2}$.
The resummed perturbative component is evaluated as
\begin{equation}
\begin{aligned}
&\mathcal{D}_{\rm resum}(\mu,b_T)={}-
\frac{\Gamma_0}{2\beta_0}\ln(1-X) \\
&\quad+\frac{a_s(\mu)}{2\beta_0(1-X)}
\left[-\frac{\beta_1\Gamma_0}{\beta_0}
\bigl(\ln(1-X)+X\bigr)+\Gamma_1X\right] \\
&\quad+\frac{a_s^2(\mu)}{(1-X)^2}\left[
\frac{\Gamma_0\beta_1^2}{4\beta_0^3}
\bigl(\ln^2(1-X)-X^2\bigr)\right. \\
&\quad+\frac{\beta_1\Gamma_1}{4\beta_0^2}
\bigl(X^2-2X-2\ln(1-X)\bigr)
+\frac{\Gamma_0\beta_2}{4\beta_0^2}X^2 \\
&\quad\left.-\frac{\Gamma_2}{4\beta_0}X(X-2)
+d^{(2,0)}\right],
\end{aligned}
\label{eq:D_resum}
\end{equation}
where $X=\beta_0a_s(\mu)\mathbf{L}_\mu$, the $\beta_i$ are defined through
$\beta(a_s)\equiv\mu^2\mathrm{d}a_s(\mu)/\mathrm{d}\mu^2
=-a_s^2(\mu)[\beta_0+\beta_1a_s(\mu)+\beta_2a_s^2(\mu)+\mathcal{O}(a_s^3)]$,
and the $\Gamma_i$ through
$\Gamma_{\rm cusp}(\mu)=\Gamma_0a_s(\mu)+\Gamma_1a_s^2(\mu)
+\Gamma_2a_s^3(\mu)+\mathcal{O}(a_s^4)$.

The initial rapidity scale is constructed as a smooth
interpolation between the perturbative and exact solutions of the
null-evolution condition,
\begin{align}
    \zeta_\mu(\mu,b_T)
    &=e^{-b_T^2/B_{\rm NP}^2}\zeta_{\rm pert}(\mu,b_T) \\
    & +\left(1-e^{-b_T^2/B_{\rm NP}^2}\right)\zeta_{\rm exact}(\mu,b_T),
    \label{eq:zeta_profile}\\
    \zeta_{\rm pert}(\mu,b_T)
    &=\frac{2\mu e^{-\gamma_E}}{b_T}\,e^{-v(\mu,b_T)}, \\
    \zeta_{\rm exact}(\mu,b_T)
    &=\mu^2\exp\left[-\frac{\mathcal{G}(\mu,b_T)}{\mathcal{D}(\mu,b_T)}\right].
\end{align}
Both solutions use the coefficients introduced above, together with the
virtual anomalous dimension defined below. The explicit QCD $\beta$-function
coefficients are
\begin{align}
\beta_0&=\frac{11}{3}C_A-\frac{4}{3}T_Fn_f, \\
\beta_1&=\frac{34}{3}C_A^2-\frac{20}{3}C_AT_Fn_f
-4C_FT_Fn_f, \\
\beta_2&=\frac{2857}{54}C_A^3
-\frac{1415}{27}C_A^2T_Fn_f \notag\\
&\quad-\frac{205}{9}C_FC_AT_Fn_f
+2C_F^2T_Fn_f \\
&\quad+\frac{158}{27}C_AT_F^2n_f^2
+\frac{44}{9}C_FT_F^2n_f^2.\notag
\end{align}
The anomalous dimensions entering the $\zeta$ prescription satisfy
\begin{equation}
\gamma_F(\mu,\zeta)
=\Gamma_{\rm{cusp}}(\mu)\ln\!\left(\frac{\mu^2}{\zeta}\right)
-\gamma_V(\mu),
\end{equation}
where the virtual anomalous dimension is expanded as
\begin{align}
\gamma_V(\mu)
&=\gamma_1a_s(\mu)+\gamma_2a_s^2(\mu)
+\mathcal{O}(a_s^3).
\end{align}
For the quark TMDs, the cusp and virtual-anomalous-dimension coefficients are
\begin{align}
\Gamma_0&=4C_F, \\
\Gamma_1&=4C_F\left[
\left(\frac{67}{9}-\frac{\pi^2}{3}\right)C_A
-\frac{20}{9}T_Fn_f\right], \\
\Gamma_2&=4C_F\Bigg[
C_A^2\left(\frac{245}{6}-\frac{134\pi^2}{27}
+\frac{11\pi^4}{45}+\frac{22}{3}\zeta_3\right) \notag\\
&\quad+C_AT_Fn_f\left(-\frac{418}{27}
+\frac{40\pi^2}{27}-\frac{56}{3}\zeta_3\right)\\
&\quad+C_FT_Fn_f\left(-\frac{55}{3}+16\zeta_3\right)
-\frac{16}{27}T_F^2n_f^2\Bigg], \notag\\
\gamma_1&=-6C_F, \\
\gamma_2&=C_F^2(-3+4\pi^2-48\zeta_3) \notag\\ 
&\quad+C_FC_A\left(-\frac{961}{27}-\frac{11\pi^2}{3}
+52\zeta_3\right) \\
&\quad+C_FT_Fn_f\left(\frac{260}{27}
+\frac{4\pi^2}{3}\right).\notag
\end{align}
The exact solution is written as
\begin{equation}
\begin{split}
    \mathcal{G}= &
    \frac{\Gamma_0}{2\beta_0^2a_s(\mu)} \times \\
    & \left[g_0(p)+a_s(\mu)g_1(p)+a_s^2(\mu)g_2(p)+\cdots\right],
    \label{eq:zeta_exact_series}
\end{split}
\end{equation}
with $p=2\beta_0\mathcal{D}(\mu,b_T)/\Gamma_0$. The coefficient functions of
the exact solution depend on $p$ as
\begin{align}
g_0(p)={}&e^{-p}-1+p, \\
g_1(p)={}&\frac{\beta_1}{\beta_0}
\left(e^{-p}-1+p-\frac{p^2}{2}\right)
\notag\\
&-\frac{\Gamma_1}{\Gamma_0}\left(e^{-p}-1+p\right)
+\frac{\beta_0\gamma_1}{\Gamma_0}p, \\
g_2(p)={}&
\left(\frac{\Gamma_1^2}{\Gamma_0^2}-\frac{\Gamma_2}{\Gamma_0}\right)
\left(\cosh p-1\right)\notag\\
&+\left(\frac{\beta_1\Gamma_1}{\beta_0\Gamma_0}
       -\frac{\beta_2}{\beta_0}\right)\left(\sinh p-p\right) \notag\\
&+\left(\frac{\beta_0\gamma_2}{\Gamma_0}
         -\frac{\beta_0\gamma_1\Gamma_1}{\Gamma_0^2}\right)
\left(e^p-1\right).
\label{eq:zeta_exact_functions}
\end{align}
The perturbative solution, in contrast, uses
\begin{equation}
\begin{aligned}
v(\mu,b_T)&=v_0+a_s(\mu)v_1(\mathbf{L}_\mu)
          +\mathcal{O}(a_s^2), 
\end{aligned}
\label{eq:zeta_pert_series}
\end{equation}
where
\begin{align}
v_0={}&\frac{\gamma_1}{\Gamma_0}, \\
v_1(\mathbf{L}_\mu)={}&\frac{\beta_0}{12}\mathbf{L}_\mu^2
-\frac{\gamma_1\Gamma_1}{\Gamma_0^2}
+\frac{\gamma_2+d^{(2,0)}}{\Gamma_0}.
\label{eq:zeta_pert_functions}
\end{align}
We work at next-to-next-to-leading logarithmic (NNLL) accuracy; the
truncation orders of the individual ingredients are summarized in Table~\ref{tab:alpha_order}.
\begin{table}[htbp]
\caption{Perturbative truncation orders used for the evolution factor and
small-$b_T$ matching.}
\label{tab:alpha_order}
\centering
\begin{tabular}{c|ccccc}
\hline\hline
Component & $\Gamma_{\rm{cusp}}$ & $\gamma_V$ & $\mathcal{D}_{\text{resum}}$ & $\zeta_\mu^{\text{pert}}$ & $\zeta_\mu^{\text{exact}}$ \\
\hline
$R$ & $\alpha_s^3$ & $\alpha_s^2$ & $\alpha_s^2$ & $\alpha_s^1$ & $\alpha_s^1$ \\
\hline
& $f_1$ & $D_1$ & $g_{1L}$ \\
$C(\mathbb{C})$ & $\alpha_s^2$ & $\alpha_s^2$ & $\alpha_s^2$ \\
\hline\hline
\end{tabular}
\end{table}
Here $C$ and $\mathbb{C}$ denote the matching coefficients for TMD~PDFs and
TMD~FFs, respectively.
In the present calculation, $v$ is retained through $a_s v_1$,
and the bracket in Eq.~\eqref{eq:zeta_exact_series} through
$a_s^2g_2$. With the overall $1/a_s$ prefactor, this retains
$\mathcal{G}$ through $\mathcal{O}(a_s)$ at fixed $p$,
consistently with Table~\ref{tab:alpha_order}. Thus $\zeta_\mu(\mu,b_T)$ follows the
null-evolution line at small $b_T$ and remains well defined after the
nonperturbative completion at large $b_T$.
\section{Splitting kernels and matching coefficients}
\label{App:matching_kernels}

This appendix collects the one-loop splitting kernels and finite matching
coefficients entering the small-$b_T$ expansions used in this work.  The
perturbative structure of the space-like unpolarized TMD~PDF coefficient is
given in Eq.~\eqref{eq:unpolarized_pdf_matching}; the rescaled TMD~FF coefficient
$z^2\mathbb{C}_{i\to j}$ follows the analogous time-like structure.
The first index labels the external TMD parton and
the second the collinear parton in the convolution.  Thus the FF arrows below
follow the time-like convention $i\to j$ of Eq.~\eqref{eq:OPE_FF}, rather than
the PDF convention $i\leftarrow j$.  For the helicity TMD~PDF, the same
structure is understood with $C_{i\leftarrow j}$ and $P_{i\leftarrow j}$
replaced by $\Delta C_{i\leftarrow j}$ and $\Delta P_{i\leftarrow j}$,
respectively.  We use
\begin{equation}
    \int_x^1 \dd{z} [h(z)]_+\varphi(z)
    =\int_0^1 \dd{z} h(z)\,
      \bigl[\Theta(z-x)\varphi(z)-\varphi(1)\bigr].
    \label{eq:plus_prescription}
\end{equation}
Only the one-loop expressions are displayed here.
The two-loop kernels $P^{(2)}$ and finite coefficients $C^{(2,0)}$ required
by the NNLO matching are not reproduced because of their length. In the
present $a_s=\alpha_s/(4\pi)$ convention, $P^{(2)}$ denotes the two-loop
DGLAP kernel.

\subsection{Unpolarized TMD~PDF \texorpdfstring{$f_1$}{f1}}

For the space-like unpolarized PDF, the one-loop splitting kernels are
\begin{align}
    P^{(1)}_{q\leftarrow q}(z)
      &=2C_F\left[\frac{1+z^2}{1-z}\right]_+, \\
    P^{(1)}_{q\leftarrow g}(z)
      &=2T_F\left[z^2+(1-z)^2\right], \\
    P^{(1)}_{g\leftarrow q}(z)
      &=2C_F\frac{1+(1-z)^2}{z}, \\
    P^{(1)}_{g\leftarrow g}(z)
      &=4C_A\left\{\left[\frac{1}{1-z}\right]_+
        +\frac{1}{z}-2+z(1-z)\right\}\notag\\
      &\quad+\beta_0\delta(1-z).
\end{align}
The corresponding finite matching coefficients are
\begin{align}
    C^{(1,0)}_{q\leftarrow q}(z)
      &=2C_F(1-z)-\frac{C_F\pi^2}{6}\delta(1-z),
    \\
    C^{(1,0)}_{q\leftarrow g}(z)
      &=4T_Fz(1-z), \\
    C^{(1,0)}_{g\leftarrow q}(z)
      &=2C_Fz,
    \\
    C^{(1,0)}_{g\leftarrow g}(z)
      &=-\frac{C_A\pi^2}{6}\delta(1-z).
    \label{eq:f1_matching_coefficients}
\end{align}
The two-loop space-like kernels can be found in
Refs.~\cite{Curci:1980uw,Furmanski:1980cm}, and the NNLO finite matching
coefficients in Ref.~\cite{Echevarria:2016scs}.

\subsection{Unpolarized TMD~FF \texorpdfstring{$D_1$}{D1}}

The time-like splitting kernels used for the unpolarized FF are
\begin{align}
    P^{(1)}_{q\to q}(z)
      &=2C_F\left[\frac{1+z^2}{1-z}\right]_+,
    \\
    P^{(1)}_{q\to g}(z)
      &=2C_F\frac{1+(1-z)^2}{z}, \\
    P^{(1)}_{g\to q}(z)
      &=2T_F\left[z^2+(1-z)^2\right],
    \\
    P^{(1)}_{g\to g}(z)
      &=4C_A\left\{\left[\frac{1}{1-z}\right]_+
        +\frac{1}{z}-2+z(1-z)\right\}\notag\\
      &\quad+\beta_0\delta(1-z).
    \label{eq:D1_splitting_kernels}
\end{align}
The associated finite matching coefficients are those of the time-like
TMD~FF OPE and contain explicit $\ln z$ terms. With the normalization
of Eq.~\eqref{eq:OPE_FF}, we display the rescaled coefficients
$z^2\mathbb{C}^{(1,0)}$:
\begin{align}
    z^2\mathbb{C}^{(1,0)}_{q\to q}(z)
      &=C_F\left[2(1-z)+4\frac{1+z^2}{1-z}\ln z\right]\notag\\
      &\quad-\frac{C_F\pi^2}{6}\delta(1-z),
    \\
    z^2\mathbb{C}^{(1,0)}_{q\to g}(z)
      &=2C_F\left[z+2\frac{1+(1-z)^2}{z}\ln z\right], \\
    z^2\mathbb{C}^{(1,0)}_{g\to q}(z)
      &=4T_F\left[z(1-z)+\left(z^2+(1-z)^2\right)\ln z\right],
    \\
    z^2\mathbb{C}^{(1,0)}_{g\to g}(z)
      &=8C_A\frac{\left(1-z(1-z)\right)^2}{z(1-z)}\ln z
        -\frac{C_A\pi^2}{6}\delta(1-z).
    \label{eq:D1_matching_coefficients}
\end{align}
The two-loop time-like kernels can be found in
Ref.~\cite{Stratmann:1996hn}, while the NNLO finite matching coefficients
can be found in Ref.~\cite{Echevarria:2016scs}.

\subsection{Helicity TMD~PDF \texorpdfstring{$g_{1L}$}{g1L}}

For the helicity TMD~PDF, the quark-to-quark splitting kernel agrees at one
loop with its unpolarized counterpart, whereas the remaining kernels differ:
\begin{align}
    \Delta P^{(1)}_{q\leftarrow q}(z)
      &=2C_F\left[\frac{1+z^2}{1-z}\right]_+,
    \\
    \Delta P^{(1)}_{q\leftarrow g}(z)
      &=2T_F(2z-1), \\
    \Delta P^{(1)}_{g\leftarrow q}(z)
      &=2C_F(2-z),
    \\
    \Delta P^{(1)}_{g\leftarrow g}(z)
      &=4C_A\left[\frac{2-3z+2z^2}{1-z}\right]_+
        +\beta_0\delta(1-z).
    \label{eq:g1L_splitting_kernels}
\end{align}
The finite polarized matching coefficients implemented in the extraction are
\begin{align}
    \Delta C^{(1,0)}_{q\leftarrow q}(z)
      &=2C_F(1-z)-\frac{C_F\pi^2}{6}\delta(1-z),
    \\
    \Delta C^{(1,0)}_{q\leftarrow g}(z)
      &=4T_F(1-z), \\
    \Delta C^{(1,0)}_{g\leftarrow q}(z)
      &=-4C_F(1-z),
    \\
    \Delta C^{(1,0)}_{g\leftarrow g}(z)
      &=-8C_A(1-z)-\frac{C_A\pi^2}{6}\delta(1-z).
    \label{eq:g1L_matching_coefficients}
\end{align}
The two-loop helicity kernels can be found in Ref.~\cite{Mertig:1995ny},
and the NNLO finite helicity matching coefficients in
Ref.~\cite{Zhu:2025gts}.
\label{App:splittingKernel}

\bibliography{ref}

@article{HERMES:2018awh,
    author = "Airapetian, A. and others",
    collaboration = "HERMES Collaboration",
    title = "{Longitudinal double-spin asymmetries in semi-inclusive deep-inelastic scattering of electrons and positrons by protons and deuterons}",
    eprint = "1810.07054",
    archivePrefix = "arXiv",
    primaryClass = "hep-ex",
    reportNumber = "DESY-18-181, DESY 18-181",
    doi = "10.1103/PhysRevD.99.112001",
    journal = "Phys. Rev. D",
    volume = "99",
    number = "11",
    pages = "112001",
    year = "2019"
}

@article{Cocuzza:2021cbi,
    author = "Cocuzza, C. and Melnitchouk, W. and Metz, A. and Sato, N.",
    collaboration = "JAM Collaboration",
    title = "{Bayesian Monte~Carlo extraction of the sea asymmetry with SeaQuest and STAR data}",
    eprint = "2109.00677",
    archivePrefix = "arXiv",
    primaryClass = "hep-ph",
    reportNumber = "JLAB-THY-21-3491",
    doi = "10.1103/PhysRevD.104.074031",
    journal = "Phys. Rev. D",
    volume = "104",
    number = "7",
    pages = "074031",
    year = "2021"
}

@article{Wang:2026nav,
    author = "Wang, Chao",
    collaboration = "STAR Collaboration",
    title = "{Measurement of the transverse momentum dependent (TMD) longitudinal single-spin asymmetry of W bosons at RHIC-STAR}",
    doi = "10.22323/1.517.0028",
    journal = "PoS",
    volume = "SPIN2025",
    pages = "028",
    year = "2026"
}

@misc{Zeng:2024gun,
    author = "Zeng, Chunhua and Dong, Hongxin and Liu, Tianbo and Sun, Peng and Zhao, Yuxiang",
    title = "{Global analysis of Sivers and Collins asymmetries within the TMD factorization}",
    eprint = "2412.18324",
    archivePrefix = "arXiv",
    primaryClass = "hep-ph",
    month = "12",
    year = "2024"
}

@misc{Zaccheddu:2026pyy,
    author = "Zaccheddu, Marco and Gamberg, Leonard and Melnitchouk, Wally and Pitonyak, Daniel and Prokudin, Alexei and Qiu, Jian-Wei and Sato, Nobuo",
    title = "{TMDs in the Lens of Generative AI: A Pixel-Based Approach to Partonic Imaging}",
    eprint = "2605.06606",
    archivePrefix = "arXiv",
    primaryClass = "hep-ph",
    reportNumber = "JLAB-THY-26-4691",
    month = "5",
    year = "2026"
}

@misc{Fernando:2026elp,
    author = "Fernando, I. P. and Keller, D.",
    title = "{Deep-neural-network extraction of unpolarized transverse-momentum-dependent parton distributions in $b_T$ space from Drell-Yan data}",
    eprint = "2608.27907",
    archivePrefix = "arXiv",
    primaryClass = "hep-ph",
    month = "8",
    year = "2026"
}

@article{Fernando:2025xzv,
    author = "Fernando, I. P. and Keller, D.",
    title = "{Deep neural network extraction of unpolarized transverse momentum distributions}",
    eprint = "2510.17243",
    archivePrefix = "arXiv",
    primaryClass = "hep-ph",
    doi = "10.1103/p3sg-k524",
    journal = "Phys. Rev. D",
    volume = "113",
    number = "9",
    pages = "096017",
    year = "2026"
}

@article{Fernando:2023obn,
    author = "Fernando, I. P. and Keller, D.",
    title = "{Extraction of the Sivers function with deep neural networks}",
    eprint = "2304.14328",
    archivePrefix = "arXiv",
    primaryClass = "hep-ph",
    reportNumber = "10.1103/PhysRevD.108.054007",
    doi = "10.1103/PhysRevD.108.054007",
    journal = "Phys. Rev. D",
    volume = "108",
    number = "5",
    pages = "054007",
    year = "2023"
}

@misc{Barry:2025glq,
    author = "Barry, P. C. and others",
    title = "{First simultaneous analysis of transverse momentum dependent and collinear parton distributions in the proton}",
    eprint = "2510.13771",
    archivePrefix = "arXiv",
    primaryClass = "hep-ph",
    reportNumber = "JLAB-THY-25-4573, IPARCOS-UCM-25-051",
    month = "10",
    year = "2025"
}

@article{DAgostini:1993arp,
    author = "D'Agostini, G.",
    title = "{On the use of the covariance matrix to fit correlated data}",
    reportNumber = "DESY-93-175",
    doi = "10.1016/0168-9002(94)90719-6",
    journal = "Nucl. Instrum. Meth. A",
    volume = "346",
    pages = "306--311",
    year = "1994"
}

@article{Ball:2009qv,
    author = "Ball, Richard D. and Del Debbio, Luigi and Forte, Stefano and Guffanti, Alberto and Latorre, Jose I. and Rojo, Juan and Ubiali, Maria",
    collaboration = "NNPDF Collaboration",
    title = "{Fitting Parton Distribution Data with Multiplicative Normalization Uncertainties}",
    eprint = "0912.2276",
    archivePrefix = "arXiv",
    primaryClass = "hep-ph",
    reportNumber = "EDINBURGH-2009-22, IFUM-950-FT, FREIBURG-PHENO-09-09, CP3-09-51",
    doi = "10.1007/JHEP05(2010)075",
    journal = "J. High Energy Phys.",
    volume = "05",
    number = "2010",
    pages = "075",
    year = "2010"
}

@article{Collins:1989gx,
    author = "Collins, John C. and Soper, Davison E. and Sterman, George F.",
    title = "{Factorization of Hard Processes in QCD}",
    eprint = "hep-ph/0409313",
    archivePrefix = "arXiv",
    reportNumber = "ITP-SB-89-31",
    doi = "10.1142/9789814503266_0001",
    journal = "Adv. Ser. Direct. High Energy Phys.",
    volume = "5",
    pages = "1--91",
    year = "1989"
}

@article{Yang:2024drd,
    author = "Yang, Ke and Liu, Tianbo and Sun, Peng and Zhao, Yuxiang and Ma, Bo-Qiang",
    collaboration = "Transverse Nucleon Tomography Collaboration",
    title = "{First Extraction of Transverse-Momentum Dependent Helicity Distributions}",
    eprint = "2409.08110",
    archivePrefix = "arXiv",
    primaryClass = "hep-ph",
    doi = "10.1103/PhysRevLett.134.121902",
    journal = "Phys. Rev. Lett.",
    volume = "134",
    number = "12",
    pages = "121902",
    year = "2025"
}

@article{Ma:1991xq,
    author = "Ma, Bo-Qiang",
    title = "{Melosh rotation: Source for the proton's missing spin}",
    eprint = "0711.2335",
    archivePrefix = "arXiv",
    primaryClass = "hep-ph",
    reportNumber = "BIHEP-TH-90-36",
    doi = "10.1088/0954-3899/17/5/001",
    journal = "J. Phys. G",
    volume = "17",
    pages = "L53--L58",
    year = "1991"
}

@article{AbdulKhalek:2021gbh,
    author = "Abdul Khalek, R. and others",
    title = "{Science Requirements and Detector Concepts for the Electron-Ion Collider}: {EIC Yellow Report}",
    eprint = "2103.05419",
    archivePrefix = "arXiv",
    primaryClass = "physics.ins-det",
    reportNumber = "BNL-220990-2021-FORE, JLAB-PHY-21-3198, LA-UR-21-20953",
    doi = "10.1016/j.nuclphysa.2022.122447",
    journal = "Nucl. Phys. A",
    volume = "1026",
    pages = "122447",
    year = "2022"
}

@article{Accardi:2009au,
    author = "Accardi, Alberto and Bacchetta, Alessandro and Melnitchouk, W. and Schlegel, Marc",
    title = "{What can break the Wandzura-Wilczek relation?}",
    eprint = "0907.2942",
    archivePrefix = "arXiv",
    primaryClass = "hep-ph",
    reportNumber = "JLAB-THY-09-1033",
    doi = "10.1088/1126-6708/2009/11/093",
    journal = "J. High Energy Phys.",
    volume = "11",
    number = "2009",
    pages = "093",
    year = "2009"
}

@article{Bacchetta:2017gcc,
    author = "Bacchetta, Alessandro and Delcarro, Filippo and Pisano, Cristian and Radici, Marco and Signori, Andrea",
    title = "{Extraction of partonic transverse momentum distributions from semi-inclusive deep-inelastic scattering, Drell-Yan and Z-boson production}",
    eprint = "1703.10157",
    archivePrefix = "arXiv",
    primaryClass = "hep-ph",
    reportNumber = "JLAB-THY-17-2437",
    doi = "10.1007/JHEP06(2017)081",
    journal = "J. High Energy Phys.",
    volume = "06",
    number = "2017",
    pages = "081",
    year = "2017",
    note = "[Erratum: JHEP 06, 051 (2019)]"
}

@article{Brodsky:2002cx,
    author = "Brodsky, Stanley J. and Hwang, Dae Sung and Schmidt, Ivan",
    title = "{Final state interactions and single spin asymmetries in semiinclusive deep inelastic scattering}",
    eprint = "hep-ph/0201296",
    archivePrefix = "arXiv",
    reportNumber = "SLAC-PUB-9135, USM-TH-121",
    doi = "10.1016/S0370-2693(02)01320-5",
    journal = "Phys. Lett. B",
    volume = "530",
    pages = "99--107",
    year = "2002"
}

@article{Bacchetta:2008af,
    author = "Bacchetta, Alessandro and Conti, Francesco and Radici, Marco",
    title = "{Transverse-momentum distributions in a diquark spectator model}",
    eprint = "0807.0323",
    archivePrefix = "arXiv",
    primaryClass = "hep-ph",
    reportNumber = "JLAB-THY-08-841",
    doi = "10.1103/PhysRevD.78.074010",
    journal = "Phys. Rev. D",
    volume = "78",
    pages = "074010",
    year = "2008"
}

@article{Avakian:2010br,
    author = "Avakian, H. and Efremov, A. V. and Schweitzer, P. and Yuan, F.",
    title = "{The transverse momentum dependent distribution functions in the bag model}",
    eprint = "1001.5467",
    archivePrefix = "arXiv",
    primaryClass = "hep-ph",
    reportNumber = "JLAB-PHY-10-1154",
    doi = "10.1103/PhysRevD.81.074035",
    journal = "Phys. Rev. D",
    volume = "81",
    pages = "074035",
    year = "2010"
}

@article{She:2009jq,
    author = "She, Jun and Zhu, Jiacai and Ma, Bo-Qiang",
    title = "{Pretzelosity $h_{1T}^\perp$ and quark orbital angular momentum}",
    eprint = "0902.3718",
    archivePrefix = "arXiv",
    primaryClass = "hep-ph",
    doi = "10.1103/PhysRevD.79.054008",
    journal = "Phys. Rev. D",
    volume = "79",
    pages = "054008",
    year = "2009"
}

@article{Deur:2018roz,
    author = "Deur, Alexandre and Brodsky, Stanley J. and De T\'eramond, Guy F.",
    title = "{The Spin Structure of the Nucleon}",
    eprint = "1807.05250",
    archivePrefix = "arXiv",
    primaryClass = "hep-ph",
    reportNumber = "JLAB-PHY-18-2760, SLAC-PUB-17279",
    doi = "10.1088/1361-6633/ab0b8f",
    journal = "Rept. Prog. Phys.",
    volume = "82",
    pages = "076201",
    year = "2019"
}

@article{Aidala:2012mv,
    author = "Aidala, Christine A. and Bass, Steven D. and Hasch, Delia and Mallot, Gerhard K.",
    title = "{The Spin Structure of the Nucleon}",
    eprint = "1209.2803",
    archivePrefix = "arXiv",
    primaryClass = "hep-ph",
    doi = "10.1103/RevModPhys.85.655",
    journal = "Rev. Mod. Phys.",
    volume = "85",
    pages = "655--691",
    year = "2013"
}

@article{Dulat:2015mca,
    author = "Dulat, Sayipjamal and Hou, Tie-Jiun and Gao, Jun and Guzzi, Marco and Huston, Joey and Nadolsky, Pavel and Pumplin, Jon and Schmidt, Carl and Stump, Daniel and Yuan, C. P.",
    title = "{New parton distribution functions from a global analysis of quantum chromodynamics}",
    eprint = "1506.07443",
    archivePrefix = "arXiv",
    primaryClass = "hep-ph",
    doi = "10.1103/PhysRevD.93.033006",
    journal = "Phys. Rev. D",
    volume = "93",
    number = "3",
    pages = "033006",
    year = "2016"
}

@article{Buckley:2014ana,
    author = {Buckley, Andy and Ferrando, James and Lloyd, Stephen and Nordstr\"om, Karl and Page, Ben and R\"ufenacht, Martin and Sch\"onherr, Marek and Watt, Graeme},
    title = "{LHAPDF6: parton density access in the LHC precision era}",
    eprint = "1412.7420",
    archivePrefix = "arXiv",
    primaryClass = "hep-ph",
    reportNumber = "GLAS-PPE-2014-05, MCNET-14-29, IPPP-14-111, DCPT-14-222",
    doi = "10.1140/epjc/s10052-015-3318-8",
    journal = "Eur. Phys. J. C",
    volume = "75",
    pages = "132",
    year = "2015"
}

@article{Butterworth:2015oua,
    author = "Butterworth, Jon and others",
    title = "{PDF4LHC recommendations for LHC Run II}",
    eprint = "1510.03865",
    archivePrefix = "arXiv",
    primaryClass = "hep-ph",
    reportNumber = "OUTP-15-17P, SMU-HEP-15-12, TIF-UNIMI-2015-14, LCTS-2015-27, CERN-PH-TH-2015-249",
    doi = "10.1088/0954-3899/43/2/023001",
    journal = "J. Phys. G",
    volume = "43",
    pages = "023001",
    year = "2016"
}

@article{Harland-Lang:2014zoa,
    author = "Harland-Lang, L. A. and Martin, A. D. and Motylinski, P. and Thorne, R. S.",
    title = "{Parton distributions in the LHC era: MMHT 2014 PDFs}",
    eprint = "1412.3989",
    archivePrefix = "arXiv",
    primaryClass = "hep-ph",
    reportNumber = "LCTS-2014-47, IPPP-14-97, DCPT-14-194",
    doi = "10.1140/epjc/s10052-015-3397-6",
    journal = "Eur. Phys. J. C",
    volume = "75",
    number = "5",
    pages = "204",
    year = "2015"
}

@article{H1:2015ubc,
    author = "Abramowicz, H. and others",
    collaboration = "H1 and ZEUS Collaborations",
    title = "{Combination of measurements of inclusive deep inelastic ${e^{\pm }p}$ scattering cross sections and QCD analysis of HERA data}",
    eprint = "1506.06042",
    archivePrefix = "arXiv",
    primaryClass = "hep-ex",
    reportNumber = "DESY-15-039",
    doi = "10.1140/epjc/s10052-015-3710-4",
    journal = "Eur. Phys. J. C",
    volume = "75",
    number = "12",
    pages = "580",
    year = "2015"
}

@article{NNPDF:2017mvq,
    author = "Ball, Richard D. and others",
    collaboration = "NNPDF Collaboration",
    title = "{Parton distributions from high-precision collider data}",
    eprint = "1706.00428",
    archivePrefix = "arXiv",
    primaryClass = "hep-ph",
    reportNumber = "EDINBURGH-2017-08, NIKHEF-2017-006, OUTP-17-04P, TIF-UNIMI-2017-3, CAVENDISH-HEP-17-06, CERN-TH-2017-077, Edinburgh 2017/08,
  Nikhef/2017-006, OUTP-17-04P,TIF-UNIMI-2017-3",
    doi = "10.1140/epjc/s10052-017-5199-5",
    journal = "Eur. Phys. J. C",
    volume = "77",
    number = "10",
    pages = "663",
    year = "2017"
}

@article{Bertone:2024taw,
    author = "Bertone, Valerio and Chiefa, Amedeo and Nocera, Emanuele R.",
    collaboration = "MAP Collaboration",
    title = "{Helicity-dependent parton distribution functions at next-to-next-to-leading order accuracy from inclusive and semi-inclusive deep-inelastic scattering data}",
    eprint = "2404.04712",
    archivePrefix = "arXiv",
    primaryClass = "hep-ph",
    doi = "10.1016/j.physletb.2025.139497",
    journal = "Phys. Lett. B",
    volume = "865",
    pages = "139497",
    year = "2025"
}

@article{Nocera:2014gqa,
    author = "Nocera, Emanuele R. and Ball, Richard D. and Forte, Stefano and Ridolfi, Giovanni and Rojo, Juan",
    collaboration = "NNPDF Collaboration",
    title = "{A first unbiased global determination of polarized PDFs and their uncertainties}",
    eprint = "1406.5539",
    archivePrefix = "arXiv",
    primaryClass = "hep-ph",
    reportNumber = "CERN-PH-TH-2014-106, IFUN-1028-FT, EDINBURGH-14-11, OUTP-14-06P",
    doi = "10.1016/j.nuclphysb.2014.08.008",
    journal = "Nucl. Phys. B",
    volume = "887",
    pages = "276--308",
    year = "2014"
}

@article{Ball:2013lla,
    author = "Ball, Richard D. and Forte, Stefano and Guffanti, Alberto and Nocera, Emanuele R. and Ridolfi, Giovanni and Rojo, Juan",
    collaboration = "NNPDF Collaboration",
    title = "{Unbiased determination of polarized parton distributions and their uncertainties}",
    eprint = "1303.7236",
    archivePrefix = "arXiv",
    primaryClass = "hep-ph",
    reportNumber = "CERN-PH-TH-2012-036, IFUM-987-FT, EDINBURGH-2013-04",
    doi = "10.1016/j.nuclphysb.2013.05.007",
    journal = "Nucl. Phys. B",
    volume = "874",
    pages = "36--84",
    year = "2013"
}

@article{deFlorian:2014yva,
    author = "de Florian, Daniel and Sassot, Rodolfo and Stratmann, Marco and Vogelsang, Werner",
    title = "{Evidence for polarization of gluons in the proton}",
    eprint = "1404.4293",
    archivePrefix = "arXiv",
    primaryClass = "hep-ph",
    doi = "10.1103/PhysRevLett.113.012001",
    journal = "Phys. Rev. Lett.",
    volume = "113",
    number = "1",
    pages = "012001",
    year = "2014"
}

@article{deFlorian:2008mr,
    author = "de Florian, Daniel and Sassot, Rodolfo and Stratmann, Marco and Vogelsang, Werner",
    title = "{Global Analysis of Helicity Parton Densities and Their Uncertainties}",
    eprint = "0804.0422",
    archivePrefix = "arXiv",
    primaryClass = "hep-ph",
    reportNumber = "BNL-NT-08-8",
    doi = "10.1103/PhysRevLett.101.072001",
    journal = "Phys. Rev. Lett.",
    volume = "101",
    pages = "072001",
    year = "2008"
}

@article{Borsa:2024mss,
    author = "Borsa, Ignacio and Stratmann, Marco and Vogelsang, Werner and de Florian, Daniel and Sassot, Rodolfo",
    title = "{Next-to-Next-to-Leading Order Global Analysis of Polarized Parton Distribution Functions}",
    eprint = "2407.11635",
    archivePrefix = "arXiv",
    primaryClass = "hep-ph",
    doi = "10.1103/PhysRevLett.133.151901",
    journal = "Phys. Rev. Lett.",
    volume = "133",
    number = "15",
    pages = "151901",
    year = "2024"
}

@article{Kanazawa:2015ajw,
    author = "Kanazawa, Koichi and Koike, Yuji and Metz, Andreas and Pitonyak, Daniel and Schlegel, Marc",
    title = "{Operator Constraints for Twist-3 Functions and Lorentz Invariance Properties of Twist-3 Observables}",
    eprint = "1512.07233",
    archivePrefix = "arXiv",
    primaryClass = "hep-ph",
    reportNumber = "RBRC-1164",
    doi = "10.1103/PhysRevD.93.054024",
    journal = "Phys. Rev. D",
    volume = "93",
    number = "5",
    pages = "054024",
    year = "2016"
}

@article{Avakian:2007mv,
    author = "Avakian, H. and Efremov, A. V. and Goeke, K. and Metz, A. and Schweitzer, P. and Teckentrup, T.",
    title = "{Are there approximate relations among transverse momentum dependent distribution functions?}",
    eprint = "0709.3253",
    archivePrefix = "arXiv",
    primaryClass = "hep-ph",
    reportNumber = "JLAB-PHY-07-736",
    doi = "10.1103/PhysRevD.77.014023",
    journal = "Phys. Rev. D",
    volume = "77",
    pages = "014023",
    year = "2008"
}

@article{Scimemi:2018mmi,
    author = "Scimemi, Ignazio and Vladimirov, Alexey",
    title = "{Matching of transverse momentum dependent distributions at twist-3}",
    eprint = "1804.08148",
    archivePrefix = "arXiv",
    primaryClass = "hep-ph",
    doi = "10.1140/epjc/s10052-018-6263-5",
    journal = "Eur. Phys. J. C",
    volume = "78",
    number = "10",
    pages = "802",
    year = "2018"
}

@article{Anderle:2021dpv,
    author = "Anderle, Daniele Paolo and Hou, Tie-Jiun and Xing, Hongxi and Yan, Mengshi and Yuan, C. -P. and Zhao, Yuxiang",
    title = "{Determining the helicity structure of the nucleon at the Electron Ion Collider in China}",
    eprint = "2103.10276",
    archivePrefix = "arXiv",
    primaryClass = "hep-ph",
    reportNumber = "JHEP08(2021)034",
    doi = "10.1007/JHEP08(2021)034",
    journal = "J. High Energy Phys.",
    volume = "08",
    number = "2021",
    pages = "034",
    year = "2021"
}

@article{Anderle:2021wcy,
    author = "Anderle, Daniele P. and others",
    title = "{Electron-ion collider in China}",
    eprint = "2102.09222",
    archivePrefix = "arXiv",
    primaryClass = "nucl-ex",
    reportNumber = "Frontiers of Physics, Volume 16 Issue (6):64701, 2021",
    doi = "10.1007/s11467-021-1062-0",
    journal = "Front. Phys. (Beijing)",
    volume = "16",
    number = "6",
    pages = "64701",
    year = "2021"
}

@article{Ma:1992sj,
    author = "Ma, Bo-Qiang and Zhang, Qi-Ren",
    title = "{The Proton spin and the Wigner rotation}",
    eprint = "hep-ph/9306241",
    archivePrefix = "arXiv",
    reportNumber = "BIHEP-TH-91-9",
    doi = "10.1007/BF01557707",
    journal = "Z. Phys. C",
    volume = "58",
    pages = "479--482",
    year = "1993"
}

@article{Melosh:1974cu,
    author = "Melosh, H. J.",
    title = "{Quarks: Currents and constituents}",
    doi = "10.1103/PhysRevD.9.1095",
    journal = "Phys. Rev. D",
    volume = "9",
    pages = "1095",
    year = "1974"
}

@article{Bacchetta:2024yzl,
    author = "Bacchetta, Alessandro and Bongallino, Alessia and Cerutti, Matteo and Radici, Marco and Rossi, Lorenzo",
    collaboration = "MAP Collaboration",
    title = "{Exploring the Three-Dimensional Momentum Distribution of Longitudinally Polarized Quarks in the Proton}",
    eprint = "2409.18078",
    archivePrefix = "arXiv",
    primaryClass = "hep-ph",
    reportNumber = "JLAB-THY-24-4204",
    doi = "10.1103/PhysRevLett.134.121901",
    journal = "Phys. Rev. Lett.",
    volume = "134",
    number = "12",
    pages = "121901",
    year = "2025"
}

@article{Moos:2023yfa,
    author = "Moos, Valentin and Scimemi, Ignazio and Vladimirov, Alexey and Zurita, Pia",
    title = "{Extraction of unpolarized transverse momentum distributions from the fit of Drell-Yan data at N$^{4}$LL}",
    eprint = "2305.07473",
    archivePrefix = "arXiv",
    primaryClass = "hep-ph",
    reportNumber = "IPARCOS-UCM-035",
    doi = "10.1007/JHEP05(2024)036",
    journal = "J. High Energy Phys.",
    volume = "05",
    number = "2024",
    pages = "036",
    year = "2024"
}

@article{Bacchetta:2024qre,
    author = "Bacchetta, Alessandro and Bertone, Valerio and Bissolotti, Chiara and Bozzi, Giuseppe and Cerutti, Matteo and Delcarro, Filippo and Radici, Marco and Rossi, Lorenzo and Signori, Andrea",
    collaboration = "MAP Collaboration",
    title = "{Flavor dependence of unpolarized quark transverse momentum distributions from a global fit}",
    eprint = "2405.13833",
    archivePrefix = "arXiv",
    primaryClass = "hep-ph",
    reportNumber = "JLAB-THY-24-4066",
    doi = "10.1007/JHEP08(2024)232",
    journal = "J. High Energy Phys.",
    volume = "08",
    number = "2024",
    pages = "232",
    year = "2024"
}

@book{Collins:2011zzd,
    author = "Collins, John",
    title = "{Foundations of Perturbative QCD}",
    doi = "10.1017/9781009401845",
    isbn = "978-1-009-40184-5, 978-1-009-40183-8, 978-1-009-40182-1",
    publisher = "Cambridge University Press",
    volume = "32",
    year = "2011"
}

@article{Zeng:2022lbo,
    author = "Zeng, Chunhua and Liu, Tianbo and Sun, Peng and Zhao, Yuxiang",
    title = "{Toward three-dimensional nucleon structures at the Electron-Ion Collider in China: A study of the Sivers function}",
    eprint = "2208.14620",
    archivePrefix = "arXiv",
    primaryClass = "hep-ph",
    doi = "10.1103/PhysRevD.106.094039",
    journal = "Phys. Rev. D",
    volume = "106",
    number = "9",
    pages = "094039",
    year = "2022"
}

@article{Bacchetta:2004jz,
    author = "Bacchetta, Alessandro and D'Alesio, Umberto and Diehl, Markus and Miller, C. Andy",
    title = "{Single-spin asymmetries: The Trento conventions}",
    eprint = "hep-ph/0410050",
    archivePrefix = "arXiv",
    reportNumber = "DESY-04-193",
    doi = "10.1103/PhysRevD.70.117504",
    journal = "Phys. Rev. D",
    volume = "70",
    pages = "117504",
    year = "2004"
}

@article{Huang:2015vpy,
    author = "Huang, Jin and Kang, Zhong-Bo and Vitev, Ivan and Xing, Hongxi",
    title = "{Spin asymmetries for vector boson production in polarized p+p collisions}",
    eprint = "1511.06764",
    archivePrefix = "arXiv",
    primaryClass = "hep-ph",
    doi = "10.1103/PhysRevD.93.014036",
    journal = "Phys. Rev. D",
    volume = "93",
    number = "1",
    pages = "014036",
    year = "2016"
}

@article{CDF:2003tlc,
    author = "Acosta, D. and others",
    collaboration = "CDF Collaboration",
    title = "{Measurement of the Polar-Angle Distribution of Leptons from $W$ Boson Decay as a Function of the W Transverse Momentum in $p\bar{p}$ Collisions at $\sqrt{s} = 1.8$ TeV}",
    eprint = "hep-ex/0311050",
    archivePrefix = "arXiv",
    reportNumber = "FERMILAB-PUB-03-381-E",
    primaryClass = "hep-ex",
    doi = "10.1103/PhysRevD.70.032004",
    journal = "Phys. Rev. D",
    volume = "70",
    pages = "032004",
    year = "2004"
}

@article{Collins:1977iv,
    author = "Collins, John C. and Soper, Davison E.",
    title = "{Angular Distribution of Dileptons in High-Energy Hadron Collisions}",
    reportNumber = "Print-77-0288 (PRINCETON)",
    doi = "10.1103/PhysRevD.16.2219",
    journal = "Phys. Rev. D",
    volume = "16",
    pages = "2219",
    year = "1977"
}

@article{Bury:2021sue,
    author = "Bury, Marcin and Prokudin, Alexei and Vladimirov, Alexey",
    title = "{Extraction of the Sivers function from SIDIS, Drell-Yan, and $W^\pm/Z$ boson production data with TMD evolution}",
    eprint = "2103.03270",
    archivePrefix = "arXiv",
    primaryClass = "hep-ph",
    reportNumber = "JLAB-THY-21-3325",
    doi = "10.1007/JHEP05(2021)151",
    journal = "J. High Energy Phys.",
    volume = "05",
    number = "2021",
    pages = "151",
    year = "2021"
}

@article{Yang:2024bfz,
    author = "Yang, Ke and Liu, Tianbo and Sun, Peng and Zhao, Yuxiang and Ma, Bo-Qiang",
    title = "{Extraction of transhelicity worm-gear distributions and opportunities at the Electron-Ion Collider in China}",
    eprint = "2403.12795",
    archivePrefix = "arXiv",
    primaryClass = "hep-ph",
    doi = "10.1103/PhysRevD.110.034036",
    journal = "Phys. Rev. D",
    volume = "110",
    number = "3",
    pages = "034036",
    year = "2024"
}

@article{Zeng:2023nnb,
    author = "Zeng, Chunhua and Dong, Hongxin and Liu, Tianbo and Sun, Peng and Zhao, Yuxiang",
    title = "{Role of sea quarks in the nucleon transverse spin}",
    eprint = "2310.15532",
    archivePrefix = "arXiv",
    primaryClass = "hep-ph",
    doi = "10.1103/PhysRevD.109.056002",
    journal = "Phys. Rev. D",
    volume = "109",
    number = "5",
    pages = "056002",
    year = "2024"
}

@article{Moos:2025sal,
    author = "Moos, Valentin and Scimemi, Ignazio and Vladimirov, Alexey and Zurita, Pia",
    title = "{Determination of unpolarized TMD distributions from the fit of Drell-Yan and SIDIS data at N$^4$LL}",
    eprint = "2503.11201",
    archivePrefix = "arXiv",
    primaryClass = "hep-ph",
    reportNumber = "IPARCOS-UCM-25-018",
    doi = "10.1007/JHEP11(2025)134",
    journal = "J. High Energy Phys.",
    volume = "11",
    number = "2025",
    pages = "134",
    year = "2025"
}

@article{Bacchetta:2022awv,
    author = "Bacchetta, Alessandro and Bertone, Valerio and Bissolotti, Chiara and Bozzi, Giuseppe and Cerutti, Matteo and Piacenza, Fulvio and Radici, Marco and Signori, Andrea",
    collaboration = "MAP Collaboration",
    title = "{Unpolarized transverse momentum distributions from a global fit of Drell-Yan and semi-inclusive deep-inelastic scattering data}",
    eprint = "2206.07598",
    archivePrefix = "arXiv",
    primaryClass = "hep-ph",
    doi = "10.1007/JHEP10(2022)127",
    journal = "J. High Energy Phys.",
    volume = "10",
    number = "2022",
    pages = "127",
    year = "2022"
}

@article{Bacchetta:2019sam,
    author = "Bacchetta, Alessandro and Bertone, Valerio and Bissolotti, Chiara and Bozzi, Giuseppe and Delcarro, Filippo and Piacenza, Fulvio and Radici, Marco",
    title = "{Transverse-momentum-dependent parton distributions up to N$^{3}$LL from Drell-Yan data}",
    eprint = "1912.07550",
    archivePrefix = "arXiv",
    primaryClass = "hep-ph",
    reportNumber = "JLAB-THY-19-3121",
    doi = "10.1007/JHEP07(2020)117",
    journal = "J. High Energy Phys.",
    volume = "07",
    number = "2020",
    pages = "117",
    year = "2020"
}

@article{deFlorian:2007ekg,
    author = "de Florian, Daniel and Sassot, Rodolfo and Stratmann, Marco",
    title = "{Global analysis of fragmentation functions for protons and charged hadrons}",
    eprint = "0707.1506",
    archivePrefix = "arXiv",
    primaryClass = "hep-ph",
    doi = "10.1103/PhysRevD.76.074033",
    journal = "Phys. Rev. D",
    volume = "76",
    pages = "074033",
    year = "2007"
}

@article{deFlorian:2017lwf,
    author = "de Florian, D. and Epele, M. and Hernandez-Pinto, R. J. and Sassot, R. and Stratmann, M.",
    title = "{Parton-to-Kaon Fragmentation Revisited}",
    eprint = "1702.06353",
    archivePrefix = "arXiv",
    primaryClass = "hep-ph",
    doi = "10.1103/PhysRevD.95.094019",
    journal = "Phys. Rev. D",
    volume = "95",
    number = "9",
    pages = "094019",
    year = "2017"
}

@article{deFlorian:2014xna,
    author = "de Florian, Daniel and Sassot, R. and Epele, Manuel and Hern\'andez-Pinto, Roger J. and Stratmann, Marco",
    title = "{Parton-to-Pion Fragmentation Reloaded}",
    eprint = "1410.6027",
    archivePrefix = "arXiv",
    primaryClass = "hep-ph",
    doi = "10.1103/PhysRevD.91.014035",
    journal = "Phys. Rev. D",
    volume = "91",
    number = "1",
    pages = "014035",
    year = "2015"
}

@article{Wigner:1939cj,
    author = "Wigner, Eugene P.",
    editor = "Kim, Y. S. and Zachary, W. W.",
    title = "{On Unitary Representations of the Inhomogeneous Lorentz Group}",
    eprint = "",
    archivePrefix = "",
    doi = "10.2307/1968551",
    journal = "Annals Math.",
    volume = "40",
    pages = "149--204",
    year = "1939"
}

@article{Collins:1981va,
    author = "Collins, John C. and Soper, Davison E.",
    title = "{Back-To-Back Jets: Fourier Transform from B to K-Transverse}",
    reportNumber = "OITS-153",
    doi = "10.1016/0550-3213(82)90453-9",
    journal = "Nucl. Phys. B",
    volume = "197",
    pages = "446--476",
    year = "1982"
}

@article{Boussarie:2023izj,
    author = "Boussarie, Renaud and others",
    title = "{TMD Handbook}",
    eprint = "2304.03302",
    archivePrefix = "arXiv",
    primaryClass = "hep-ph",
    reportNumber = "JLAB-THY-23-3780, LA-UR-21-20798, MIT-CTP/5386",
    journal = "",
    volume = "",
    pages = "",
    month = "4",
    year = "2023"
}

@article{Bacchetta:2006tn,
    author = "Bacchetta, Alessandro and Diehl, Markus and Goeke, Klaus and Metz, Andreas and Mulders, Piet J. and Schlegel, Marc",
    title = "{Semi-inclusive deep inelastic scattering at small transverse momentum}",
    eprint = "hep-ph/0611265",
    archivePrefix = "arXiv",
    reportNumber = "DESY-06-204",
    doi = "10.1088/1126-6708/2007/02/093",
    journal = "J. High Energy Phys.",
    volume = "02",
    number = "2007",
    pages = "093",
    year = "2007"
}

@article{Wiringa:1994wb,
    author = "Wiringa, Robert B. and Stoks, V. G. J. and Schiavilla, R.",
    title = "{An Accurate nucleon-nucleon potential with charge independence breaking}",
    eprint = "nucl-th/9408016",
    archivePrefix = "arXiv",
    reportNumber = "PHY-7742-TH-94, CEBAF-TH-94-19",
    doi = "10.1103/PhysRevC.51.38",
    journal = "Phys. Rev. C",
    volume = "51",
    pages = "38--51",
    year = "1995"
}

@article{CLAS:2017yrm,
    author = "Jawalkar, S. and others",
    collaboration = "CLAS Collaboration",
    title = "{Semi-Inclusive $\pi_0$ target and beam-target asymmetries from 6 GeV electron scattering with CLAS}",
    eprint = "1709.10054",
    archivePrefix = "arXiv",
    primaryClass = "nucl-ex",
    reportNumber = "JLAB-PHY-17-2567",
    doi = "10.1016/j.physletb.2018.06.014",
    journal = "Phys. Lett. B",
    volume = "782",
    pages = "662--667",
    year = "2018"
}

@article{COMPASS:2016klq,
    author = "Adolph, C. and others",
    collaboration = "COMPASS Collaboration",
    title = "{Azimuthal asymmetries of charged hadrons produced in high-energy muon scattering off longitudinally polarised deuterons}",
    eprint = "1609.06062",
    archivePrefix = "arXiv",
    primaryClass = "hep-ex",
    reportNumber = "CERN-EP-2016-245",
    doi = "10.1140/epjc/s10052-018-6379-7",
    journal = "Eur. Phys. J. C",
    volume = "78",
    number = "11",
    pages = "952",
    year = "2018",
    note = "[Erratum: Eur.Phys.J.C 80, 298 (2020)]"
}

@article{Scimemi:2019cmh,
    author = "Scimemi, Ignazio and Vladimirov, Alexey",
    title = "{Non-perturbative structure of semi-inclusive deep-inelastic and Drell-Yan scattering at small transverse momentum}",
    eprint = "1912.06532",
    archivePrefix = "arXiv",
    primaryClass = "hep-ph",
    doi = "10.1007/JHEP06(2020)137",
    journal = "J. High Energy Phys.",
    volume = "06",
    number = "2020",
    pages = "137",
    year = "2020"
}

@article{Bertone:2019nxa,
    author = "Bertone, Valerio and Scimemi, Ignazio and Vladimirov, Alexey",
    title = "{Extraction of unpolarized quark transverse momentum dependent parton distributions from Drell-Yan/Z-boson production}",
    eprint = "1902.08474",
    archivePrefix = "arXiv",
    primaryClass = "hep-ph",
    doi = "10.1007/JHEP06(2019)028",
    journal = "J. High Energy Phys.",
    volume = "06",
    number = "2019",
    pages = "028",
    year = "2019"
}

@article{Ebert:2022cku,
    author = "Ebert, Markus A. and Michel, Johannes K. L. and Stewart, Iain W. and Sun, Zhiquan",
    title = "{Disentangling long and short distances in momentum-space TMDs}",
    eprint = "2201.07237",
    archivePrefix = "arXiv",
    primaryClass = "hep-ph",
    reportNumber = "MIT--CTP 5391, MPP--2021--217",
    doi = "10.1007/JHEP07(2022)129",
    journal = "J. High Energy Phys.",
    volume = "07",
    number = "2022",
    pages = "129",
    year = "2022"
}

@article{Du:2024sjt,
    author = "Du, Yong",
    title = "{Parity violation on longitudinal single-spin asymmetries at the EicC}",
    eprint = "2412.20469",
    archivePrefix = "arXiv",
    primaryClass = "hep-ph",
    doi = "10.1103/b8tr-9mcw",
    journal = "Phys. Rev. D",
    volume = "111",
    number = "11",
    pages = "116026",
    year = "2025"
}

@article{Huang:2025wdq,
    author = "Huang, Yuan-Yuan and Cao, Xu and Feng, Taifu and Kumeri{\v{c}}ki, Kre{\v{s}}imir and Lu, Yu",
    title = "{Assessing the impact of the Electron-Ion Collider in China on deeply virtual Compton scattering*}",
    eprint = "2512.19145",
    archivePrefix = "arXiv",
    primaryClass = "hep-ph",
    doi = "10.1088/1674-1137/ae457e",
    journal = "Chin. Phys. C",
    volume = "50",
    number = "5",
    pages = "053110",
    year = "2026"
}

@article{Xiao:2026tbs,
    author = "Xiao, Bo-Wen and Zhao, Yuxiang and Zhou, Jian",
    title = "{Physics of the Electron{\textendash}Ion Collider in China}",
    eprint = "2608.11712",
    archivePrefix = "arXiv",
    primaryClass = "hep-ph",
    doi = "10.1016/j.ppnp.2026.104264",
    journal = "Prog. Part. Nucl. Phys.",
    volume = "151",
    pages = "104264",
    year = "2026"
}

@inproceedings{Burkert:2008rj,
    author = "Burkert, Volker D.",
    title = "{CLAS12 and its initial Science Program at the Jefferson Lab Upgrade}",
    booktitle = "{CLAS 12 RICH Detector Workshop}",
    eprint = "0810.4718",
    archivePrefix = "arXiv",
    primaryClass = "hep-ph",
    reportNumber = "JLAB-PHY-08-884",
    month = "10",
    year = "2008"
}

@article{CLAS:2021opg,
    author = "Diehl, S. and others",
    collaboration = "CLAS Collaboration",
    title = "{Multidimensional, High Precision Measurements of Beam Single Spin Asymmetries in Semi-inclusive $\pi^{+}$~Electroproduction off Protons in the Valence Region}",
    eprint = "2101.03544",
    archivePrefix = "arXiv",
    primaryClass = "hep-ex",
    reportNumber = "JLAB-PHY-21-3259",
    doi = "10.1103/PhysRevLett.128.062005",
    journal = "Phys. Rev. Lett.",
    volume = "128",
    number = "6",
    pages = "062005",
    year = "2022"
}

@article{STAR:2018fty,
    author = "Adam, Jaroslav and others",
    collaboration = "STAR Collaboration",
    title = "{Measurement of the longitudinal spin asymmetries for weak boson production in proton-proton collisions at $\sqrt{s}$ = 510 GeV}",
    eprint = "1812.04817",
    archivePrefix = "arXiv",
    primaryClass = "hep-ex",
    doi = "10.1103/PhysRevD.99.051102",
    journal = "Phys. Rev. D",
    volume = "99",
    number = "5",
    pages = "051102",
    year = "2019"
}

@article{STAR:2010xwx,
    author = "Aggarwal, M. M. and others",
    collaboration = "STAR Collaboration",
    title = "{Measurement of the parity-violating longitudinal single-spin asymmetry for $W^{\pm}$ boson production in polarized proton-proton collisions at $\sqrt{s} = 500 GeV$}",
    eprint = "1009.0326",
    archivePrefix = "arXiv",
    primaryClass = "hep-ex",
    doi = "10.1103/PhysRevLett.106.062002",
    journal = "Phys. Rev. Lett.",
    volume = "106",
    pages = "062002",
    year = "2011"
}

@article{Curci:1980uw,
    author = "Curci, G. and Furmanski, W. and Petronzio, R.",
    title = "{Evolution of Parton Densities Beyond Leading Order: The Nonsinglet Case}",
    reportNumber = "CERN-TH-2815",
    doi = "10.1016/0550-3213(80)90003-6",
    journal = "Nucl. Phys. B",
    volume = "175",
    pages = "27--92",
    year = "1980"
}

@article{Furmanski:1980cm,
    author = "Furmanski, W. and Petronzio, R.",
    title = "{Singlet Parton Densities Beyond Leading Order}",
    reportNumber = "CERN-TH-2933",
    doi = "10.1016/0370-2693(80)90636-X",
    journal = "Phys. Lett. B",
    volume = "97",
    pages = "437--442",
    year = "1980"
}

@article{Echevarria:2016scs,
    author = "Echevarria, Miguel G. and Scimemi, Ignazio and Vladimirov, Alexey",
    title = "{Unpolarized Transverse Momentum Dependent Parton Distribution and Fragmentation Functions at next-to-next-to-leading order}",
    eprint = "1604.07869",
    archivePrefix = "arXiv",
    primaryClass = "hep-ph",
    doi = "10.1007/JHEP09(2016)004",
    journal = "J. High Energy Phys.",
    volume = "09",
    number = "2016",
    pages = "004",
    year = "2016"
}

@article{Stratmann:1996hn,
    author = "Stratmann, M. and Vogelsang, W.",
    title = "{Next-to-leading order evolution of polarized and unpolarized fragmentation functions}",
    eprint = "hep-ph/9612250",
    archivePrefix = "arXiv",
    reportNumber = "DO-TH-96-23, RAL-TR-96-097",
    doi = "10.1016/S0550-3213(97)00182-X",
    journal = "Nucl. Phys. B",
    volume = "496",
    pages = "41--65",
    year = "1997"
}

@article{Mertig:1995ny,
    author = "Mertig, R. and van Neerven, W. L.",
    title = "{The Calculation of the two loop spin splitting functions $P_{ij}^{(1)}(x)$}",
    eprint = "hep-ph/9506451",
    archivePrefix = "arXiv",
    reportNumber = "INLO-PUB-6-95, NIKHEF-H-95-031",
    doi = "10.1007/s002880050138",
    journal = "Z. Phys. C",
    volume = "70",
    pages = "637--654",
    year = "1996"
}

@article{Zhu:2025gts,
    author = "Zhu, Yu Jiao",
    title = "{The N$^{3}$LO twist-2 matching of helicity TMDs and SIDIS q$_{*}$ spectrum}",
    eprint = "2509.01655",
    archivePrefix = "arXiv",
    primaryClass = "hep-ph",
    reportNumber = "MPP-2025-159",
    doi = "10.1007/JHEP02(2026)115",
    journal = "J. High Energy Phys.",
    volume = "02",
    number = "2026",
    pages = "115",
    year = "2026"
}

@article{Bollweg:2025ecn,
    author = "Bollweg, Dennis and Gao, Xiang and Mukherjee, Swagato and Zhao, Yong",
    title = "{Lattice QCD Benchmark of Proton Helicity and Flavor-Dependent Unpolarized Transverse Momentum-Dependent Parton Distribution Functions at Physical Quark Masses}",
    eprint = "2505.18430",
    archivePrefix = "arXiv",
    primaryClass = "hep-lat",
    doi = "10.1103/tb2w-3tks",
    journal = "Phys. Rev. Lett.",
    volume = "135",
    number = "20",
    pages = "201901",
    year = "2025"
}

@article{LatticePartonCollaborationLPC:2022myp,
    author = {He, Jin-Chen and Chu, Min-Huan and Hua, Jun and Ji, Xiangdong and Sch{\"a}fer, Andreas and Su, Yushan and Wang, Wei and Yang, Yi-Bo and Zhang, Jian-Hui and Zhang, Qi-An},
    collaboration = "Lattice Parton Collaboration (LPC)",
    title = "{Unpolarized transverse momentum dependent parton distributions of the nucleon from lattice QCD}",
    eprint = "2211.02340",
    archivePrefix = "arXiv",
    primaryClass = "hep-lat",
    doi = "10.1103/PhysRevD.109.114513",
    journal = "Phys. Rev. D",
    volume = "109",
    number = "11",
    pages = "114513",
    year = "2024"
}

@article{Avkhadiev:2023poz,
    author = "Avkhadiev, Artur and Shanahan, Phiala E. and Wagman, Michael L. and Zhao, Yong",
    title = "{Collins-Soper kernel from lattice QCD at the physical pion mass}",
    eprint = "2307.12359",
    archivePrefix = "arXiv",
    primaryClass = "hep-lat",
    reportNumber = "MIT-CTP/5587, FERMILAB-PUB-23-375-T",
    doi = "10.1103/PhysRevD.108.114505",
    journal = "Phys. Rev. D",
    volume = "108",
    number = "11",
    pages = "114505",
    year = "2023"
}

@article{Yoon:2017qzo,
    author = {Yoon, Boram and Engelhardt, Michael and Gupta, Rajan and Bhattacharya, Tanmoy and Green, Jeremy R. and Musch, Bernhard U. and Negele, John W. and Pochinsky, Andrew V. and Sch{\"a}fer, Andreas and Syritsyn, Sergey N.},
    title = "{Nucleon Transverse Momentum-dependent Parton Distributions in Lattice QCD: Renormalization Patterns and Discretization Effects}",
    eprint = "1706.03406",
    archivePrefix = "arXiv",
    primaryClass = "hep-lat",
    reportNumber = "LA-UR-17-24472",
    doi = "10.1103/PhysRevD.96.094508",
    journal = "Phys. Rev. D",
    volume = "96",
    number = "9",
    pages = "094508",
    year = "2017"
}
\end{document}